\documentstyle[preprint,aps,epsf,amssymb,amsbsy,amstext,subeqnarray,amsfonts]{revtex}
\draft
\tighten

\begin{document}

\def\Journal#1#2#3#4{{#1} {\bf #2}, #3 (#4)}

\def\NCA{\em Nuovo Cimento}
\def\NIM{\em Nucl. Instrum. Methods}
\def\NIMA{{\em Nucl. Instrum. Methods} A}
\def\NPB{{\em Nucl. Phys.} B}
\def\PLB{{\em Phys. Lett.}  B}
\def\PRL{\em Phys. Rev. Lett.}
\def\PRD{{\em Phys. Rev.} D}
\def\ZPC{{\em Z. Phys.} C}
\def\SJNP{{\em Sov.J.Nucl.Phys.}}
\def\PPNP{{\em Prog.Part.Nucl.Phys.}}
\def\PREP{{\em Phys. Rep.}}

\title{QED radiative corrections to virtual Compton scattering}
\author{M. Vanderhaeghen$^{1}$, J.M. Friedrich$^{1}$, 
D. Lhuillier$^{2}$, D. Marchand$^{3}$, 
L. Van Hoorebeke$^{4}$ and J. Van de Wiele$^{3}$.}
\address{\it $^1$ Institut f\"ur Kernphysik, Johannes Gutenberg Universit\"at, D-55099 Mainz, Germany}
\address{\it $^2$ CEA/Saclay, DAPNIA/SPhN, F-91191 Gif-sur-Yvette, France}
\address{\it $^3$ Institut de Physique Nucl\'eaire, F-91406 Orsay, France}
\address{\it $^4$ FWO-Vlaanderen, RUG, Proeftuinstraat 86,
  B-9000 Gent, Belgium}
\date{\today}

\maketitle

\begin{abstract}
The QED radiative corrections to virtual Compton scattering 
(reaction $e p \to e p \gamma$) are calculated 
to first order in $\alpha_{em} \equiv e^2 / 4 \pi$. 
A detailed study is presented for the one-loop virtual corrections and
for the first order soft-photon emission contributions. 
Furthermore, a full numerical
calculation is given for the radiative tail, corresponding with photon
emission processes, where the photon energy is not very small compared
with the lepton momenta. We compare our results with existing works on
elastic electron-proton scattering, and show for the $e p \to e p \gamma$
reaction how the observables are modified due to these first order QED
radiative corrections. We show results for both unpolarized and
polarized observables of the virtual Compton scattering in the low
energy region (where one is sensitive to the generalized
polarizabilities of the nucleon), as well as for the deeply virtual Compton
scattering. 
 
\end{abstract}


\newpage

\section{Introduction}

Virtual Compton scattering (VCS) 
has become in recent years a new and versatile tool in the study of nucleon
structure and has triggered an important activity on both the 
theoretical and experimental side (see e.g. \cite{clermont,vcsreview}).
VCS, which is accessed through the  
$(e, e' \gamma)$ reaction, is studied now in various kinematical domains. 
\newline
\indent
At low energy, below pion production threshold, it allows to access
generalized polarizabilities of the proton
\cite{Guichon95,Drechsel98}. These response functions, which
constitute new nucleon structure observables, have been calculated in
various approaches and models 
\cite{Guichon95,MVdh96,metz97,HHKS97,PS98,HHKD99}. 
To extract this nucleon structure information from VCS below pion production 
threshold, a considerable experimental effort  is taking place 
at various electron laboratories. The first few events of VCS 
were observed in \cite{slac}. 
The first dedicated VCS experiment has been
performed at MAMI and for the first time, two combinations of
generalized polarizabilities have been determined at a four-momentum 
squared $Q^2$ = 0.33 GeV$^2$ \cite{mami}. An experiment at
higher $Q^2$ (1 - 2 GeV$^2$) 
at JLab \cite{jlab} has already been performed, which is under
analysis at the time of writing, and a further experiment at lower
$Q^2$ is planned at MIT-Bates \cite{bates}. 
\newline
\indent
The VCS is also studied vigorously in the Bjorken regime (where the  
photon virtuality $Q^2$ and the photon-proton c.m. energy $\sqrt{s}$ 
are both large, with $Q^2 / s$ finite), which
is referred to as deeply virtual Compton scattering (DVCS).  
In this region, the DVCS amplitude is parametrized 
at leading order in $Q$ in terms of six 
generalized parton distributions \cite{Ji97,Rady96,Jireview}, 
commonly denoted as skewed parton distributions (4 quark helicity
conserving functions and 2 which involve a quark helicity flip). 
This field has generated by now a whole theoretical industry, 
and first experiments of DVCS and related hard electroproduction 
reactions are being performed, analyzed 
or planned both at JLab \cite{bertindvcs}, HERMES/HERA
\cite{ryckbosch}, and COMPASS \cite{loicompass}. 
\newline
\indent
The first absolute measurement of the VCS cross section on the nucleon 
performed at MAMI \cite{mami}, indicates that QED radiative corrections 
provide an important contribution to the 
$e p \rightarrow e p \gamma$ reaction (of the order of 20\% in
the kinematics considered in \cite{mami}). 
The $e p \rightarrow e p \gamma$ 
reaction is particular in comparison with other
electron scattering reactions because the photon can be emitted from both
the proton side (this is the VCS process which contains the nucleon structure
information of interest) or can be emitted from one of the electrons 
(which is the parasitic Bethe-Heitler process). 
The radiative corrections to the Bethe-Heitler process are formally
different compared with the case of electron scattering. 
The importance of a very good understanding of the radiative corrections 
is indispensable if one wants to extract nucleon structure information 
from the $e p \to e p \gamma$ reaction, especially in those
kinematical situations where the Bethe-Heitler process is not negligible. 
The calculation of these QED radiative corrections to the 
$e p \rightarrow e p \gamma$ reaction 
to first order in $\alpha_{em} \equiv e^2 / 4 \pi \approx 1 /
137.036$, is the subject of this paper.  
\newline
\indent
Radiative corrections were first  
calculated by {\it Schwinger} for potential scattering \cite{schwinger49}.  
{\it Tsai} \cite{Tsai61} extented the calculation 
of the radiative corrections to electron-proton scattering.
The field has a long history and we refer to   
the standard review papers \cite{mo69,maximon69}, 
which were used in the interpretation of many 
electron scattering experiments.
\newline
\indent
The outline of the present paper is as follows. In section
\ref{sec:lowest}, we introduce the kinematics and notations used for
the $e p \to e p \gamma$ reaction, and give the lowest order
amplitudes. 
\newline
\indent
In section \ref{sec:first}, we give the first order QED radiative
corrections to the $e p \to e p \gamma$ reaction. We first calculate, 
in section~\ref{sec:virtual}, the one-loop virtual radiative
corrections originating from the lepton side, to the $e p \to e p
\gamma$ reaction. Our strategy used to evaluate the rather complicated
loop integrals, is to solve first simpler loop integrals, which
contain entirely the ultraviolet (UV) and infrared (IR) divergences,
and in which the lowest order amplitudes factorize. 
These simpler loop integrals are evaluated analytically. The finite
remainder with respect to the original amplitude, is then expressed through
Feynman parameter integrals, which are calculated numerically in this
work. 
\newline
\indent
In section~\ref{sec:real}, we calculate the soft photon emission
contributions from the lepton side, to the $e p \to e p \gamma$
reaction. We discuss the similarities and differences with the bremsstrahlung
contribution to elastic electron-nucleon scattering.  
These bremsstrahlung processes contain IR divergences which are shown
to cancel exactly the IR divergences from the virtual photon processes.
\newline
\indent
In section~\ref{sec:int}, the numerical method to evaluate the
remaining finite Feynman parameter integrals is presented. We discuss
subsequently the cases where the integrand is regular or singular, the
latter originating from the propagation of on-shell intermediate
states in the one-loop corrections to the 
$e p \to e p \gamma$ reaction. In particular, we discuss
the different numerical checks performed and the accuracy of the
calculation. 
\newline
\indent
In section~\ref{sec:protonside}, we discuss the radiative corrections
at the proton side and the two-photon exchange corrections, by
referring to the recent work of {\it Maximon and Tjon} \cite{MTj99}. 
\newline
\indent
In section~\ref{sec:radtail}, we give a full numerical
calculation for elastic electron-proton scattering 
of the photon emission processes where the photon energy
is not very small compared with the lepton momenta, and which makes
up the radiative tail. We compare this full calculation with an
approximate procedure based on the angular peaking approximation, and
show to what extent the full calculation validates the approximate
method for the case of elastic electron-nucleon scattering. The
approximate method will be seen to be realistic enough to apply it
next to the calculation of the radiative tail in the case of VCS. 
\newline
\indent
In section~\ref{sec:results}, we start by briefly discussing the
radiative corrections to elastic electron-proton scattering. We apply the
radiative corrections to elastic scattering data on the proton. 
We next give our results for the $e p \to e p
\gamma$ reaction, and indicate how the observables are modified due to
the first order QED radiative corrections. We discuss first the
polarizability region for the $e p \to e p \gamma$ reaction,
corresponding with a low outgoing photon energy. We show results for
both unpolarized and polarized cross sections in MAMI and JLab
kinematics. Subsequently, we give the effect of the first order QED
radiative corrections to the DVCS cross section and the electron single
spin asymmetry. 
\newline
\indent
Finally, we give our conclusions in section~\ref{sec:conclusion}. 
\newline
\indent
We present technical details needed in the calculations, in two
appendices. In appendix~\ref{app:elast}, we 
calculate the radiative corrections to elastic lepton-nucleon
scattering, which serves as a point of comparison with the $e p
\to e p \gamma$ reaction. In particular, 
we present the details of the calculation of
the soft photon emission contributions, and perform analytically the 
phase space integral over the soft photon in an exact way. We compare
with other calculations in the literature. 
In appendix~\ref{app:b}, we present some technical details on the 
integration method used to evaluate singular Feynman parameter integrals.


\section{Lowest order amplitudes of the $e p \rightarrow e p \gamma$ reaction}
\label{sec:lowest}

\begin{figure}[Hbtp]
\begin{center}
\leavevmode
\hbox{%
\epsfysize=11.0truecm
\epsfbox{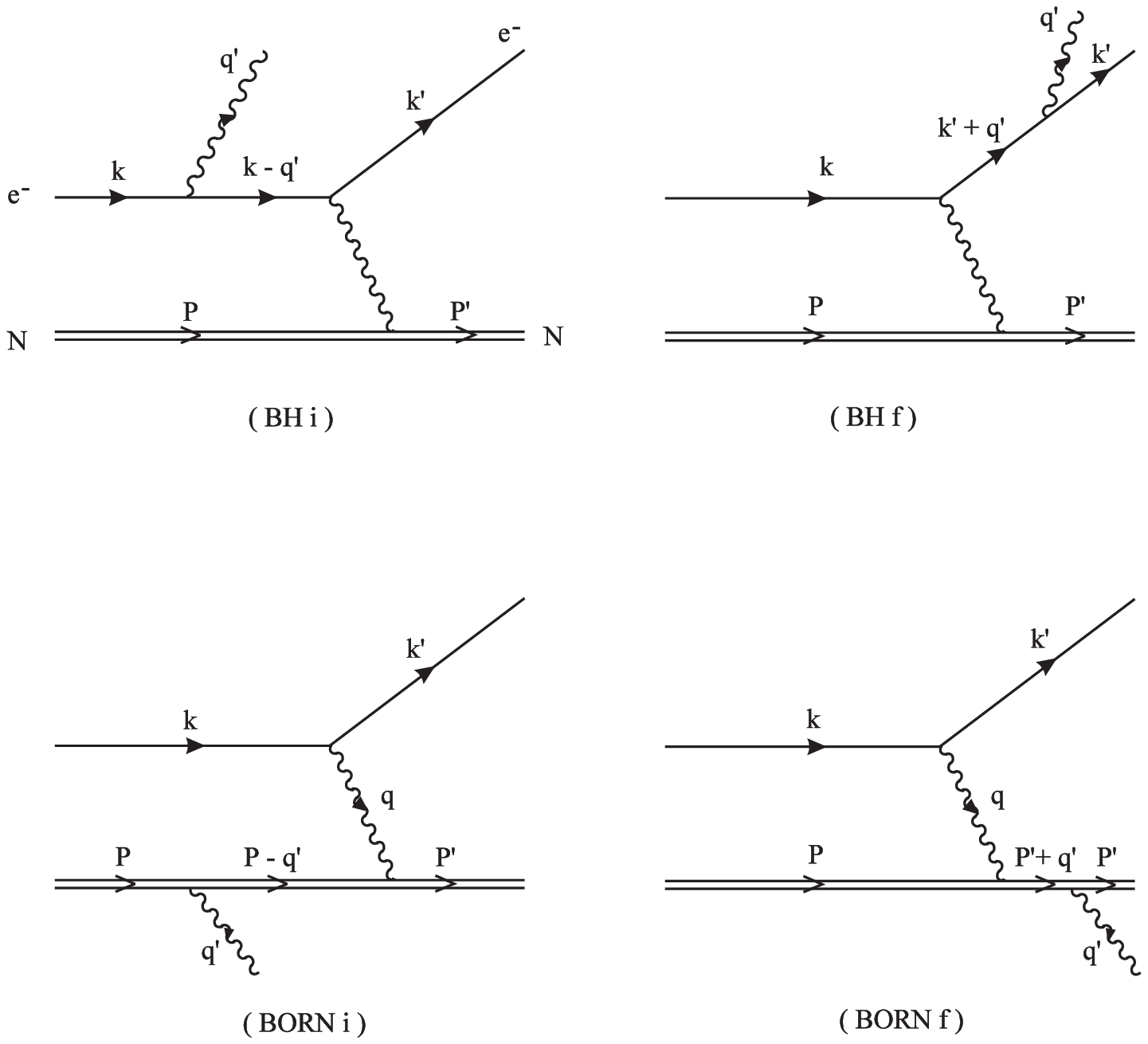} }
\end{center}
\centerline{\small \it  }
\caption{Tree level contributions to the 
$e p \to e p \gamma$ reaction:
Bethe-Heitler diagrams (a) and (b); nucleon Born diagrams (c) and (d).}
\label{fig:vcstree}
\end{figure}

The lowest order (in $\alpha_{em}$),  
contributions to the $e p \rightarrow e p \gamma$
reaction are given by the one-photon exchange processes. 
We denote in this work the four-momenta of the initial and final
electrons by $k(E_e, \vec k_e)$ and $k'(E'_e, \vec k^{\; '}_e)$; 
the four-momenta of the initial and final
protons by $p(E_N, \vec p_N)$ and $p'(E'_N, \vec p^{\; '}_N)$; 
and the four-momentum of the outgoing photon by 
$q'(|\vec q^{\; '}|, \vec q^{\; '})$. Furthermore, we denote 
$q \equiv k - k' = p' - p + q'$ and $Q^2 = -q^2 > 0$. 
The masses of the electron and proton are denoted by $m$ and $M_N$
respectively. 
The helicities of the 
initial (final) electrons are denoted by $h (h')$; the spins of   
initial (final) protons by $s_p (s'_p)$; and the polarization four-vector
of the outgoing photon by $\varepsilon$. 
The spinors of initial and final electrons are denoted by 
$u( k , h )$ and $u( k' , h' )$; whereas the spinors of initial and
final protons are denoted by $N( p , s_p )$ and $N( p' , s'_p )$.
Throughout this work, 
we follow the conventions of {\it Bjorken and Drell} \cite{BD64}. 
\newline
\indent
In Figs.~\ref{fig:vcstree} (BHi) and 
(BHf), which are known as the Bethe-Heitler (BH) diagrams, a photon is emitted 
by either the incident or final electrons. The expressions for 
Figs.~\ref{fig:vcstree} (BHi) and (BHf) are respectively given by~: 
\begin{eqnarray}
\label{eq:bhi}
M_{BH}^i \,=\,&&i \,e^3 \,\bar u( k^{'} , h' )\,\gamma ^\nu\,
{{\left( {\not k-\not q^{'}+m} \right)} \over {-2k.q^{'}}}\;\not \varepsilon ^*
\, u( k , h )\; 
{1 \over {\left( {p^{'} - p} \right)^2}} \;
\bar N( p^{'} , s_p^{'} )\,\Gamma_{\nu}\left(p^{'} , p\right)\,
N( p , s_p ) , \\
M_{BH}^f \,=\,&&i \,e^3 \,\bar u( k^{'} , h' )\,\not \varepsilon ^*\,
{{\left( {\not k^{'}+\not q^{'}+m} \right)} \over {2k^{'}.q^{'}}}\,
\gamma ^\nu \, u( k , h ) \; 
{1 \over {\left( {p^{'}-p} \right)^2}} \;
\bar N( p^{'} , s_p^{'} )\,\Gamma_{\nu}\left(p^{'} , p\right)\,
N( p , s_p ) ,
\label{eq:bhf}
\end{eqnarray}
where the electron charge is given by $(-e)$ (i.e. $e > 0$ in this
work).   
The on-shell electromagnetic vertex at the hadron side $\Gamma_{\nu}$ 
in Eqs.~(\ref{eq:bhi},\ref{eq:bhf}) is given by 
\begin{equation}
\Gamma_{\nu}\left(p^{'} , p\right)\;=\; F_1 \left((p^{'}-p)^2 \right) 
\, \gamma_{\nu} + 
F_2 \left((p^{'}-p)^2 \right) 
\, i \sigma_{\nu \lambda} {{(p^{'}-p)^{\lambda}} \over {2 M_N}} \;,
\label{eq:hadroncurrent}
\end{equation}
where $F_1$ and $F_2$ are respectively the Dirac and Pauli electromagnetic 
(on-shell) form factors of the nucleon. 
The four-momentum squared of the virtual photon in the 
BH processes is $t = (p' - p)^2$, in contrast to $q^2$, which is the 
four-momentum squared for the VCS process $\gamma^* p \to \gamma p$, where the
final photon is emitted from the hadron side. This latter part 
contains the nucleon structure information. 
\newline
\indent
The amplitude of the VCS contribution 
to the $e^- p \to e^- p \gamma$ reaction is given by~:
\begin{equation}
M_{VCS} \,=\, -i \,e^3 \;\bar u( k^{'} , h' )\,\gamma_\nu\,
u( k , h ) \; {1 \over {q^2}} \;
\varepsilon^*_{\mu} \; H^{\mu \nu} \, .
\label{eq:m_vcs} 
\end{equation}
Remark that for a positive lepton, the VCS amplitude changes sign. 
In Eq.~(\ref{eq:m_vcs}), 
the gauge-invariant, hadronic tensor $H^{\mu \nu}$ is defined by~:
\begin{equation}
H^{\mu \nu }=-i\int d^{4}x\, {\rm e}^{-iq.x}
< p \, '|T\left[ j^{\nu }(x),j^{\mu }(0)\right] | p > \;,
\label{eq:hadtensor}
\end{equation}
where $T$ represents the time ordering, and $j^\nu$ the
electromagnetic current operator.  
\newline
\indent
For the DVCS process in the Bjorken limit, the hadronic tensor of
Eq.~(\ref{eq:hadtensor}) is parametrized in terms of six leading twist
skewed parton distributions (see e.g. \cite{Jireview}).
\newline
\indent
For the VCS process at low energy, 
as investigated experimentally in \cite{mami,jlab,bates}, 
an important contribution to the tensor of Eq.(\ref{eq:hadtensor})
originates from the nucleon pole contributions shown in 
Figs.~\ref{fig:vcstree} (BORNi) and (BORNf). 
The contributions of the Born diagrams to the hadronic tensor are given by~: 
\begin{eqnarray}
\label{eq:borni}
H^{\mu \nu}_{BORN, i} \,=\,&& 
\bar N( p^{'} , s_p^{'} )\,\Gamma^{\nu}\left(p^{'} , p-q^{'}\right)\,
{{\left( {\not p-\not q^{'}+M_N} \right)} \over {-2p.q^{'}}} \,
\Gamma^{\mu}\left(p-q^{'} , p\right) \, N( p , s_p ) ,\\
H^{\mu \nu}_{BORN, f} \,=\,&&
\bar N( p^{'} , s_p^{'} )\,
\Gamma^{\mu}\left(p^{'} , p^{'}+q^{'}\right) \,
{{\left( {\not p^{'}+\not q^{'}+M_N} \right)} \over {2p^{'}.q^{'}}} \,
\Gamma^{\nu} \left(p^{'}+q^{'} , p\right) \, N( p , s_p ) ,
\label{eq:bornf}
\end{eqnarray}
where the vertex $\Gamma^{\mu}$ is now evaluated for off mass-shell
values of one of its arguments. In Ref.~\cite{Guichon95}, the Born
diagrams were evaluated by using the vertex of
Eq.~(\ref{eq:hadroncurrent}). Doing so, the Born diagrams are
separately gauge invariant. All nucleon structure effects are then
absorbed in a non-Born amplitude which is regular in $q'$ and
for which the Low Energy Theorem (LET) tells that it starts at order
$q'$. The nucleon structure effects to the VCS tensor 
(Eq.~(\ref{eq:hadtensor})) below pion threshold, 
are then parametrized at order $q'$ in terms of six generalized 
(i.e. $Q^2$ dependent) nucleon polarizabilities \cite{Guichon95,Drechsel98}.


\section{First order radiative corrections to the 
$e p \rightarrow e p \gamma$  reaction}
\label{sec:first}

\subsection{Virtual radiative corrections}
\label{sec:virtual}

In this section, we calculate the one-loop QED virtual radiative corrections to
the $e p \rightarrow e p \gamma$  reaction, which are represented in 
Fig.~\ref{fig:radcorr}. In the present section, we consider only the
corrections originating from the electron side as they can be
calculated model-independently. The corrections originating from the
hadronic side, for which a nucleon structure model is needed, will be
discussed and estimated in section \ref{sec:protonside}. 

The virtual radiative corrections to the BH process contain 
vertex corrections : Figs.~\ref{fig:radcorr} (V1i - V3i) and (V1f - V3f); 
electron self-energy corrections : Figs.~\ref{fig:radcorr} (Si, Sf); 
and vacuum polarization corrections : Figs.~\ref{fig:radcorr} (P1i, P1f). 
We indicate in our notation of the different diagrams whether the
photon in the $e p \to e p \gamma$ reaction 
is emitted from the initial (i) electron or from the final (f)
electron. 

The part of the virtual radiative corrections 
to the VCS process (i.e. where the
photon in the reaction $e p \to e p \gamma$ is emitted from the
hadronic side) which can be calculated model-independently, consists of the
vertex diagram shown in Fig.~\ref{fig:radcorr} (V4) and the 
vacuum polarization diagram shown in Fig.~\ref{fig:radcorr} (P2). 
The blob in those figures represents the VCS process. 
For VCS below pion threshold, the blob is given by the Born diagrams 
(Fig.~\ref{fig:vcstree} (BORNi) and (BORNf)) + non-Born diagrams, 
which describe the nucleon polarizability effects. 
For DVCS, the blob is given in leading order by the so-called handbag
diagrams, where the photon hits a quark in the proton 
\cite{Ji97,Rady96,vcsreview}. 

The calculation of the virtual radiative corrections to the VCS
process is similar to that for electron scattering. 
The virtual radiative corrections to the Bethe-Heitler process are
different, but involve the same one-loop building blocks, i.e. electron
vertex, electron self-energy and photon self-energy. 
Therefore, we give in appendix~\ref{app:elast} 
(sections \ref{app:elast1} - \ref{app:elast4}) the derivation and the
expressions for these basic building blocks, and we apply it to
elastic electron-nucleon scattering. 
In our calculations, we use the dimensional
regularization method to treat both ultraviolet (UV) and infrared (IR)
divergences. This amounts to evaluate all loop integrals in $D$
dimensions. The divergences then show up (when one takes $D \to 4$) as
poles of the form $1/\varepsilon$, 
where $\varepsilon \equiv 2 - D/2$. UV divergences
are regularized by taking $D < 4$ (i.e. $\varepsilon_{UV} = 2 - D/2 >
0$), whereas IR divergences are regularized by taking $D > 4$
(i.e. $\varepsilon_{IR} = 2 - D/2 < 0$). Care has to be taken as to
isolate the UV and IR divergent parts in the loop integrals first, 
as two different limits are understood when one takes $D = 4$ at the end.  
The technical details of our calculational method can also be found in 
appendix \ref{app:elast}. We apply it here to calculate the diagrams 
of Fig.~\ref{fig:radcorr} to the $e p \to e p \gamma$ reaction.

\subsubsection{Vertex correction diagrams of Figs.~\ref{fig:radcorr}
  (V1i) and (V1f)} 

The amplitude corresponding to Fig.~\ref{fig:radcorr} (V1i) is given by 
\begin{eqnarray}
\label{eq:v1_i1}
&&M_{V1}^i=\,{{e^5} \over {\left( {p^{'}-p} \right)^2}}\;
\bar N( p^{'} , s_p^{'} )\, \Gamma_{\nu}\left(p^{'} , p\right)\,
N( p , s_p )  \nonumber\\
&&\times \bar u( k^{'} , h' )\,\gamma ^\nu {{\left( 
{\not k-\not q^{'}+m}\right)} \over {-2k.q^{'}}}
\, \mu^{4 -  D} \int {{{d^Dl} \over {\left( {2\pi }
\right)^D}}}\,{{\gamma ^\alpha \,\left( {\not k-\not q^{'}-\not l+m}
\right) \not \varepsilon^* \left( {\not k-\not l+m} \right) \gamma
_\alpha } \over {\left[ {l^2} \right]\,\,\left[ {l^2-2l.k}
\right]\,\left[ {l^2-2l.\left( {k-q^{'}} \right)-2k.q^{'}} \right]}}
\,u( k , h ) ,
\end{eqnarray}
where a mass scale $\mu$ (renormalization scale) is introduced
when passing to $D \neq 4$ dimensions in order to keep the coupling
constant dimensionless.  
One sees by inspection that the loop integral in Eq.~(\ref{eq:v1_i1}),
when taking $D = 4$,  
is IR finite ($l \to 0$ behavior), but has an UV divergence ($l \to
\infty$ behavior). Our strategy to evaluate a complicated loop integral
as in Eq.~(\ref{eq:v1_i1}), is to solve first a simpler loop integral
which contains entirely the UV divergence and which can be done
analytically more easily. We observe from Eq.~(\ref{eq:v1_i1}) that 
only the term in the numerator proportional to 
$\gamma^\alpha \not l \not \varepsilon ^* \not l\gamma _\alpha$ 
is responsible for the UV divergence. To evaluate it, we add a similar term 
by replacing one factor in the denominator 
and evaluate this term analytically. In order to obtain the 
equivalence with $M_{v1}^i$, we have to subtract the added term 
again from the expression of Eq.~(\ref{eq:v1_i1}). This leads to 
\begin{eqnarray}
\label{eq:v1_i2}
&&M_{V1}^i=\;{{e^5} \over {\left( {p^{'}-p} \right)^2}}\;
\bar N( p^{'} , s_p^{'} )\,\Gamma_{\nu}\left(p^{'} , p\right)\,
N( p , s_p ) \nonumber\\
&&\times \bar u( k^{'} , h' )\,\gamma ^\nu {{\left( 
{\not k-\not q^{'}+m}\right)} \over {-2k.q^{'}}} \, 
\left\{ \mu^{4 - D} {\int {{{d^Dl} \over {\left( {2\pi } \right)^D}}}\,
{{\gamma^\alpha \not l \not \varepsilon ^* \not l\gamma _\alpha } \over
{\left[ {l^2} \right]\,\,\left[ {l^2-2l.k} \right]\,\left[ {l^2-2l.k^{'}}
\right]}}} \right.\nonumber\\
&&\hspace{1.5cm}+ 2\int {{{d^4l} \over {\left( {2\pi } \right)^4}}}\,
{ {\not l \not \varepsilon ^* \left( {\not k-\not q^{'}}
\right) + \not k \not \varepsilon ^* \not l\,-\,m^2 \not
\varepsilon^* 
+ 4m\,\varepsilon ^*.\left( k - l \right) - \not k \not \varepsilon
^* \left( {\not k-\not q^{'}} \right) } 
\over {\left[{l^2} \right]\,\,\left[ {l^2-2l.k} \right]\,
\left[ {l^2-2l.\left( {k-q^{'}}\right)-2k.q^{'}} \right]}} \nonumber\\
&&\hspace{1.5cm}\left.{+ 2\int {{{d^4l} \over {\left( {2\pi } \right)^4}}}\,
{{-2 \not l \not \varepsilon ^* \not l\,\left[ {l.\left( {q-q^{'}} \right)
+ k.q^{'}}\right]} \over {\left[ {l^2} \right]\,\,\left[ {l^2-2l.k}
\right]\,\left[ {l^2-2l.k^{'}} \right]\,\left[ {l^2-2l.\left( {k-q^{'}}
\right)-2k.q^{'}} \right]}} }\right\} u( k , h ) \;.
\end{eqnarray}
It should be remarked that only the added term in Eq.~(\ref{eq:v1_i2}) 
(first term within curly brackets of Eq.~(\ref{eq:v1_i2}) ) 
is UV divergent and has therefore to be evaluated in $D$ dimensions using 
the dimensional regularization method. 
The third term within the curly brackets in Eq.~(\ref{eq:v1_i2}) is the 
difference between the term proportional to 
$\gamma^\alpha \not l \not \varepsilon ^* \not l\gamma _\alpha$ 
in Eq.~(\ref{eq:v1_i1}) and the added term (which has one different factor 
in the denominator). As can be seen by power counting, this term is UV 
finite and can therefore readily be evaluated for $D = 4$. 
The denominator in the UV divergent first term of Eq.~(\ref{eq:v1_i2}), was 
chosen so that it corresponds with the vertex correction which appears
in electron scattering. Therefore, this UV divergent term can be 
calculated analytically along a similar way 
as was performed in appendix \ref{app:elast}. The result is given by~: 
\begin{eqnarray}
\label{eq:v1_i3}
&&{\mu^{4 - D} \int {{{d^Dl} \over {\left( {2\pi } \right)^D}}}\;{{\gamma
^\alpha \not l \not \varepsilon ^* \not l\gamma _\alpha } \over
{\left[ {l^2} \right]\,\,\left[ {l^2-2l.k} \right]\,\left[ {l^2-2l.k^{'}}
\right]}}} \nonumber\\
&&={i \over {\left( {4\pi } \right)^2}} \left\{ {\not \varepsilon
^* \left[ {{1 \over {\varepsilon _{UV}}}-\gamma_E +\ln \left( {{{4\pi \mu
^2} \over {m^2}}} \right) + 1 - v \ln \left( {{{v+1} \over {v-1}}} \right)}
\right] - {1 \over {Q^2}}\not q \not \varepsilon ^*\not q}
\right.\nonumber\\
&&\left. \hspace{1.5cm} {+{1 \over{Q^2 v}}\,\ln \left( {{{v+1} \over 
{v-1}}} \right) \left[ {\not k \not \varepsilon ^*\not k^{'}
+ \not k^{'} \not \varepsilon ^*\not k\,
+\left( {{{v^2+1} \over 2}} \right) 
\not q \not \varepsilon ^*\not q} \right] } \right\} \;,
\end{eqnarray}
where $v$ is defined as
\begin{equation}
v^2 \; \equiv \; 1 + {{4 m^2} \over Q^2} \;. 
\label{eq:v1_v}
\end{equation}
The UV divergence in Eq.~(\ref{eq:v1_i3}) is removed by the 
corresponding vertex counterterm as given by 
Eqs.~(\ref{eq:countertermL},\ref{eq:a_v7}) 
\begin{equation}
\label{eq:v1_i4}
(CT)_{V1}^i= M_{BH}^i {{(- e^2)} \over {\left( {4\pi } \right)^2}}
\left\{ \left[ {{1 \over {\varepsilon _{UV}}}-\gamma_E 
+\ln \left( {{4\pi \mu^2} \over {m^2}} \right)} \right] 
+ 2 \left[{1 \over {\varepsilon _{IR}}} -\gamma_E 
+ \ln \left( {{4\pi \mu^2} \over {m^2}} \right) \right]
+ 4 \right\} ,
\end{equation}
where we have used the expression of Eq.~(\ref{eq:bhi}) for the BH
amplitude $M_{BH}^i$. 
Adding the counterterm of Eq.~(\ref{eq:v1_i4}) to Eq.~(\ref{eq:v1_i2}) 
and introducing a Feynman parametrization in the second and third terms 
of Eq.~(\ref{eq:v1_i2}) in order to perform the integrals over $l$, yields 
the total, UV finite result~:
\begin{eqnarray}
\label{eq:v1_i5}
&&M_{V1}^i + (CT)_{V1}^i 
= M_{BH}^i\,{{e^2} \over {\left( {4\pi } \right)^2}}
\left\{ {- 2 \left[ {{1 \over {\varepsilon _{IR}}}-\gamma_E 
+\ln \left( {{4\pi \mu^2} \over {m^2}} \right)} \right] - 3 - v\,
\ln \left( {{{v+1} \over {v-1}}} \right)}\right\}\nonumber\\
&&+{{i e^5} \over {\left( {4\pi } \right)^2}}\,{1 \over {\left( {p^{'}-p} 
\right)^2}}\,
\bar N( p^{'} , s_p^{'} )\,\Gamma_{\nu}\left(p^{'} , p\right)\,
N( p , s_p ) \nonumber\\
&&\times \bar u( k^{'} , h' )\,
\gamma ^\nu {{\left( {\not k-\not q^{'}+m}
\right)} \over {-2k.q^{'}}}\, \nonumber\\
&&\times \left\{  {1 \over {Q^2}} \left[ {\left( {-1 +
{{v^2+1} \over {2v}}\,\ln \left( {{{v+1} \over {v-1}}} \right)  }\right)
\not q \not \varepsilon^* \not q } 
{ {\,+ {{1 \over v} \ln \left( {{{v+1} \over {v-1}}} 
\right)} \left\{ {\not k \not \varepsilon ^*\not k^{'}
+\not k^{'} \not \varepsilon ^*\not k } \right\} } }
\right] \right. \nonumber\\
&&-\,2 \int\limits_0^1 {dy \int\limits_0^1 {dx}\,
{1 \over {B^i_1}}\,\left[ {y \left( {\not k-\not q^{'}x} \right)}
\right.}\not \varepsilon ^*\left( {\not k-\not q^{'}} \right)+y \not
k \not \varepsilon ^*\left( {\not k-\not q^{'}x} \right)\nonumber\\
&&\hspace{3.5cm}\left. {+\,4m \left( {\varepsilon ^*.k} \right)
\left( {1-y} \right) -\not k\not \varepsilon ^*\left( {\not k-\not q^{'}}
\right) - m^2\not \varepsilon ^*} \right] \nonumber\\
&&-\,4 \int\limits_0^1 {dx_3\,x_3^2} \int\limits_0^1
{dx_2}\,x_2 \int\limits_0^1 {dx_1}
\left[ {\left( {{1 \over {A^i}}\not \varepsilon ^*
+{1 \over {\left( {A^i}\right)^2}}\not P^i 
\not \varepsilon ^*\not P^i} \right)}
\right.\left( {P^i.\left( {q-q^{'}} \right)+k.q^{'}} \right)\nonumber\\
&&\hspace{5cm} \left.{\left. {-\,{1 \over {2A^i}}\,
\left( {\left( {\not q-\not q^{'}}\right)
\not \varepsilon ^*\not P^i + \not P^i \not
\varepsilon ^*\left( {\not q-\not q^{'}} \right)} \right)} \right]
}\right\}\,u( k , h ) ,
\end{eqnarray}
with the four-vector $P^i$ defined by 
\begin{equation}
\label{eq:v1_i6}
P^i\;\equiv\;\left( {k - q^{'}} \right) (1 - x_3)\,+
\,\left( k - q \, x_1 \right)\,x_2\, x_3\;,
\end{equation}
and the scalars $A^i$ and $B^i$ defined by
\begin{equation}
\label{eq:v1_i7}
A^i\;\equiv\;2k.q^{'} \,(1 - x_3)\,+\,\left( {P^i} \right)^2\;,
\end{equation}
and
\begin{equation}
\label{eq:v1_i8}
B^i_1\;\equiv\;2k.q^{'} \,x (1 - y)\,+\,m^2 y\;.
\end{equation}
Remark that although Eq.~(\ref{eq:v1_i5}) is UV finite, it contains
now an IR divergence through the vertex counterterm of
Eq.~(\ref{eq:v1_i4}) as shown in appendix A (Eq.~(\ref{eq:a_v7})). 
We will demonstrate however in section \ref{sec:real}, that all IR
divergences, arising from the one-loop corrections to the $e p \to e p
\gamma$ reaction, are cancelled when adding the corresponding soft
photon emission contributions. 
\newline
\indent
The Feynman parameter integrals in Eq.~(\ref{eq:v1_i5}) 
which orginate from the finite integrals in Eq.~(\ref{eq:v1_i2}) remain to 
be evaluated. As an analytical calculation of these integrals is 
rather complicated, we will evaluate them numerically in this paper, 
which will be discussed in section \ref{sec:int}.
\newline
\indent
In a completely similar way as for Fig.~\ref{fig:radcorr} (V1i), 
the total amplitude including the counterterm 
corresponding to Fig.~\ref{fig:radcorr} (V1f) yields~: 
\begin{eqnarray}
\label{eq:v1_f1}
&&M_{V1}^f + (CT)_{V1}^f
= M_{BH}^f\,{{e^2} \over {\left( {4\pi } \right)^2}}\,
\left\{ - 2 \,\left[ {1 \over {\varepsilon _{IR}}}-\gamma_E 
+\ln \left( {{4\pi \mu^2} \over {m^2}} \right) \right]
- 3 - \,v\,\ln \left( {{v+1} \over {v-1}} \right)  \right\} \nonumber\\
&&+{{i e^5} \over {\left( {4\pi } \right)^2}}\,{1 \over {\left( {p^{'}-p} 
\right)^2}}\,\bar N( p^{'} , s_p^{'} )\,\Gamma_{\nu}\left(p^{'} , p\right)\,
N( p , s_p ) \nonumber\\
&&\times \bar u( k' , h' )
\left\{ { {1 \over {Q^2}} \left[ {\left( {-1 +
{{v^2+1} \over {2v}} \ln \left( {{{v+1} \over {v-1}}} \right)  }\right)
\not q \not \varepsilon ^*\not q} \right.} 
{+ {{1 \over v} \ln \left( {{{v+1} \over {v-1}}} 
\right)} \left\{ {\not k \not \varepsilon ^*\not k^{'}
+\not k^{'} \not \varepsilon ^*\not k } \right\} } \right] \nonumber\\
&&\hspace{1.3cm}-\,2 \int\limits_0^1 {dy \int\limits_0^1 {dx}\,
{1 \over {B^f_1}}\,\left[ {y\,\left( {\not k^{'}+\not q^{'}} \right)}
\right.}\not \varepsilon ^*\left( {\not k^{'}+\not q^{'}x} \right)
+y\,\left( {\not k^{'}+\not q^{'}x} \right) \not \varepsilon ^* \not k^{'}
\nonumber\\
&&\hspace{4.5cm}\left. {+\,4m\,\left( {\varepsilon ^*.k^{'}} \right)
\left( {1-y} \right)\,
-\left( {\not k^{'}+\not q^{'}}\right) \not \varepsilon ^* \not k^{'}\,
-m^2\not \varepsilon ^*} \right]\;\nonumber\\
&&\hspace{1.3cm}+\,4 \int\limits_0^1 {dx_3\,x_3^2} \int\limits_0^1
{dx_2}\,x_2 \int\limits_0^1 {dx_1}  
\left[ {\left( {{1 \over
{A^f}}\not \varepsilon ^*+{1 \over {\left( {A^f}\right)^2}}\not P^f 
\not \varepsilon ^*\not P^f} \right)}
\right.\left( {P^f.\left( {q-q^{'}} \right)+k^{'}.q^{'}} \right)\nonumber\\
&&\hspace{6.cm}  {\left. {-\,{1 \over {2A^f}}\,
\left( {\left( {\not q-\not q^{'}}\right)
\not \varepsilon ^*\not P^f\,+\,\not P^f\,\not
\varepsilon ^*\left( {\not q-\not q^{'}} \right)} \right)} \right]
}  \nonumber\\
&&\hspace{1.3cm} \left. {{} \over {} }\right\} 
{{\left( {\not k^{'}+\not q^{'}+m} \right)} \over {2k^{'}.q^{'}}}
\,\gamma ^\nu \,u( k , h )\;,
\end{eqnarray}
with the four-vector $P^f$ defined by 
\begin{equation}
\label{eq:v1_f2}
P^f\;\equiv\;\left( {k^{'}\,+\,q^{'}} \right)\,(1 - x_3)\;+
\;\left( {k^{'}\,+\,q}\,x_1 \right)\,x_2\, x_3\;,
\end{equation}
and the scalars $A^f$ and $B^f_1$ defined by
\begin{equation}
\label{eq:v1_f3}
A^f\;\equiv\;-2k^{'}.q^{'} \,(1 - x_3)\;+\;\left( {P^f} \right)^2\;,
\end{equation}
and
\begin{equation}
\label{eq:v1_f4}
B^f_1\;\equiv\;-2k^{'}.q^{'} \,x (1 - y)\;+\;m^2 y\;.
\end{equation}

\subsubsection{Vertex correction diagrams of Figs.~\ref{fig:radcorr}
  (V2i) and (V2f)} 

The amplitude corresponding to Fig.~\ref{fig:radcorr} (V2i) is given by 
\begin{eqnarray}
\label{eq:v2_i0}
&&M_{V2}^i=\,{{e^5} \over {\left( {p^{'}-p} \right)^2}}\;
\bar N( p^{'} , s_p^{'} )\, \Gamma_{\nu}\left(p^{'} , p\right)\,
N( p , s_p )  \nonumber\\
&&\times \bar u( k^{'} , h' )
\, \mu^{4 -  D} \int {{{d^Dl} \over {\left( {2\pi }
\right)^D}}} {{\gamma^\alpha \left( {\not k' -\not l+m} \right) 
\gamma ^\nu \left( {\not k-\not q^{'}-\not l+m}
\right)   \gamma_\alpha } \over {\left[ {l^2} \right]\,\,\left[ {l^2-2l.k'}
\right]\,\left[ {l^2-2l.\left( {k-q^{'}} \right)-2k.q^{'}} \right]}}
 {{\left( {\not k-\not q^{'}+m}\right)} \over {-2k.q^{'}}}
\not \varepsilon^* \,u( k , h ) .
\end{eqnarray}
One sees from Eq.~(\ref{eq:v2_i0}) that again only the term in the numerator
proportional to $\gamma^\alpha \not l \gamma^\nu \not l \gamma_\alpha$
contains an UV divergence for $D = 4$. 
To evaluate the loop integral of Eq.~(\ref{eq:v2_i0}), we 
therefore apply a similar trick as used before in
Eq.~(\ref{eq:v1_i2}). This amounts to adding and subtracting a term in
Eq.~(\ref{eq:v2_i0}) by replacing $(l^2-2l.\left( {k-q'}
\right)-2k.q')$ in the denominator by $(l^2-2l.k')$, and which
contains entirely the UV divergence. The further steps are then
analogous to those following Eq.~(\ref{eq:v1_i2}), and yield the
  following result for Fig.~\ref{fig:radcorr} (V2i)~:
\begin{eqnarray}
\label{eq:v2_i1}
&&M_{V2}^i + (CT)_{V2}^i 
= M_{BH}^i\,{{e^2} \over {\left( {4\pi } \right)^2}}\,
\left\{ {-2 \left[ {1 \over {\varepsilon _{IR}}}-\gamma_E 
+\ln \left( {{4\pi \mu^2} \over {m^2}} \right) \right] \,- 3 -\,v\,
\ln \left( {{{v+1} \over {v-1}}} \right)}\right\}\nonumber\\
&&+{{i e^5} \over {\left( {4\pi } \right)^2}}\,{1 \over {\left( {p^{'}-p} 
\right)^2}}\,\bar N( p^{'} , s_p^{'} )\,\Gamma_{\nu}\left(p^{'} , p\right)\,
N( p , s_p ) \nonumber\\
&&\times \, \bar u( k^{'} , h' )
\left\{ { {1 \over {Q^2}} \left[ {\left( {-1 +
{{v^2+1} \over {2v}} \ln \left( {{{v+1} \over {v-1}}} \right)  }\right)
 \not q \gamma^\nu \not q}  \right.} 
+ {{1 \over v} \ln \left( {{{v+1} \over {v-1}}} 
\right)} \left\{ {\not k \gamma^\nu \not k^{'}
+\not k^{'} \gamma^\nu \not k} \right\} \right\} \nonumber\\
&&\hspace{1.5cm}-\,2 \int\limits_0^1 {dy} \int\limits_0^1 {dx}\,
{1 \over {B^i_2}} \left[{ {y \left( {\not k-\not q^{'}} \right)}
\gamma^\nu \left( {\not k^{'} + \left({\not q - \not q^{'}}\right) x} \right)
+y \left( {\not k^{'}+\left( {\not q - \not q^{'}}\right) x}  
\right) \gamma^\nu \not k^{'} }\right.\nonumber\\
&&\hspace{5.cm}\left. {+\,4m\,{\left( k^{'} \right)}^\nu
\left( {1-y} \right) -\left( {\not k-\not q^{'}}\right) 
\gamma^\nu \not k^{'}\, -m^2 \gamma^\nu} \right] \nonumber\\
&&\hspace{1.5cm}-\,4 \int\limits_0^1 {dx_3\,x_3^2} \int\limits_0^1
{dx_2}\,x_2 \int\limits_0^1 {dx_1} \left[ {\left( {{1 \over
{A^i}}\gamma^\nu +{1 \over {\left( {A^i}\right)^2}}\not P^i 
\gamma^\nu \not P^i} \right)}
\right.\left( {q^{'}.\left( {k-P^i} \right)} \right)\nonumber\\
&&\hspace{6.7cm}\left. +\,{1 \over {2A^i}}\,
\left( {\not q^{'} \gamma^\nu \not P^i\,
+\,\not P^i \gamma^\nu \not q^{'} } \right) \right]   \nonumber\\
&&\hspace{1.5cm} \left. {{} \over {} }\right\} 
{{\left( {\not k-\not q^{'}+m} \right)} \over {-2k.q^{'}}}
\not \epsilon^* \,u( k , h )\;,
\end{eqnarray}
where $A^i$ is given as in Eq.~(\ref{eq:v1_i7}) and where 
\begin{equation}
\label{eq:v2_i2}
B^i_2\;\equiv\;m^2 y\;+\;x^2 y {\left( {q - q^{'}} \right)}^2 
+\,2x\,k.q^{'} \,+ 2 x y \,k^{'}.{\left( {q - q^{'}} \right)} \;.
\end{equation}
\newline
\indent
In an analogous way, 
the amplitude corresponding to Fig.~\ref{fig:radcorr} (V2f) 
can be calculated, and yields as result~: 
\begin{eqnarray}
\label{eq:v2_f1}
&&M_{V2}^f + (CT)_{V2}^f
= M_{BH}^f\,{{e^2} \over {\left( {4\pi } \right)^2}}\,
\left\{ {-2 \left[ {1 \over {\varepsilon _{IR}}}-\gamma_E 
+\ln \left( {{4\pi \mu^2} \over {m^2}} \right) \right] \,-3-\,v\,
\ln \left( {{{v+1} \over {v-1}}} \right)}\right\}\nonumber\\
&&+{{i e^5} \over {\left( {4\pi } \right)^2}}\,{1 \over {\left( {p^{'}-p} 
\right)^2}}\,\bar N( p^{'} , s_p^{'} )\,\Gamma_{\nu}\left(p^{'} , p\right)\,
N( p , s_p ) \nonumber\\
&&\times \, \bar u( k^{'} , h' )\, \not \epsilon^*\, 
{{\left( {\not k^{'}+\not q^{'}+m} \right)} \over {2k^{'}.q^{'}}} \nonumber\\
&&\times \left\{ { {1 \over {Q^2}} \left[ {\left( {-1 + 
{{v^2+1} \over {2v}}\,\ln \left( {{{v+1} \over {v-1}}} \right)  }\right)
\,\not q \gamma^\nu \not q}  \right.} 
\,+ {{1 \over v}\,\ln \left( {{{v+1} \over {v-1}}} 
\right)} \,\left\{ {\not k \gamma^\nu \not k^{'}
+\not k^{'} \gamma^\nu \not k} \right\} \right\} \nonumber\\
&&\;\;-\,2\,\int\limits_0^1 {dy}\,\int\limits_0^1 {dx}\,
{1 \over {B^i_2}}\,\left[{ {y\,\not k}
\gamma^\nu \left( {\not k - \left({\not q - \not q^{'}}\right) x} \right)
+y\,\left( {\not k -\left( {\not q - \not q^{'}}\right) x}  
\right) \gamma^\nu \left( {\not k^{'} + \not q^{'}} \right)}\right.\nonumber\\
&&\hspace{3.5cm}\left. {+\,4m\,{\left( k \right)}^\nu
\left( {1-y} \right)\,-\not k \gamma^\nu 
\left( {\not k^{'}+\not q^{'}}\right)\,-m^2 \gamma^\nu} \right] \;\nonumber\\
&&\;\;+\,4\,\int\limits_0^1 {dx_3\,x_3^2}\,\int\limits_0^1
{dx_2}\,x_2\,\int\limits_0^1 {dx_1}\,\left[ {\left( {{1 \over
{A^f}}\gamma^\nu +{1 \over {\left( {A^f}\right)^2}}\not P^f 
\gamma^\nu \not P^f} \right)}
\right.\left( {q^{'}.\left( {k^{'}-P^f} \right)} \right)\nonumber\\
&&\hspace{5.5cm}\left. \left. +\,{1 \over {2A^f}}\,
\left( {\not q^{'} \gamma^\nu \not P^f\,+\,\not P^f \gamma^\nu \not q^{'} } 
\right)  \right]  \right\}  \,u( k , h )\;,
\end{eqnarray}
where $A^f$ is given as in Eq.~(\ref{eq:v1_f3}) and where 
\begin{equation}
\label{eq:v2_f2}
B^f_2\;\equiv\;m^2 y\;+\;x^2 y {\left( {q - q^{'}} \right)}^2 
-\,2x\,k^{'}.q^{'} \,- 2 x y \,k.{\left( {q - q^{'}} \right)} \;.
\end{equation}

\subsubsection{Vertex correction diagrams of Figs.~\ref{fig:radcorr}
  (V3i) and (V3f)} 

The amplitude $M_{V3}^i$ corresponding to Fig.~\ref{fig:radcorr} (V3i) 
is given by 
\begin{eqnarray}
\label{eq:v3_i1}
&&M_{V3}^i = {{e^5} \over {\left( {p^{'}-p} \right)^2}}\, 
\bar N( p^{'} , s_p^{'} )\,\Gamma_{\nu}\left(p^{'} , p\right)\,
N( p , s_p )  \nonumber\\
&&\times\,\bar u( k^{'} , h' )\, \mu^{4 -  D} 
\int {{{d^Dl} \over {\left( {2\pi }
\right)^D}}} {{\gamma ^\alpha \left( {\not k^{'}+\not l+m}
\right)\gamma ^\nu \left( {\not k-\not q^{'}+\not l+m} \right)\not
\varepsilon ^*\left( {\not k+\not l+m} \right)\gamma _\alpha } \over
{\left[ {l^2} \right]\,\left[ {l^2+2l.k^{'}} \right]\,\left[ {l^2+2l.k}
\right]\,\left[ {l^2+2l.\left( {k-q^{'}} \right)-2k.q^{'}} \right]}}\,
u( k , h ) .
\end{eqnarray}
Remark that the loop integral in Eq.~(\ref{eq:v3_i1}) 
is UV finite but contains an IR divergence for $D = 4$. 
This is because in Fig.~\ref{fig:radcorr} (V3i), a soft virtual 
photon ($l \to 0$) couples to two on-shell electron lines. 
To isolate the IR divergence, we first decompose the numerator in 
Eq.~(\ref{eq:v3_i1}) by using the relations  
$\bar u( k^{'} , h' )\,
\gamma ^\alpha \left( {\not k^{'}+m}\right)\,=\,
\bar u( k^{'} , h' )\,2 k^{'}_\alpha$ and 
$\left( {\not k+m} \right) \gamma _\alpha
\,u( k , h )\,=\, 2 k_\alpha \,u( k , h )$. 
This yields~:
\begin{eqnarray}
\label{eq:v3_i2}
&&M_{V3}^i = {{e^5} \over {\left( {p^{'}-p} \right)^2}}\,
\bar N( p^{'} , s_p^{'} )\,\Gamma_{\nu}\left(p^{'} , p\right)\,
N( p , s_p )  \nonumber\\
&&\times\,\bar u( k^{'} , h' )\,\mu^{4 - D}
\int {{{d^Dl} \over {\left( {2\pi }\right)^D}}} {{1} \over
{\left[ {l^2} \right]\,\left[ {l^2+2l.k^{'}} \right]\,\left[ {l^2+2l.k}
\right]\,\left[ {l^2+2l.\left( {k-q^{'}} \right)-2k.q^{'}} \right]}}\nonumber\\
&&\times \left\{ {4 (k.k') \,\gamma ^\nu 
\left( {\not k-\not q^{'}+m} \right) \not \varepsilon^* 
+ 4 (k.k') \,\gamma ^\nu \not l \not \varepsilon ^*} 
+ 2 \gamma ^\nu \left( {\not k-\not q^{'}+\not l+m} \right)
\not\varepsilon^*\not l \not k^{'} \right.\nonumber\\
&&\;\;\left.+ 2 \not k \not l \gamma ^\nu
\left( {\not k-\not q^{'}+\not l+m} \right) \not \varepsilon^*
+ \gamma ^\alpha {\not l\,\gamma ^\nu \left(
{\not k-\not q^{'}+\not l+m} \right) \not \varepsilon ^*\not l\,\gamma
_\alpha }\right\} u( k , h ) .
\end{eqnarray}
In Eq.~(\ref{eq:v3_i2}), only the term in the numerator which is
$l$-independent (the first term within the curly brackets) contains an IR
divergence, whereas all the other terms are finite. As before, instead of
aiming at an analytical formula for a rather complicated integral, we
evaluate the IR divergent part of the integral in Eq.~(\ref{eq:v3_i2})
by adding and subtracting a term that contains the divergence and
that can be performed analytically rather easily. 
In constructing this term, we are looking for a denominator which contains 
the same dependence as the basic BH process in order that this 
BH amplitude can be factored from this IR divergent term.  
This yields the following expression, which is by construction identical to 
Eq.~(\ref{eq:v3_i2})~:
\begin{eqnarray}
\label{eq:v3_i3}
&&M_{V3}^i = {{e^5} \over {\left( {p^{'}-p} \right)^2}}\,
\bar N( p^{'} , s_p^{'} )\,\Gamma_{\nu}\left(p^{'} , p\right)\,
N( p , s_p )  \nonumber\\
&&\times\,\bar u( k^{'} , h' )
\left\{\mu^{4 - D}  \int {{{d^Dl} \over {\left( {2\pi }\right)^D}}}\,
{ {4 (k.k') \gamma ^\nu 
\left( {\not k-\not q^{'}+m} \right)\,\not \varepsilon ^*} \over
{\left[ {l^2} \right]\,\left[ {l^2+2l.k^{'}} \right]\,\left[ {l^2+2l.k}
\right]\,\left[ {-2k.q^{'}} \right]}} \right. \nonumber\\
&&\hspace{1.3cm}+\,\int {{{d^4l} \over {\left( {2\pi }\right)^4}}}\,{{1} \over
{\left[ {l^2} \right]\,\left[ {l^2+2l.k^{'}} \right]\,\left[ {l^2+2l.k}
\right]\,\left[ {l^2+2l.\left( {k-q^{'}} \right)-2k.q^{'}} \right]}}\nonumber\\
&&\hspace{1.8cm}\times \left[ {4 ( k.k' ) \gamma ^\nu 
\left( {\not k-\not q^{'}+m} \right) \not \varepsilon ^*\, 
{{-l^2 -2l.\left( {k-q^{'}} \right)}\over{-2k.q^{'}}}
+\,4 (k.k') \gamma ^\nu \not l \,\not \varepsilon ^*}
\right. \nonumber\\
&&\hspace{2cm}+\,2\,\gamma ^\nu \left( {\not k-\not q^{'}+\not l+m} \right)
\not\varepsilon ^*\not l \not k^{'} +\,2 \not k \not l \gamma ^\nu
\left( {\not k-\not q^{'}+\not l+m} \right) \not \varepsilon ^*\nonumber\\
&&\hspace{2cm}+\,\gamma ^\alpha \left. {\left. {\not l \gamma ^\nu \left(
{\not k-\not q^{'}+\not l+m} \right) \not \varepsilon ^*\not l \gamma
_\alpha }\right]}  \right\}  \,u( k , h )\,.
\end{eqnarray}
Remark that the added term (first term of Eq.~(\ref{eq:v3_i3})) contains
the IR divergence whereas the other terms of Eq.~(\ref{eq:v3_i3})) do not
have any divergences so that the corresponding integrals may be
performed directly in four dimensions as indicated. For the first term
of Eq.~(\ref{eq:v3_i3}) we furthermore see that the $l$-independent part of
the energy denominator is the same as the one occuring in the
corresponding Bethe-Heitler diagram (Fig.~\ref{fig:vcstree}a). 
The $l$-dependent
part of the energy denominator for this term is the same as the one for
the vertex correction to elastic electron scattering, 
Eq.~(\ref{eq:a_v2}). The corresponding integral may therefore be evaluated
analytically in a similar way as was done in appendix A. 
This yields for the IR divergent term in Eq.~(\ref{eq:v3_i3})~: 
\begin{eqnarray}
\label{eq:v3_i4}
&&{{e^5} \over {\left( {p^{'}-p} \right)^2}}\,
\bar N( p^{'} , s_p^{'} )\,\Gamma_{\nu}\left(p^{'} , p\right)\,
N( p , s_p )  \nonumber\\
&&\times \bar u( k^{'} , h' )\, 
\gamma ^\nu\,{{ \left( {\not k-\not q^{'}+m} \right) } 
\over {- 2\,k.q^{'}} } \not \varepsilon ^* \, u( k , h )\,
\mu^{4 - D} \int {{{d^Dl} \over {\left( {2\pi }\right)^D}}}\,
{{4\left( {k.k^{'}} \right)} \over
{\left[ {l^2} \right]\,\left[ {l^2+2l.k^{'}} \right]\,\left[ {l^2+2l.k}
\right]}}\;\nonumber\\
&&= M_{BH}^i\,{{e^2} \over {\left( {4\pi } \right)^2}} \left\{ 
\left[ {1 \over {\varepsilon _{IR}}}-\gamma_E 
+\ln \left( {{4\pi \mu^2} \over {m^2}} \right) \right]
{{v^2+1} \over v}\,\ln \left( {{{v+1}
\over {v-1}}} \right) \right.\nonumber\\
&&\hspace{1.8cm}\left. + {{v^2+1} \over {2v}} \ln \left( {{{v+1}
\over {v-1}}}\right) \ln \left( {{{v^2-1} \over {4v^2}}}\right)
+ {{v^2+1} \over v} \left[ {Sp\left( {{{v+1} \over
{2v}}} \right) - Sp\left( {{{v-1} \over {2v}}} \right)} \right] \right\} .
\end{eqnarray}
The evaluation of the finite four-dimensional integral in
Eq.~(\ref{eq:v3_i3}) can be performed at the expense of the introduction
of three Feynman parameter integrals due to the four energy denominators~:
\begin{eqnarray}
\label{eq:v3_i5}
&&{{1} \over
{\left[ {l^2} \right]\,\left[ {l^2+2l.k^{'}} \right]\,\left[ {l^2+2l.k}
\right]\,\left[ {l^2+2l.\left( {k-q^{'}} \right)-2k.q^{'}} \right]}} 
\nonumber\\
&&= \, 6\int\limits_0^1 {dy\,y^2} \int\limits_0^1 {dx_2}\,x_2 \int\limits_0^1
{dx_1}\,{1 \over {\left[ {\left( {l+y\,P^i_{x_1 x_2}}
\right)^2 - y\,C^i} \right]^4}} ,
\end{eqnarray}
with the four-vector $P_{x_1 x_2}^i$ defined by 
\begin{equation}
\label{eq:v3_i6}
P^i_{x_1 x_2}\;\equiv\;\left( {q\,-\,x_1\,q^{'}} \right)\,x_2\;+\;k^{'}\;,
\end{equation}
and the scalar $C^i$ defined by
\begin{equation}
\label{eq:v3_i7}
C^i\;\equiv\;2k.q^{'} \,x_1x_2\;+\;y\,\left( {{P^i_{x_1 x_2}}} \right)^2\;.
\end{equation}
The final result for the amplitude $M_{V3}^i$ is then given by
\begin{eqnarray}
\label{eq:v3_i8}
&&M_{V3}^i = M_{BH}^i\,{{e^2} \over {\left( {4\pi } \right)^2}} \left\{ 
\left[ {1 \over {\varepsilon _{IR}}}-\gamma_E 
+\ln \left( {{4\pi \mu^2} \over {m^2}} \right) \right]
\,{{v^2+1} \over v}\,\ln \left( {{{v+1}
\over {v-1}}} \right) \right. \nonumber\\
&&\hspace{2.8cm}\left. 
+ {{v^2+1} \over {2v}}\,\ln \left( {{{v+1}\over {v-1}}}\right)\,
\ln \left( {{{v^2-1} \over {4v^2}}}\right) 
+ {{v^2+1} \over v} \left[ {Sp\left( {{{v+1} \over{2v}}} \right)
- Sp\left( {{{v-1} \over {2v}}} \right)} \right] \right\}\nonumber\\
&&+{{i e^5} \over {\left( {4\pi } \right)^2}}\,{1 \over {\left( {p^{'}-p} 
\right)^2}}\,\bar N( p^{'} , s_p^{'} )\,\Gamma_{\nu}\left(p^{'} , p\right)\,
N( p , s_p ) \,
\int\limits_0^1 {dy\,y}\,\int\limits_0^1
{dx_2}\,x_2\,\int\limits_0^1 {dx_1}\nonumber\\
&&\times\,\bar u( k^{'} , h' )
\left\{ {\gamma ^\nu }
{{\left( {\not k-\not q^{'}+m} \right)} \over {-2k.q^{'}}}
\not \varepsilon ^*\,4 \, k.k^{'} 
\left[ {{2 \over {C^i}}+{1 \over
{\left({C^i} \right)^2}}\left( {-y\,\left( {P^i_{x_1 x_2}}
\right)^2+2\,P^i_{x_1 x_2}.(k-q^{'})} \right)} \right] \right. \nonumber\\
&&{2 \over {C^i}} \left[ {\gamma ^\nu } \not
\varepsilon ^*\not k^{'}+\not k \gamma ^\nu \not \varepsilon
^*-4m\,\varepsilon ^{*\nu } 
{- \gamma ^\nu\left( {\not k-\not q^{'}-y 
\not P^i_{x_1 x_2}}\right)\not \varepsilon ^*
+ y \not \varepsilon ^*\gamma ^\nu \not P^i_{x_1 x_2}
+ y \not P^i_{x_1 x_2} \not \varepsilon ^*\gamma ^\nu }\right]\nonumber\\
&&+{1 \over {\left( {C^i} \right)^2}}
\left[ {-4\,(k.k^{'})\,\gamma^\nu } \right.\not P^i_{x_1 x_2}
\not \varepsilon ^*
-2\,\gamma ^\nu \left( {\not k-\not q^{'}-y \not
P^i_{x_1 x_2}+m} \right)\not\varepsilon ^*\not P^i_{x_1 x_2}
\not k^{'} \nonumber\\
&&\hspace{1.5cm}-2 \not k \not P^i_{x_1 x_2} 
\gamma ^\nu \left( {\not k-\not q^{'}-y \not P^i_{x_1 x_2}+m} \right)\not
\varepsilon ^*\nonumber\\
&&\hspace{1.5cm} \left. \left. {+ y \not P^i_{x_1 x_2} \left( {-2 \not
\varepsilon ^*\left( {\not k-\not q^{'}-y \not P^i_{x_1 x_2}} \right) \gamma
^\nu +\,4m\,\varepsilon ^{*\nu }} \right)
\not P^i_{x_1 x_2}\,} \right] \right\} u( k , h ) .
\end{eqnarray}
The Feynman parameter integrals in Eq.~(\ref{eq:v3_i8}) will be 
performed numerically as explained in Section \ref{sec:int}. 
\newline
\indent
In an analogous way, the result for the amplitude $M_{V3}^f$ 
corresponding to Fig.~\ref{fig:radcorr} (V3f) can be calculated, 
and yields as result~:
\begin{eqnarray}
\label{eq:v3_f1}
&&M_{V3}^f = M_{BH}^f\,{{e^2} \over {\left( {4\pi } \right)^2}} \left\{ 
\left[ {1 \over {\varepsilon _{IR}}}-\gamma_E 
+\ln \left( {{4\pi \mu^2} \over {m^2}} \right) \right]
\,{{v^2+1} \over v} \, \ln \left( {{{v+1}
\over {v-1}}} \right) \right.\nonumber\\
&&\left. \hspace{3cm}
+ {{v^2+1} \over {2v}} \ln \left( {{{v+1}\over {v-1}}}\right)
\ln \left( {{{v^2-1} \over {4v^2}}}\right) 
+ {{v^2+1} \over v} \left[ {Sp\left( {{{v+1} \over{2v}}} \right)
- Sp\left( {{{v-1} \over {2v}}} \right)} \right] \right\}\nonumber\\
&&+{{i e^5} \over {\left( {4\pi } \right)^2}}\,{1 \over {\left( {p^{'}-p} 
\right)^2}}\,\bar N( p^{'} , s_p^{'} )\,\Gamma_{\nu}\left(p^{'} , p\right)\,
N( p , s_p ) \,
\int\limits_0^1 {dy\,y}\,\int\limits_0^1
{dx_2}\,x_2\,\int\limits_0^1 {dx_1}\nonumber\\
&&\times\,\bar u( k^{'} , h' )
\left\{ {\not \varepsilon ^*}
{{\left( {\not k^{'} +\not q^{'}+m} \right)} \over {2k^{'}.q^{'}}}
\gamma ^\nu \,4 \, k.k^{'}  
\left[ {{2 \over {C^f}}+{1 \over
{\left({C^f} \right)^2}}\left( {-y\,\left( {P^f_{x_1 x_2}}
\right)^2+2\,P^f_{x_1 x_2}.(k^{'}+q^{'})} \right)} \right] 
\right. \nonumber\\
&&+{2 \over {C^f}} \left[ {\not \varepsilon ^*} 
\gamma ^\nu \not k^{'}+\not k \not \varepsilon
^* \gamma ^\nu - 4m\,\varepsilon ^{*\nu } 
{- \not \varepsilon ^*\left( {\not k^{'}+\not q^{'}-y
\not P^f_{x_1 x_2}}\right) \gamma ^\nu
+ y \gamma ^\nu \not \varepsilon ^* \not P^f_{x_1 x_2} 
+ y \not P^f_{x_1 x_2} \gamma ^\nu \not \varepsilon ^*}\right]\nonumber\\
&&+{1 \over {\left( {C^f} \right)^2}}
\left[ {-4\,(k.k^{'}) \not \varepsilon ^*} 
\right.\not P^f_{x_1 x_2} \gamma^\nu 
-2 \not\varepsilon ^* \left( {\not k^{'}+\not q^{'}-y \not
P^f_{x_1 x_2}+m} \right)\gamma ^\nu \not P^f_{x_1 x_2} \not k^{'} \nonumber\\
&&\hspace{1.5cm}-2\,\not k \not P^f_{x_1 x_2}
\not \varepsilon ^* \left( {\not k^{'}+\not q^{'}-y \not P^f_{x_1 x_2}+m} 
\right) \gamma ^\nu \nonumber\\
&&\hspace{1.5cm}\left.\left. {+\,y\,\not P^f_{x_1 x_2}\,\left( {-2 \gamma^\nu\,
\left( {\not k^{'}+\not q^{'}-y\,\not P^f_{x_1 x_2}} \right)
\not \varepsilon ^* +\,4m \varepsilon ^{*\nu }} 
\right)\not P^f_{x_1 x_2}} \right] \right\} \, u( k , h ) ,
\end{eqnarray}
with the four-vector $P^f_{x_1 x_2}$ defined by 
\begin{equation}
\label{eq:v3_f2}
P^f_{x_1 x_2}\;\equiv\;-\left( {q\,-\,x_1\,q^{'}} \right)\,x_2\;+\;k\;,
\end{equation}
and the scalar $C^f$ defined by
\begin{equation}
\label{eq:v3_f3}
C^f\;\equiv\;-2k^{'}.q^{'} \,x_1x_2\;+\;y\,\left({{P^f_{x_1 x_2}}} \right)^2\;.
\end{equation}
\newline
\indent
Remark that in the vertex correction diagrams where the photon couples to 
the final electron (diagrams of Fig.~\ref{fig:radcorr} denoted by $f$), 
the invariant mass of the virtual $(e^- +\gamma^*)$ state in the loop is 
given by $m^2 + 2 k^{'}.q^{'} \;\geq\; m^2$. This means that an 
on-shell propagation is possible for the $(e^- + \gamma^*)$ state. 
This translates mathematically into the presence of integrable singularities 
in the corresponding Feynman parameter integrals of Eqs.~(\ref{eq:v1_f1}, 
\ref{eq:v2_f1}) and (\ref{eq:v3_f1}), and yields an imaginary part for
the corresponding amplitude. In contrast, in the 
vertex correction diagrams where the photon couples to the initial electron 
(diagrams of Fig.~\ref{fig:radcorr} denoted by $i$), 
the invariant mass of the virtual $(e^- + \gamma^*)$ system in the loop is 
given by $m^2$ which means that the corresponding integrals contain no 
singularities. The numerical treatment of those singular 
Feynman parameter integrals will be discussed in section \ref{sec:int}.

\begin{figure}[Hbtp]
\begin{center}
\leavevmode
\hbox{%
\epsfysize=16.0truecm
\epsfbox{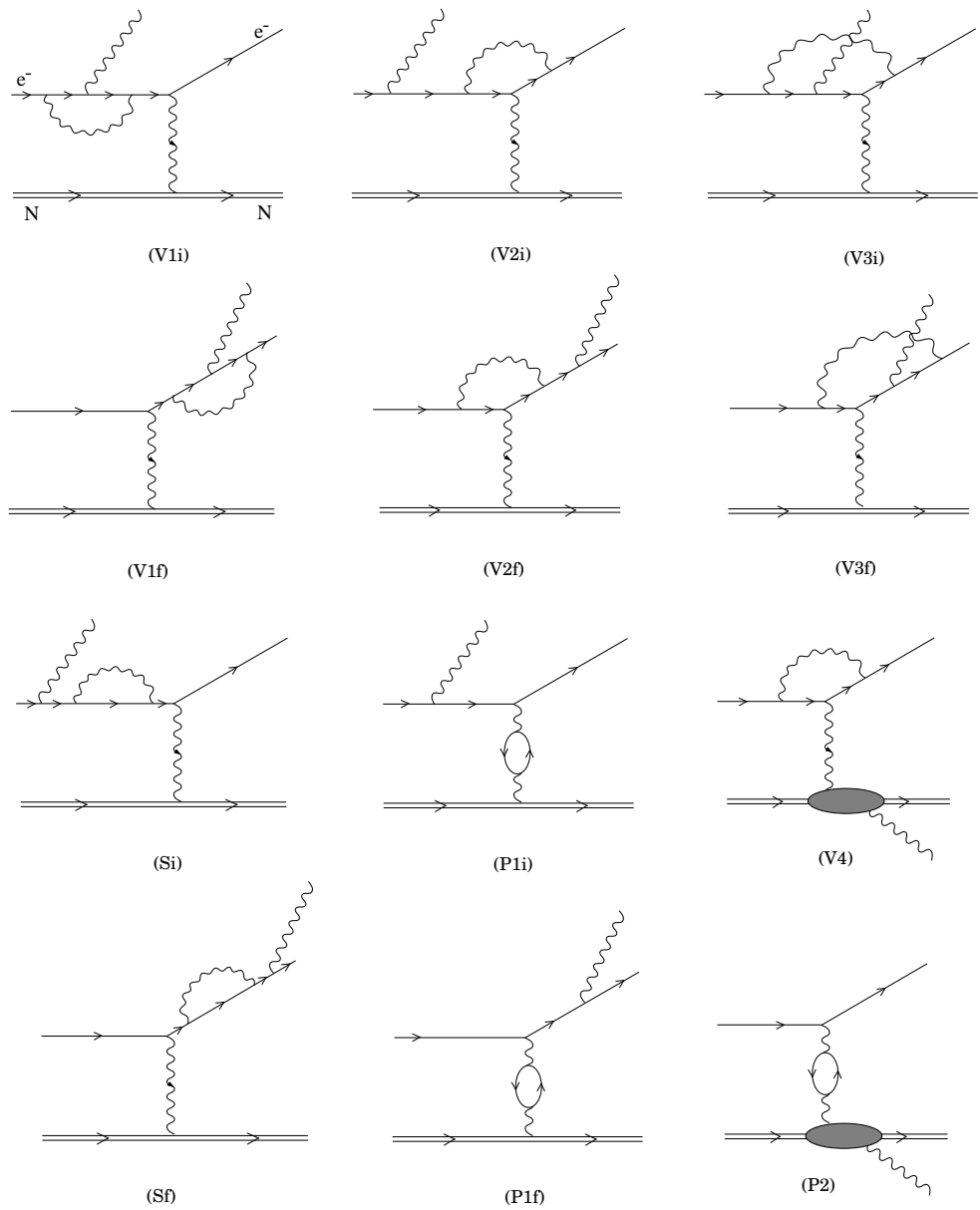} }
\end{center}
\centerline{\small \it  }
\caption{First order virtual photon radiative corrections to the $e p
  \to e p \gamma$ reaction.}
\label{fig:radcorr}
\end{figure}

\subsubsection{Electron self-energy diagrams of Figs.~\ref{fig:radcorr}
  (Si) and (Sf)} 

We next evaluate the electron self-energy diagrams of 
Figs.~\ref{fig:radcorr} (Si) and (Sf). We only have to consider those 
diagrams where a photon is emitted and re-absorbed by an intermediate
electron line. The diagrams with a loop on the initial or final
electron lines are already absorbed in the wavefunction and electron
mass renormalization, and therefore do not yield an additional
correction. This can also be seen from the expression 
Eq.~(\ref{eq:a_self10}) for the renormalized lepton self-energy, 
which vanishes on-shell. 

The amplitude corresponding to Fig.~\ref{fig:radcorr} (Si) is then given
by~:
\begin{eqnarray}
\label{eq:self_i1}
M_{Si} &=& i e^3 \,{1 \over {\left( {p^{'}-p} \right)^2}}\,
\bar N( p^{'} , s_p^{'} )\,\Gamma_{\nu}\left(p^{'} , p\right)\,
N( p , s_p ) \, \nonumber\\
&\times&\,\bar u( k^{'} , h' ) {\gamma ^\nu }
{{\left( {\not k-\not q^{'}+m} \right)} \over {-2k.q^{'}}} 
\,\tilde \Sigma(k - q') \, 
{{\left( {\not k-\not q^{'}+m} \right)} \over {-2k.q^{'}}} 
\not \varepsilon^* u( k , h ) ,
\end{eqnarray}
where the renormalized self-energy 
is denoted by $\tilde \Sigma$ and is given by 
Eq.~(\ref{eq:a_self10}). Remark that the UV divergence in the loop
integral of Fig.~\ref{fig:radcorr} (Si) has been removed through the
renormalization of the electron field and electron mass. The UV finite
renormalized self-energy $\tilde \Sigma$ contains however an IR
divergence from the counterterms. Inserting the expression for 
$\tilde \Sigma$ (Eq.~(\ref{eq:a_self10})) into Eq.(\ref{eq:self_i1}),
yields~:
\begin{eqnarray}
\label{eq:self_i2}
M_{Si} &=& M_{BH}^i\,{{e^2} \over {( 4 \pi )^2}} \,2\,
\left[ {1 \over {\varepsilon _{IR}}} - \gamma_E 
+\ln \left( {{4\pi \mu^2} \over {m^2}} \right) \right] \nonumber \\
&+& {{i e^5} \over {(4 \pi)^2}} 
\,{1 \over {\left( {p^{'}-p} \right)^2}}\,
\bar N( p^{'} , s_p^{'} )\,\Gamma_{\nu}\left(p^{'} , p\right)\,
N( p , s_p ) \, \nonumber\\
&&\times\,\bar u( k^{'} , h' ) \, {\gamma ^\nu }
{{\left( {\not k-\not q^{'}+m} \right)} \over {-2k.q^{'}}} 
\left\{ {{m \left(\not k - \not q' \right)} \over {m^2 - 2k.q'}}
\left[ 1 + {{-2 m^2 + 6 k.q'} \over {m^2 - 2 k.q'}} 
\ln\left( {{2 k.q'} \over {m^2}}\right) \right] \right. \nonumber\\
&&\left.\hspace{5cm} + \left[ 3 - {{2 m^2 + 2 k.q'} \over {m^2 - 2 k.q'}} 
\ln\left( {{2 k.q'} \over {m^2}}\right) \right] 
\right\} \not \varepsilon^* u( k , h ) .
\end{eqnarray}

The amplitude corresponding to Fig.~\ref{fig:radcorr} (Sf) is given by~:
\begin{eqnarray}
\label{eq:self_f1}
M_{Sf} &=& i e^3 \,{1 \over {\left( {p^{'}-p} \right)^2}}\,
\bar N( p^{'} , s_p^{'} )\,\Gamma_{\nu}\left(p^{'} , p\right)\,
N( p , s_p ) \, \nonumber\\
&\times&\,\bar u( k^{'} , h' ) \not \varepsilon^* 
{{\left( {\not k'+\not q'+m} \right)} \over {2k'.q'}} 
\,\tilde \Sigma(k' + q') \, 
{{\left( {\not k'+\not q'+m} \right)} \over {2k.q'}} 
{\gamma ^\nu } u( k , h ) ,
\end{eqnarray}
which can be worked out analogously as before and yields~:
\begin{eqnarray}
\label{eq:self_f2}
M_{Sf} &=& M_{BH}^f\,{{e^2} \over {( 4 \pi )^2}} \,2\,
\left[ {1 \over {\varepsilon _{IR}}} - \gamma_E 
+\ln \left( {{4\pi \mu^2} \over {m^2}} \right) \right] \nonumber \\
&+& {{i e^5} \over {(4 \pi)^2}} 
\,{1 \over {\left( {p^{'}-p} \right)^2}}\,
\bar N( p^{'} , s_p^{'} )\,\Gamma_{\nu}\left(p^{'} , p\right)\,
N( p , s_p ) \, \nonumber\\
&&\times\,\bar u( k^{'} , h' ) \not \varepsilon^*
{{\left( {\not k' + \not q'+m} \right)} \over {2 k'.q'}} 
\left\{ {{m \left(\not k' + \not q' \right)} \over {m^2 + 2 k'.q'}}
\left[ 1 + {{-2 m^2 - 6 k'.q'} \over {m^2 + 2 k'.q'}} 
\ln\left( {{- 2 k'.q'} \over {m^2}}\right) \right] \right. \nonumber\\
&&\left.\hspace{5cm} + \left[ 3 - {{2 m^2 - 2 k'.q'} \over {m^2 + 2 k'.q'}} 
\ln\left( {{- 2 k'.q'} \over {m^2}}\right) \right] 
\right\}  {\gamma ^\nu } \, u( k , h ) .
\end{eqnarray}
Note that in Fig.~\ref{fig:radcorr} (Sf), the four-momentum squared of
the $(e^- + \gamma^*)$ state in the loop is given by 
$(k' + q')^2 = m^2 + 2 k'.q' \geq m^2$. Therefore, the self-energy and
the amplitude for Fig.~\ref{fig:radcorr} (Sf) is complex, as was also
noted for the vertex diagrams of Fig.~\ref{fig:radcorr} where the
photon is emitted from the final electron (denoted by $f$). 
Eq.~(\ref{eq:self_f2}) yields indeed a complex amplitude because 
$\ln\left( - 2 k'.q' / m^2 \right) = 
\ln\left( 2 k'.q' / m^2 \right) + i \pi$, for $k'.q' > 0$.

\subsubsection{Vertex correction diagram of Fig.~\ref{fig:radcorr} (V4)} 

The vertex correction to the VCS process is given by 
Fig.~\ref{fig:radcorr} (V4), and its calculation is the same as the one
for elastic electron scattering. This yields for the renormalized 
vertex correction~:
\begin{equation}
M_{V4} = -i \,e^3 \,\bar u( k^{'} , h' )
\left[ \left( {F(Q^2) - F(Q^2 = 0) }\right) \,\gamma_\nu\,
-\,G(Q^2)\,i \sigma_{\nu \kappa} {{q^\kappa} \over {2 m}} \right]  
u( k , h ) \, {1 \over {q^2}} \, \varepsilon^*_{\mu} \, H^{\mu \nu} .
\label{eq:vcsvertex}
\end{equation}
In Eq.~(\ref{eq:vcsvertex}), $F(Q^2) \,-\,F(Q^2 = 0)$ is given by
Eq.~(\ref{eq:a_v8}) and reduces in the ultrarelativistic limit 
($Q^2 >> m^2$) to Eq.~(\ref{eq:vertexhighs}). 
The magnetic correction $G(Q^2)$ is given by Eq.~(\ref{eq:a_v5}), and vanishes
in the ultrarelativistic limit.

\subsubsection{Vacuum polarization diagrams of Figs.~\ref{fig:radcorr}
  (P1i, P1f) and (P2)} 

The vacuum polarization corrections of Figs.~\ref{fig:radcorr} 
(P1i, P1f) and (P2) involve the renormalized photon self-energy 
$\tilde \Pi(Q^2)$, which has been calculated in appendix
\ref{app:elast}. Therefore, we get for the vacuum polarization
correction to the BH process (Figs.~\ref{fig:radcorr} (P1i, P1f))~:
\begin{equation}
M_{P1}^i = M_{BH}^i \, {1 \over {1 - \tilde \Pi(-t)}} \, ,
\hspace{1cm} {\mathrm and} \hspace{1cm} 
M_{P1}^f = M_{BH}^f \, {1 \over {1 - \tilde \Pi(-t)}} \, ,
\end{equation}
with $t = (p' - p)^2$.
\indent
Similarly, we get for the vacuum polarization correction to the VCS
process (Fig.~\ref{fig:radcorr} (P2))~:
\begin{equation}
M_{P2} = M_{VCS} \, {1 \over {1 - \tilde \Pi(Q^2)}} \, .
\end{equation}
In the ultrarelativistic limit ($Q^2 >> m^2$), $\tilde \Pi(Q^2)$ is
obtained from Eq.~(\ref{eq:photonpol})
\begin{equation}
\tilde \Pi(Q^2) \, = \, {{e^2} \over {(4 \pi)^2}} \, 
{4 \over 3} \left\{ -{5 \over 3} \,+\, 
\ln\left( {Q^2 \over m^2} \right) \right\} \, .
\end{equation}

 
\subsection{Soft-photon emission contributions 
and cancellation of IR divergences}
\label{sec:real}

After removing the UV divergences from the virtual photon corrections to
the $e p \to e p \gamma$ reaction in the last section, the resulting
expressions still contain IR divergences.
Both the corrections to the BH process of 
Figs.~\ref{fig:radcorr} (V1i, V1f, V2i, V2f, V3i, V3f, Si and Sf)
and the vertex correction of Fig.~\ref{fig:radcorr} (V4) to the 
VCS process contain IR divergences. 
It is known for QED since a long time \cite{Bloch37,Jauch55}, that
these IR divergences are cancelled at the cross section level by 
soft photon emission contributions. 
These soft photons are emitted from the charged particle lines 
and can have energies up to some maximal value $\Delta E_s$ which is 
related to the finite resolution of the detector. 
In appendix~\ref{app:elast} (section \ref{app:elast5}), we calculate
the soft bremsstrahlung contribution to electron scattering 
by performing the phase space integral over the soft photon in an
exact way, and give the
finite correction (after cancellation of all IR divergences) to the
elastic electron scattering cross section. In this section, we
generalize the result of appendix~\ref{app:elast} to the case of the
$e p \to e p \gamma$ reaction. The diagrams for the $e p \to e p \gamma$ 
reaction with one additional soft photon are shown in
Fig.~\ref{fig:brems}, where the hard photon of the $e p \to e p
\gamma$ process is indicated by its four-momentum $q'$. 
In this section, we will show that the 
soft photon emission contributions of Fig.~\ref{fig:brems} 
contain IR divergences which exactly
cancel the IR divergences appearing in the virtual photon correction
diagrams of Fig.~\ref{fig:radcorr}. 
The process where the energy $\Delta E_s$ of the additionally emitted
photon is not very small compared with the lepton momenta in the
process, makes up the radiative tail to the $e p \to e p \gamma$
reaction. Its calculation will be discussed in section \ref{sec:radtail}.

\subsubsection{Factorization of amplitude for soft-photon emission processes}
 
\begin{figure}[Hbtp]
\begin{center}
\leavevmode
\hbox{%
\epsfysize=16.0truecm
\epsfbox{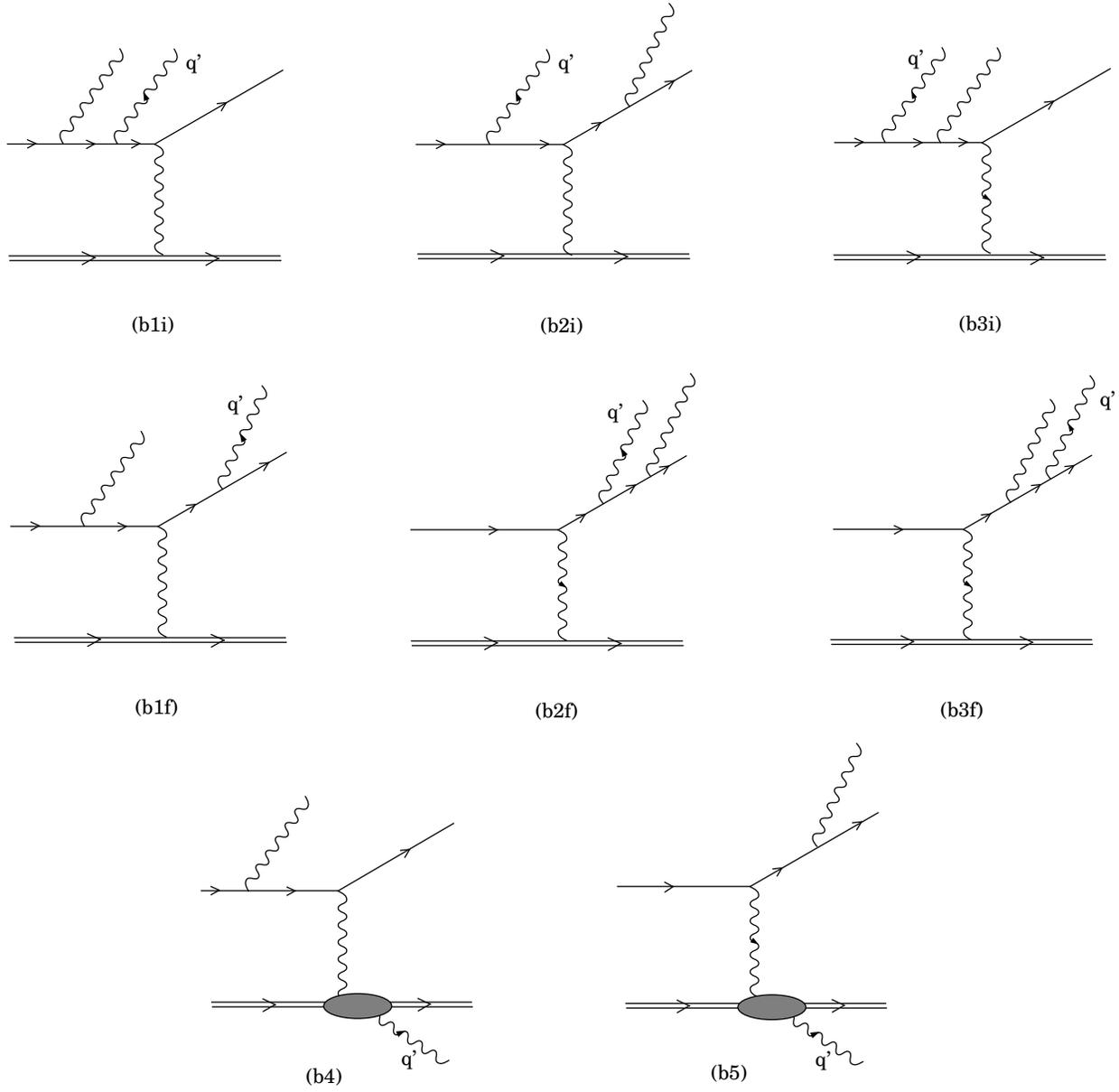} }
\end{center}
\centerline{\small \it  }
\caption{First order soft photon emission contributions to the $e p
  \to e p \gamma$ reaction.}
\label{fig:brems}
\end{figure}

Here, we evaluate the diagrams of Fig.~\ref{fig:brems} in the soft photon
limit, i.e. when the second emitted photon has an energy much
smaller than the initial and final lepton energies and also smaller
than the hard photon (denoted by $q'$) in order to distinguish both
photons. We will see that only the diagrams where a soft
photon couples to an on-shell lepton 
contain IR divergences and lead to a finite logarithmic correction 
in $\Delta E_s$. 

The amplitude corresponding with Fig.~\ref{fig:brems} (b1i) is given by~:
\begin{eqnarray}
\label{eq:b1i}
M_{b1i} \,=\,&&i \,e^3 \,\bar u( k^{'} , h' )\,\gamma ^\nu\,
{{\left( {\not k-\not q'-\not l+m} \right)} 
\over {-2k.q' -2 l.(k-q')}}\,\not \varepsilon^*(q') \,
{{\left( {\not k-\not l+m} \right)} 
\over {-2k.l}}\,\left( -e \not \varepsilon^*(l) \right) 
\, u( k , h ) \nonumber\\ 
&&\times\;{1 \over {\left( {p^{'} - p} \right)^2}} \;
\bar N( p^{'} , s_p^{'} )\,\Gamma_{\nu}\left(p^{'} , p\right)\, 
N( p , s_p ) \, , 
\end{eqnarray}
where $l$ is the four-momentum of the soft photon. 
In the soft photon limit ($l \to 0$), Eq.~(\ref{eq:b1i}) simplifies by using 
$\left( \not k-\not l+m \right) \gamma^\alpha \, u( k , h )
\,=\, \left( 2 k^\alpha - \not l \gamma^\alpha \right) \, u( k , h )
\, \approx\, 2  k^\alpha \, u( k , h )$, 
which yields for Eq.~(\ref{eq:b1i}) in the soft photon limit~ :
\begin{equation}
M_{b1i} \,=\, M_{BH}^i \, (-e) \, \varepsilon_\alpha^*(l) 
\left[ - {{k^\alpha} \over {k.l}}\right] \, ,
\end{equation}
where $M_{BH}^i$ is the Bethe-Heitler amplitude of Eq.~(\ref{eq:bhi}) -
corresponding with photon emission from the initial lepton. 
Similarly, we can derive the amplitude for Figs.~\ref{fig:brems} (b2i,
b1f and b2f) which yields in the soft photon limit~:
\begin{eqnarray}
\label{eq:bremsbhi}
M_{b1i} + M_{b2i} \,&=&\, M_{BH}^i \, (-e) \, \varepsilon_\alpha^*(l) 
\left[{{k^{' \alpha}} \over {k'.l}} - {{k^\alpha} \over {k.l}}\right] \, ,\\
M_{b1f} + M_{b2f} \,&=&\, M_{BH}^f \, (-e) \, \varepsilon_\alpha^*(l) 
\left[{{k^{' \alpha}} \over {k'.l}} - {{k^\alpha} \over {k.l}}\right] \, ,
\label{eq:bremsbhf}
\end{eqnarray}
where $M_{BH}^f$ is the Bethe-Heitler amplitude of Eq.~(\ref{eq:bhf}) -
corresponding with photon emission from the final lepton. 

Figs.~\ref{fig:brems} (b3i) and (b3f) contain the contributions where
the soft photon couples to an off-shell lepton line. 
The amplitude corresponding with Fig.~\ref{fig:brems} (b3i) is given by~:
\begin{eqnarray}
\label{eq:b3i}
M_{b3i} \,=\,&&- i \,e^4 \,\bar u( k^{'} , h' )\,\gamma ^\nu\,
{{\left( {\not k-\not q'-\not l+m} \right)} 
\over {-2k.q' -2 l.(k-q')}}\,\not \varepsilon^*(l) \,
{{\left( {\not k-\not q'+m} \right)} \over {-2k.q'}}\,
\not \varepsilon^*(q') \, u( k , h ) \nonumber\\ 
&&\times\;{1 \over {\left( {p^{'} - p} \right)^2}} \;
\bar N( p^{'} , s_p^{'} )\,\Gamma_{\nu}\left(p^{'} , p\right)\, 
N( p , s_p ) \, , 
\end{eqnarray}
In the soft photon limit, Eq.~(\ref{eq:b3i}) can be simplified by using~:
\begin{eqnarray}
{{\left( {\not k-\not q'-\not l+m} \right)} 
\over {-2k.q' -2 l.(k-q')}}\, \gamma^\alpha \,
{{\left( {\not k-\not q'+m} \right)} \over {-2k.q'}}\,
&\approx&{{\left( {\not k-\not q'+m} \right)} 
\over {-2k.q'}}\, \gamma^\alpha \,
{{\left( {\not k-\not q'+m} \right)} \over {-2k.q'}}\, \nonumber\\
&=& {{\left( {\not k-\not q'+m} \right)} \over {-2k.q'}}\,
{{(k - q')^\alpha} \over {-k.q'}} \,-\,{{\gamma^\alpha} \over {-2
    k.q'}} \, .  
\end{eqnarray}
Consequently, the amplitude of Eq.~(\ref{eq:b3i}) is given by~:
\begin{eqnarray}
\label{eq:b3i2}
M_{b3i} \,&=&\, M_{BH}^i \, (-e)\, \varepsilon_\alpha^*(l) 
{{(k - q')^\alpha} \over {-k.q'}} \nonumber\\
&+&\, i \,e^4 \,\bar u( k^{'} , h' )\, 
{{\gamma ^\nu\,\not \varepsilon^*(l) \,\not \varepsilon^*(q')} \over 
{-2k.q'}}\,u( k , h ) \, 
{1 \over {\left( {p^{'} - p} \right)^2}} \;
\bar N( p^{'} , s_p^{'} )\,\Gamma_{\nu}\left(p^{'} , p\right)\, 
N( p , s_p ) \, . 
\end{eqnarray}
Similarly, the amplitude corresponding with 
Fig.~\ref{fig:brems} (b3f) is given by~:
\begin{eqnarray}
\label{eq:b3f}
M_{b3f} \,&=&\, M_{BH}^f \, (-e)\, \varepsilon_\alpha^*(l) 
{{(k' + q')^\alpha} \over {k'.q'}} \nonumber\\
&+&\, i \,e^4 \,\bar u( k^{'} , h' )\, 
{{\not \varepsilon^*(q')\, \not \varepsilon^*(l) \, \gamma ^\nu} \over 
{2k'.q'}}\,u( k , h ) \,
{1 \over {\left( {p^{'} - p} \right)^2}} \;
\bar N( p^{'} , s_p^{'} )\,\Gamma_{\nu}\left(p^{'} , p\right)\, 
N( p , s_p ) \, . 
\end{eqnarray}

In complete analogy to Eqs.~(\ref{eq:bremsbhi},\ref{eq:bremsbhf}), we
can also calculate the soft photon emission contributions to the VCS
process. They are shown in Figs.~\ref{fig:brems} (b4) and (b5), and
their calculation in the soft photon limit yields~:
\begin{equation}
M_{b4} + M_{b5} \,=\, M_{VCS} \, (-e) \, \varepsilon_\alpha^*(l) 
\left[{{k^{' \alpha}} \over {k'.l}} - {{k^\alpha} \over {k.l}}\right] \, ,
\label{eq:bremsvcs}
\end{equation}
where $M_{VCS}$ is the VCS amplitude of Eq.~(\ref{eq:m_vcs}).
\newline
\indent
We see from
Eqs.~(\ref{eq:bremsbhi},\ref{eq:bremsbhf} and \ref{eq:bremsvcs}), 
that for the diagrams of Fig.~\ref{fig:brems} where the soft photon
couples to an on-shell lepton, the original amplitude factorizes~: 
in Eqs.~(\ref{eq:bremsbhi},\ref{eq:bremsbhf}) the BH amplitude factorizes, 
and in Eq.~(\ref{eq:bremsvcs}) the VCS amplitude factorizes. 
The resulting amplitudes are proportional to $1/l$, which leads to
a logarithmic divergence when integrating over the phase space of the
soft photon. In contrast, the amplitudes of
Eqs.~(\ref{eq:b3i2},\ref{eq:b3f}) where the photon couples to an
off-shell lepton line are finite when $l \to 0$, and the corresponding
phase space integral becomes vanishingly small in the limit $l \to 0$. 

\subsubsection{Radiative correction due to soft-photon emission processes}

In the soft-photon limit, we therefore need only to keep the bremsstrahlung
corrections of Eqs.~(\ref{eq:bremsbhi},\ref{eq:bremsbhf} and
\ref{eq:bremsvcs}), where the BH and VCS amplitudes factorize. 
To first order in $\alpha_{em}$ (relative 
to the BH + VCS cross section) the bremsstrahlung correction therefore
amounts to calculate the phase space integral of the form~: 
\begin{eqnarray}
d \sigma \sim   
{{d^3 \vec k^{\; '}_e} \over {\left( {2\pi }\right)^3 \, 2 E'_e}} 
{{d^3 \vec q^{\; '}} \over {\left( {2\pi }\right)^3 \, 2 |\vec q^{\, '}|}} 
&&{{d^3 \vec p_N^{\; '}} \over {\left( {2\pi }\right)^3 \, 2 E'_N}} 
{{d^3 \vec l} \over {\left( {2\pi }\right)^3 \, 2 {\mathrm l}}} 
\; (2 \pi)^4 \delta^4(k + p - k' - q' - p' - l) \nonumber\\
&&\times\, |M_{BH} + M_{VCS}|^2 \, \left( - e^2 \right)  
\left[{ {k^{'}_\mu \over {k^{'}.l}} - {k_\mu \over {k.l}}}
  \right] . 
\left[{ {k'^\mu \over {k^{'}.l}} - {k^\mu \over {k.l}}}
  \right] ,
\label{eq:bremsint1}
\end{eqnarray}
where ${\mathrm l} \equiv | \vec l|$ denotes the soft photon energy,
and where the total BH amplitude is given by $M_{BH} = M^i_{BH} + M^f_{BH}$.
The calculation of the bremsstrahlung integral of
Eq.~(\ref{eq:bremsint1}) goes along similar lines as the corresponding 
integral for elastic scattering, for which the technical details can
be found in appendix \ref{app:elast} (section \ref{app:elast5}). 
We will point out in this section the differences which arise for the
$e p \to e p \gamma$ reaction. 
\newline
\indent
There are two practical ways to measure the $e p \to e p \gamma$
reaction, by measuring two particles in the final state. 
One can either measure the outgoing electron in coincidence
with the recoiling nucleon : this is the ideal technique when measuring
the $e p \to e p \gamma$ reaction at low outgoing photon energy as is done in
\cite{mami,jlab,bates}. 
The alternative is to measure the outgoing electron in
coincidence with the photon : this is the technique when doing a very
inelastic experiment, such as deeply virtual Compton scattering, where
the photon is produced with a large energy. 
We discuss here first the case where one detects the outgoing electron and
photon, and indicate at the end the changes which apply when measuring
the outgoing electron and recoiling nucleon. 
\newline
\indent
If one measures the $e p \to e p \gamma$ reaction  
by detecting the outgoing electron 
and photon, one eliminates in Eq.~(\ref{eq:softel0a}) the
integral over $\vec p_N^{\;'}$ with the momentum conserving
$\delta$-function, which gives~:
\begin{eqnarray}
d \sigma \sim   
&&{{d^3 \vec k^{\; '}_e} \over {\left( {2\pi }\right)^3 \, 2 E'_e}} 
{{d^3 \vec q^{\; '}} \over {\left( {2\pi }\right)^3 \, 2 |\vec q^{\, '}|}} 
{{d^3 \vec l} \over {\left( {2\pi }\right)^3 \, 2 {\mathrm l}}} 
\, {1 \over {2 E'_N}} \, \nonumber\\
&&\times \,(2 \pi) \delta\left(E_e + E_N - E'_e - |\vec q^{\, '}| - 
\sqrt{(\vec q + \vec p_N - \vec q^{\, '} - \vec l)^2 + M_N^2}
- {\mathrm l} \right) \nonumber\\
&&\times\, | M_{BH} + M_{VCS}|^2 \, \left( - e^2 \right)  
\left[{ {k^{'}_\mu \over {k^{'}.l}} - {k_\mu \over {k.l}}}
  \right] . 
\left[{ {k'^\mu \over {k^{'}.l}} - {k^\mu \over {k.l}}}
  \right] .
\label{eq:bremsint2}
\end{eqnarray}
Due to the energy conserving $\delta$-function in
Eq.~(\ref{eq:softel0b}), the upper limit in the 
integration over the soft photon phase space depends on the angle. 
Therefore, this integration volume 
has a complicated ellipsoidal shape in the {\it lab} system. 
In order for the soft-photon phase space integration volume 
to be spherical, one has to perform
the calculation in the c.m. system ${\mathcal S}_1$ of the (recoiling nucleon +
soft-photon), generalizing the procedure of appendix \ref{app:elast}
for elastic scattering to the $e p \to e p \gamma$ reaction. 
The system ${\mathcal S}_1$ is defined by~: 
$\vec p_N^{\; '} + \vec l = \vec p_N + \vec q - \vec q^{\; '} = 0$. 
In the system ${\mathcal S}_1$, 
the energy conserving delta function in Eq.~(\ref{eq:bremsint2})
is independent of the soft-photon
angles, and the maximal soft photon energy is isotropic. 
The integral over the soft-photon momentum (up to some maximum value
$\Delta E_s$) can then be performed independently from the integration
over the soft photon emission angles. 
If $\Delta E_s$ is sufficiently small, one can furthermore 
neglect the soft photon energy with respect to the other energies 
in the $\delta$-function, and perform the integral over
the photon momentum $|\vec q^{\; '}|$ in Eq.~(\ref{eq:bremsint2}) 
to obtain the correction to the fivefold differential $e p \to e p
\gamma$ cross section. We indicate in the following only how the
squared matrix element for the $e p \to e p \gamma$ reaction is
modified due to soft photon emission. This correction due to soft 
bremsstrahlung is given by~:
\begin{equation}
\label{eq:bremsint3}
| M^{SOFT \gamma}_{e p \to e p \gamma} |^2
\,=\, | M_{BH} + M_{VCS} |^2 \nonumber\\
\left( -\,e^2 \right)  
\int {{{d^3 \vec l} \over {\left( {2\pi }\right)^3 \, 2 {\mathrm l}}}} 
\left[{ {k^{'}_\mu \over {k^{'}.l}} - {k_\mu \over {k.l}}}
  \right] . 
\left[{ {k'^\mu \over {k^{'}.l}} - {k^\mu \over {k.l}}}
  \right] .
\end{equation}
The correction factor multiplying $| M_{BH} + M_{VCS} |^2$ gives 
immediately the correction factor to the fivefold $e p \to e p \gamma$
cross section. In Eq.~(\ref{eq:bremsint3}), the soft-photon phase
space integral is understood to be performed in
the system ${\mathcal S}_1$, where the integration volume is spherical. 
Its calculation was already performed in appendix \ref{app:elast}. 
One sees that the integral in Eq.~(\ref{eq:bremsint3}) has a logarithmic 
IR divergence, corresponding with the emission of photons with zero energy. 
To evaluate it, one has to regularize it, which is done in this work 
by using dimensional regularization. 
This amounts to evaluate the integral (in the system ${\mathcal S}_1$)
in $D - 1$ dimensions ($D \rightarrow 4$ corresponds to the physical limit).  
This calculation is performed in appendix \ref{app:elast} 
and yields (similar to Eq.~(\ref{eq:softel})) as result~: 
\begin{eqnarray}
| M^{SOFT \gamma}_{e p \to e p \gamma} |^2
\,&=&\, | M_{BH} + M_{VCS} |^2 \nonumber\\
&\times&\, \left\{ {{e^2} \over {4 \pi^2}} \, 
 \left[ - {1 \over {\varepsilon _{IR}}}+\gamma_E 
-\ln \left( {{4\pi \mu^2} \over {m^2}} \right)\, \right]  \, 
\left[ {{{v^2+1} \over {2 v}}\,\ln \left( {{{v+1} \over {v-1}}}
    \right) \,-\, 1} \right] \,+\,\delta_R \right\} ,
\label{eq:softvcs}
\end{eqnarray}
In Eq.~(\ref{eq:softvcs}), $\delta_R$ is the finite part 
of the real radiative correction
corresponding with soft photon emission, and is given as in 
appendix~\ref{app:elast} (Eq.~(\ref{eq:radcorrreal})) by~:
\begin{eqnarray}
\delta_R &\stackrel{Q^2 >> m^2}{\longrightarrow}&
 {\alpha_{em} \over \pi} \left\{
\ln \left( {{ (\Delta E_s)^2} \over { \tilde E_e \tilde E_e^{'}}} \right)  
\left[ \ln \left( {Q^2 \over m^2} \right) \,-\, 1 \right] \right. \nonumber\\
&&\hspace{.8cm}\left. - {1 \over 2} \ln^2 \left( {{\tilde E_e} 
\over {\tilde E_e^{'}}}  
\right) \,+\, {1 \over 2} \ln^2 \left( {Q^2 \over m^2} \right)
\,-\, {{\pi^2} \over 3} \,+\, Sp\left( \cos^2 {\tilde \theta_e \over 2}\right) 
\right\} \;,
\label{eq:radcorrrealbis}
\end{eqnarray}
\indent
In Eq.~(\ref{eq:radcorrrealbis}), we next have to express 
the kinematical variables ($\tilde E_e, \tilde E_e', \cos \tilde \theta_e$) 
in the system ${\mathcal S}_1$ (denoted by tilded quantities)  
in terms of the {\it lab} quantities, which we denote
by untilded quantities ($E_e, E_e', \cos \theta_e$). 
To make the transformation between the system ${\mathcal S}_1$ and the 
{\it lab} system, we first introduce the 
missing four-momentum $p_{m1} \equiv p_N^{\; '} + l$. 
The system ${\mathcal S}_1$ is defined by $\vec p_{m1} = \vec 0$, 
and the missing mass $M_{m1}$ of the system 
($p'+l$) is defined by~:
\begin{equation}
M^2_{m1} = (p' + l)^2 = (p + q - q')^2 \;.
\label{eq:missmass1}
\end{equation}
We can then easily express the electron energies and angle 
in the system ${\mathcal S}_1$ in terms
of {\it lab} quantities~:
\begin{eqnarray}
&&\tilde E_e = {{k . p_{m1}} \over {M_{m1}}} 
= {1 \over M_{m1}} k . (p + q - q')
= {{M_N} \over {M_{m1}}} 
\left(E_e - {{Q^2} \over {2 M_N}} - {{k \cdot q'} \over {M_N}} \right) \, , 
\label{eq:vcstransf1} \\
&&\tilde E_e' = {{k' . p_{m1}} \over {M_{m1}}} 
= {1 \over M_{m1}} k' . (p + q - q')
= {{M_N} \over {M_{m1}}} 
\left(E_e' + {{Q^2} \over {2 M_N}} - {{k' \cdot q'} \over {M_N}} \right) \, , 
\label{eq:vcstransf2} \\
&&\sin^2 \tilde \theta_e/2 = {{E_e E_e'} \over {\tilde E_e \tilde E_e'}}
\sin^2 \theta_e/2 \;.
\label{eq:vcstransf3} 
\end{eqnarray}
The maximal soft-photon energy $\Delta E_s$ in the system ${\mathcal S}_1$, 
is given by~:
\begin{equation}
\Delta E_s = {{M^2_{m1} - M^2_N} \over {2 M_{m1}}} \;.  
\label{eq:deltae1vcs}
\end{equation}
\indent
If one measures the $e p \to e p \gamma$ reaction by detecting the
outgoing electron and recoiling proton, the derivation goes along
similar lines as above. 
One starts now by eliminating in  
Eq.~(\ref{eq:softel0a}) the integral over $\vec q^{\;'}$. Then one
goes into the c.m. system ${\mathcal S}_2$ of the (VCS photon $q'$ +
soft photon), 
where the energy conserving $\delta$-function is independent of the soft-photon
angles, and where the maximal soft photon energy is isotropic. 
This system ${\mathcal S}_2$ is defined by~: 
$\vec q^{\; '} + \vec l = \vec p_N + \vec q - \vec p_N^{\; '} = 0$. 
The calculation of the soft-photon emission integral 
is then completely similar as above, and leads to the finite
correction of Eq.(\ref{eq:radcorrrealbis}), where the 
the kinematical variables ($\tilde E_e, \tilde E_e', \cos \tilde \theta_e$) 
are now understood in the system ${\mathcal S}_2$.  
To make the transformation between the system ${\mathcal S}_2$ and the 
{\it lab} system, we first introduce the 
missing four-momentum $p_{m2} \equiv q^{\; '} + l$. 
The system ${\mathcal S}_2$ is defined by $\vec p_{m2} = \vec 0$, and
the missing mass $M_{m2}$ of the system ($q'+l$) is defined by~:
\begin{equation}
M^2_{m2} = (q' + l)^2 = (p + q - p')^2 \;.
\label{eq:missmass2}
\end{equation}
We can then easily express the electron energies and angle 
in the system ${\mathcal S}_2$ in terms
of {\it lab} quantities~:
\begin{eqnarray}
&&\tilde E_e = {{k . p_{m2}} \over {M_{m2}}} 
= {1 \over M_{m2}} k . (p + q - p')
= {{M_N} \over {M_{m2}}} \left(E_e - {{Q^2} \over {2 M_N}} 
- {{k \cdot p'} \over {M_N}} \right) \, , 
\label{eq:vcstransf1b} \\
&&\tilde E_e' = {{k' . p_{m2}} \over {M_{m2}}} 
= {1 \over M_{m2}} k' . (p + q - p')
= {{M_N} \over {M_{m2}}} \left(E_e' + {{Q^2} \over {2 M_N}} 
- {{k' \cdot p'} \over {M_N}} \right) \, , 
\label{eq:vcstransf2b} \\
&&\sin^2 \tilde \theta_e/2 = {{E_e E_e'} \over {\tilde E_e \tilde E_e'}}
\sin^2 \theta_e/2 \;.
\label{eq:vcstransf3b} 
\end{eqnarray}
The maximal soft-photon energy $\Delta E_s$ in the system ${\mathcal S}_2$, 
is given by~:
\begin{equation}
\Delta E_s = {{M_{m2} \over 2}} \;, \\ 
\label{eq:deltae2vcs}
\end{equation}

\subsubsection{Cancellation of IR divergences}

We can now demonstrate for the $e p \to e p \gamma$ reaction, that the
IR divergences from the soft photon emission corrections exactly
cancel against the IR divergences from the virtual radiative
corrections, calculated in section \ref{sec:virtual}. 
Concentrating here only on the IR divergent parts 
of the virtual radiative corrections, we found in section
\ref{sec:virtual} that the amplitudes of 
Eqs.~(\ref{eq:v1_i5},\ref{eq:v1_f1},\ref{eq:v2_i1},\ref{eq:v2_f1},\ref{eq:v3_i8},\ref{eq:v3_f1},\ref{eq:self_i2},\ref{eq:self_f2},
and \ref{eq:vcsvertex}) contain IR divergences. 
Those IR divergent parts are given by~:
\begin{eqnarray}
&&M^i_{V1} + (CT)^i_{V1} + M^f_{V1} + (CT)^f_{V1}
\,\to\, M_{BH} \,  {{e^2} \over {4 \pi^2}} 
\, \left( {{-1} \over 2} \right) 
 \left[ {1 \over {\varepsilon _{IR}}}-\gamma_E 
+\ln \left( {{4\pi \mu^2} \over {m^2}} \right)\, \right]  \, , 
\label{eq:bhir1} \\
&&M^i_{V2} + (CT)^i_{V2} + M^f_{V2} + (CT)^f_{V2}
\,\to\, M_{BH} \, {{e^2} \over {4 \pi^2}} 
\, \left( {{-1} \over 2} \right)
 \left[ {1 \over {\varepsilon _{IR}}}-\gamma_E 
+\ln \left( {{4\pi \mu^2} \over {m^2}} \right)\, \right]  \, , 
\label{eq:bhir2} \\
&&M^i_{V3} + M^f_{V3}
\,\to\, M_{BH} \,  {{e^2} \over {4 \pi^2}} \, \left({1 \over 2} \right) 
{{v^2+1} \over {2 v}}\,\ln \left( {{{v+1} \over {v-1}}} \right) \; 
 \left[ {1 \over {\varepsilon _{IR}}}-\gamma_E 
+\ln \left( {{4\pi \mu^2} \over {m^2}} \right)\, \right]  \, , 
\label{eq:bhir3} \\
&&M_{Si} + M_{Sf} 
\,\to\, M_{BH} \, {{e^2} \over {4 \pi^2}}  
\, \left( {1 \over 2} \right)
 \left[ {1 \over {\varepsilon _{IR}}}-\gamma_E 
+\ln \left( {{4\pi \mu^2} \over {m^2}} \right)\, \right]  \, ,
\label{eq:bhir4} \\
&&M_{V4} + (CT)_{V4}
\to M_{VCS} \,  {{e^2} \over {4 \pi^2}} 
 \left( {1 \over 2} \right)  
\left[ {{v^2+1} \over {2 v}}\,\ln \left( {{{v+1} \over {v-1}}} \right)  
- 1\right]   \left[ {1 \over {\varepsilon _{IR}}}-\gamma_E 
+\ln \left( {{4\pi \mu^2} \over {m^2}} \right) \right]  .
\end{eqnarray}
Adding them all up gives the following correction to the squared
amplitude for the virtual radiative corrections~:
\begin{eqnarray}
&&|\, M_{BH} + M_{DVCS} + M^{VIRTUAL \gamma}_{e p \to e p \gamma}
\,|^2 \nonumber\\
&&= |M_{BH} + M_{VCS}|^2 \left\{ 1 + {{e^2} \over {4 \pi^2}} 
\left[ {{v^2+1} \over {2 v}}\,\ln \left( {{{v+1} \over {v-1}}} \right)  
- 1\right]   \left[ {1 \over {\varepsilon _{IR}}}-\gamma_E 
+\ln \left( {{4\pi \mu^2} \over {m^2}} \right)\, \right] \right\} +
... ,
\label{eq:irdivvirtual}
\end{eqnarray}
where the ellipses denote the finite first order virtual radiative correction 
to the $e p \to e p \gamma$ reaction as was calculated and can be
found in section \ref{sec:virtual}.  
Adding the virtual (Eq.(\ref{eq:irdivvirtual})) and real 
(Eq.(\ref{eq:softvcs})) radiative corrections to the 
$e p \to e p \gamma$ reaction, one verifies
that the IR divergences in the sum exactly cancel, showing QED at work! 
Note that this cancellation is different as
compared to the case of elastic electron scattering. 
Indeed, for the virtual photon correction diagrams 
to the Bethe Heitler process, there are 3 types of vertex
diagrams (Eqs.~(\ref{eq:bhir1},\ref{eq:bhir2} and \ref{eq:bhir3})) 
and the self energy diagram (Eq.~(\ref{eq:bhir4})), and the 
corresponding counterterms, which have an IR divergence.
On the other hand, for the virtual radiative corrections to elastic electron
scattering, there is only one vertex diagram which is IR divergent.


\subsection{Integration method for the virtual photon corrections}
\label{sec:int}

At this stage of the calculation of the first order QED radiative corrections 
to the $e p \to e p \gamma$ reaction, the treatment of 
all UV and IR divergences, resulting from the radiative corrections at
the electron side, has been performed. 
The UV divergences have been removed by the renormalization procedure 
whereas the IR divergences were shown to cancel at the cross section level 
when adding the soft photon emission processes. 
Now, the evaluation of the remaining Feynman parameter 
integrals in the finite terms 
such as in Eq.~(\ref{eq:v1_i5}) still has to be done. 
\newline
\indent
Among the one-loop virtual radiative 
corrections to the $e p \to e p \gamma$ reaction shown in 
Fig.~\ref{fig:radcorr}, six give rise to 
simple analytical formulas. For the six vertex diagrams, 
denoted by V1i,V2i,V3i,V1f,V2f and V3f, the trick consisting of 
adding and subtracting the divergent term for each of them 
(as explained in Section \ref{sec:virtual}) gives rise to Feynman parameter 
integrals that are rather complicated to be done analytically.  
Therefore, we will evaluate them in this work by a numerical procedure. 
Although these Feynman parameter integrals are by construction finite,
appropriate numerical methods are needed to perform them. 
Two main difficulties are encountered in these numerical integrations. 
Firstly, the variations of the integrated functions are always extremely 
sharp near the integration limits. In fact, a 
typical behaviour is a rather flat dependence in the middle of the domain 
and two pronounced rises when approaching 
$0$ or $1$ for the Feynman parameters 
with a width of the order $m/E_e$. The contribution of 
these two peaks has to be evaluated carefully  
in order to obtain a good precision for the final result. 
Secondly, we know that the virtual radiative corrections to the 
$e p \to e p \gamma$ reaction allow the 
propagation of on-shell states (see section \ref{sec:virtual}). This 
is mathematically expressed by the presence of integrable 
singularities in the Feynman parameter integrals which require an analytical 
continuation into the complex plane and gives rise to an imaginary part 
for the amplitude.
\newline
\indent
To evaluate the Feynman parameter integrals, 
our strategy is to perform the first integration analytically. 
The last integrations will then be performed numerically using the 
Gauss-Legendre integration routine. 
The analytical calculation of the first integration 
provides a shorter calculational time and a higher precision. 
The main advantage however is that in the case of a 
singularity, the pole is avoided by deforming the integration contour
into the complex plane, using analytical continuation. 
In this way, one removes the difficulties for the remaining  
integrations along the real axis.
\newline
\indent
To classify the Feynman parameter integrals that occur in the six
vertex diagrams under study, we start by 
factorizing all the Dirac $\gamma$ matrices and decomposing the 
components of the four-vectors. 
All resulting integrals then reduce to the generic form~:
\begin{equation}
\int\!\!\!\!\int\!\!\!\!\int\limits_0^1\ dx_1 dx_2 dx_3\ \frac{P(x_1,x_2,x_3)}{Q(x_1,x_2,x_3)}\;,
\label{generic_int}
\end{equation}
where $P$ and $Q$ are polynomials in three (real) 
Feynman parameters $x_1, x_2, x_3$. 
Let's choose $x_1$ to be the more internal variable. 
Then the  first integration is either of the form~:
\begin{equation}
\int\limits_0^1\ \frac{x_1^m\ dx_1}{(\alpha x_1+\beta )^n} \;,
\label{eq:first_int1}
\end{equation}
or 
\begin{equation}
\int\limits_0^1\ \frac{x_1^m\ dx_1}{(\alpha x_1^2+\beta x_1+\gamma)^n}\;,
\label{eq:first_int2}
\end{equation}
where $\alpha$, $\beta$ and $\gamma$ are polynomials in $x_2$ and $x_3$ 
with coefficients that are functions of kinematical variables. 
In Eqs.~(\ref{eq:first_int1},\ref{eq:first_int2}),
$m$ varies from $0$ to $4$ and $n$ is equal to $1$ or $2$, 
to accomodate all cases appearing in section \ref{sec:virtual}. 
These successive decompositions increase the number of terms to 
calculate but they have the advantage to provide two simple 
classes of integrals without any 
vector or matrix dependence. The possibility of poles 
in the integrands of Eqs.~(\ref{eq:first_int1},\ref{eq:first_int2}) 
naturally splits the problem into two parts, whether the integrand is 
regular or singular.

\subsubsection{Regular integrand}

When the denominator doesn't have any singularities, some recurrence 
relations exist for these integrals and can be 
found in Ref.\cite{vdw}. Unfortunately for small values of $\alpha$ 
as compared to $\beta$ or to $\gamma$, it has been seen  
that these relations are numerically unstable. This has thus led us 
to use several methods of integration with each a 
different domain of validity. For small ratios $r$ 
($r = \alpha /\beta$ for Eq.~(\ref{eq:first_int1}) or 
$r = \alpha /\gamma$ for Eq.~(\ref{eq:first_int2}) )
as compared to 1, we perform 
a Taylor expansion of the integral and tune the order of each 
development to complete a fixed criterium of 
convergence (for example we require that the ratio between the last and the 
first terms is of the order of the numerical precision in double precision). 
For $r > 1$ the recurrence relations \cite{vdw} are used as they are stable 
in this range. In the intermediate zone $(0.2 \leq r \leq 1)$, 
we use the Gauss-Legendre numerical integration method. 

\subsubsection{Singular integrand}

In the case of the propagation of on-shell intermediate states, the 
polynomials of the denominators in Eqs.
(\ref{eq:first_int1},\ref{eq:first_int2})
acquire one (or two) roots in the domain of integration. 
Some simple physical considerations have 
shown that among the six diagrams numerically evaluated, the three processes 
where the photon in the $e p \to e p \gamma$ reaction is emitted from 
the initial electron line are free of poles (section \ref{sec:virtual}). 
In contrast, the three vertex graphs 
where the photon is emitted from the final electron line 
were seen to contain singularities. 
The corresponding integrals are then defined by an analytical continuation 
into the complex plane and take the form~:
\begin{equation}
\int\limits_0^1\ \frac{x_1^m\ dx_1}{(\alpha x_1+\beta \pm i\epsilon)^n} 
\qquad  {\mathrm or} \qquad  
\int\limits_0^1\ 
\frac{x_1^m\ dx_1}{(\alpha x_1^2+\beta x_1+\gamma \pm i\epsilon)^n}\;.
\label{complex_int}
\end{equation}
The prescription for on-shell propagation is of course already taken into 
account in the propagators and determines the sign in 
front of the $i\epsilon$ (which can also be obtained by 
applying the simple trick $m\rightarrow m-i\epsilon/2$). Complications can 
occur from the possibility of two 
distinct roots in the interval $[0,1]$ for the second order polynomial. 
An important remark then concerns the variable of 
integration. In Eq.~(\ref{generic_int}), the choice of $x_1$ as the 
more internal dimension was purely arbitrary. In 
fact all the decompositions in the three parameters have been derived and it 
has been shown that it was always 
possible to find an expansion providing at most one singularity.
\newline
\indent
In appendix \ref{app:b}, we give the analytical results for the
integrals of Eq.~(\ref{complex_int}). We checked these results with a
numerical method, where one pole along the interval [0,1] is avoided
by analytically continuing the integrand into the complex plane. In
this way, the integral along [0,1] is replaced by an integration along
a semi-circle (with origin at $0.5 + 0 \, i$ and radius 1/2) in the
opposite complex half-plane with respect to the pole. 
A comparison between the two methods shows a perfect agreement. Only
in the special cases where one pole comes close to an edge of the
domain of integration [0,1] (typically within a distance  
$m^2/E_e^2$ to 0 or 1), one has to increase the number of integration
points of the numerical method to obtain the same precision.

\subsubsection{Numerical checks and accuracy}

\begin{figure}[h]
\begin{center}
\leavevmode
\hbox{
\epsfysize=15.0truecm
\epsfbox{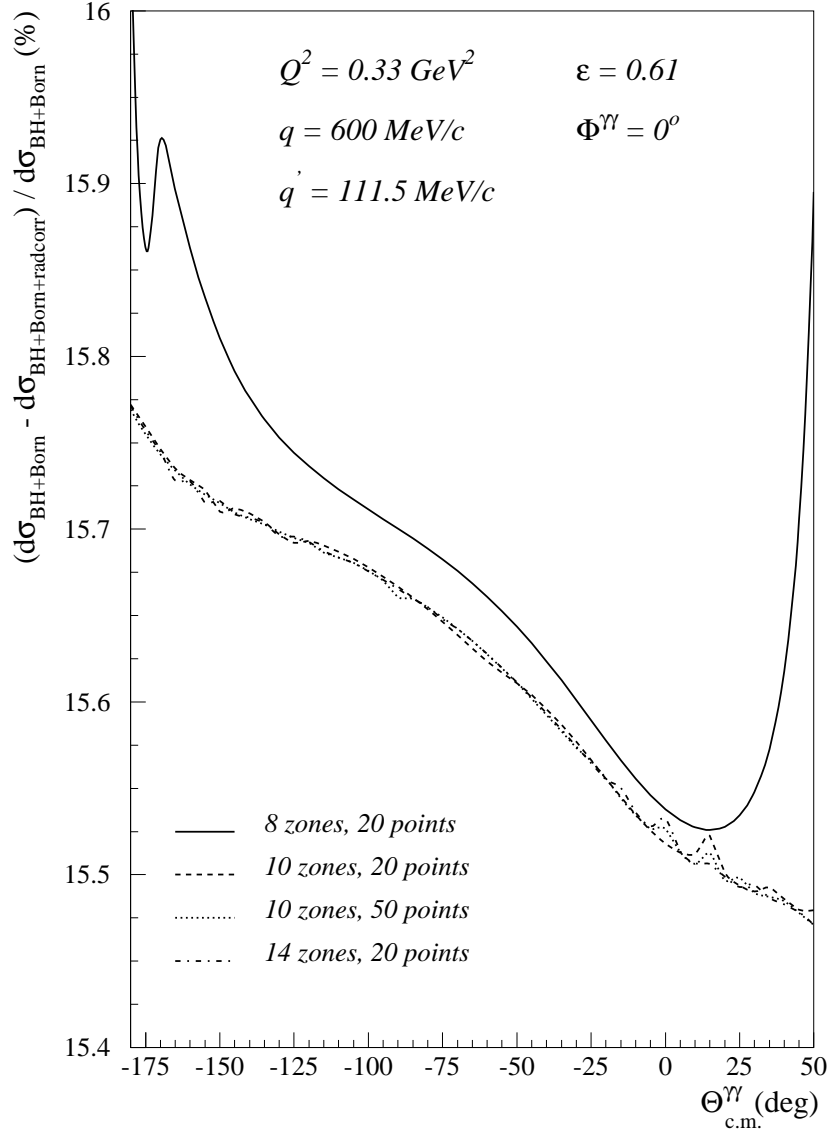} }
\end{center}
\centerline{\small \it  }
\vspace{-0.75cm}
\caption{Test of convergence : The relative effect of the virtual
  radiative corrections to the BH + Born cross section is shown as a
  function of the angle between the two photons $(q,q')$ for a typical
  MAMI kinematics. The curves correspond to tests performed with different
  densities of integration zones and points near the edge of the integration
  domain. They show the good numerical convergence obtained.}
\label{fig:conv_mami}
\end{figure}

\begin{figure}[h]
\begin{center}
\leavevmode
\hbox{
\epsfysize=15.0truecm
\epsfbox{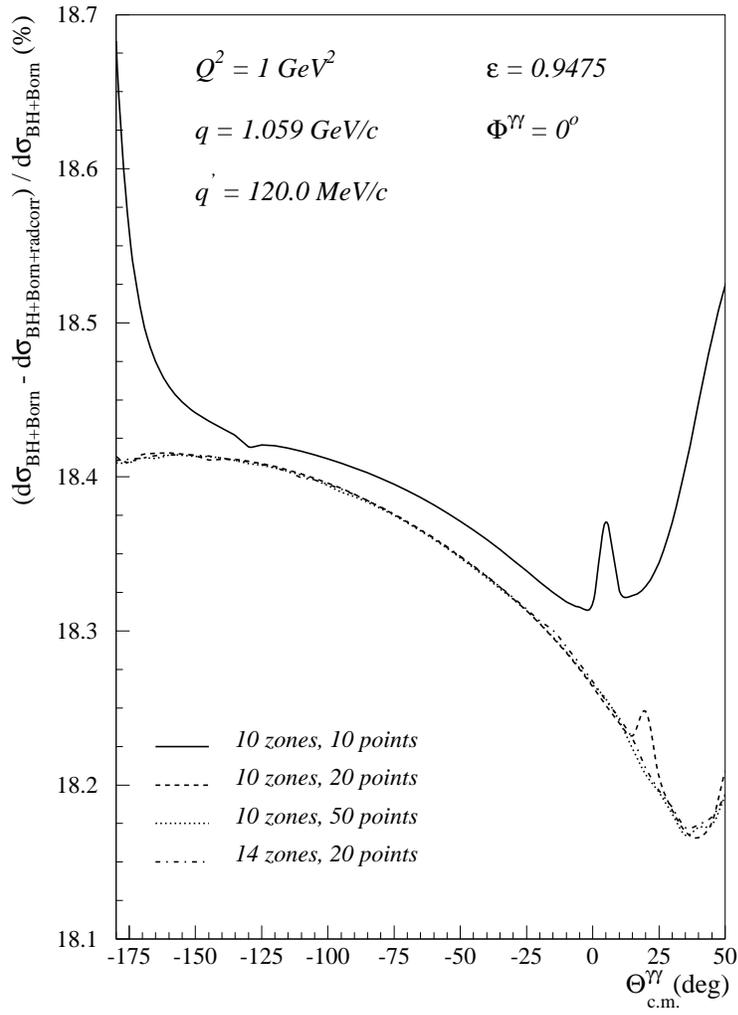} }
\end{center}
\centerline{\small \it  }
\vspace{-0.75cm}
\caption{Analogous test of convergence as in Fig.~\ref{fig:conv_mami}
but for JLab kinematics.}
\label{fig:conv_cebaf}
\end{figure}

\begin{figure}[h]
\begin{center}
\leavevmode
\hbox{
\epsfysize=15.0truecm
\epsfbox{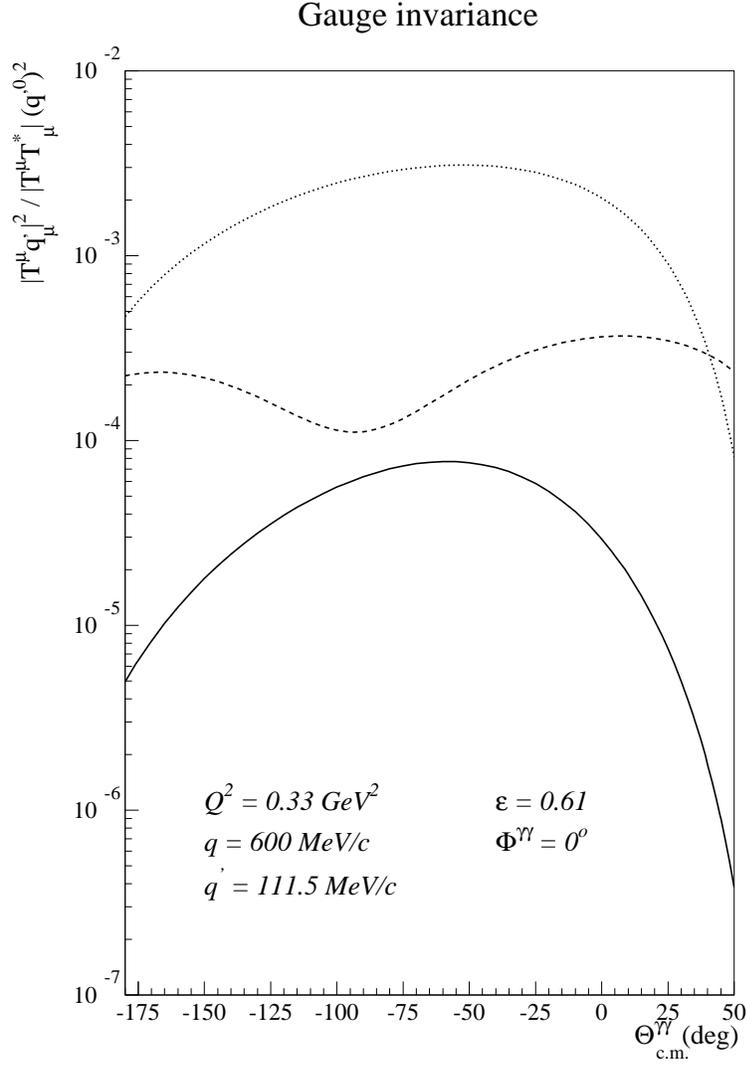} }
\end{center}
\centerline{\small \it  }
\vspace{-0.75cm}
\caption{Test of gauge invariance for MAMI kinematics. 
The dotted curve shows (for illustrative purpose only) the result when
only the diagrams (V2i) and (V2f) of Fig.~\ref{fig:radcorr} are
included. The dashed curve is the result of all analytically
calculated virtual radiative corrections. The full curves show the
result when also the numerical contributions (Feynman parameter
integrals) are included.}
\label{fig:gauge_inv}
\end{figure}

\begin{figure}[Hbtp]
\begin{center}
\leavevmode
\hbox{
\epsfysize=15.0truecm
\epsfbox{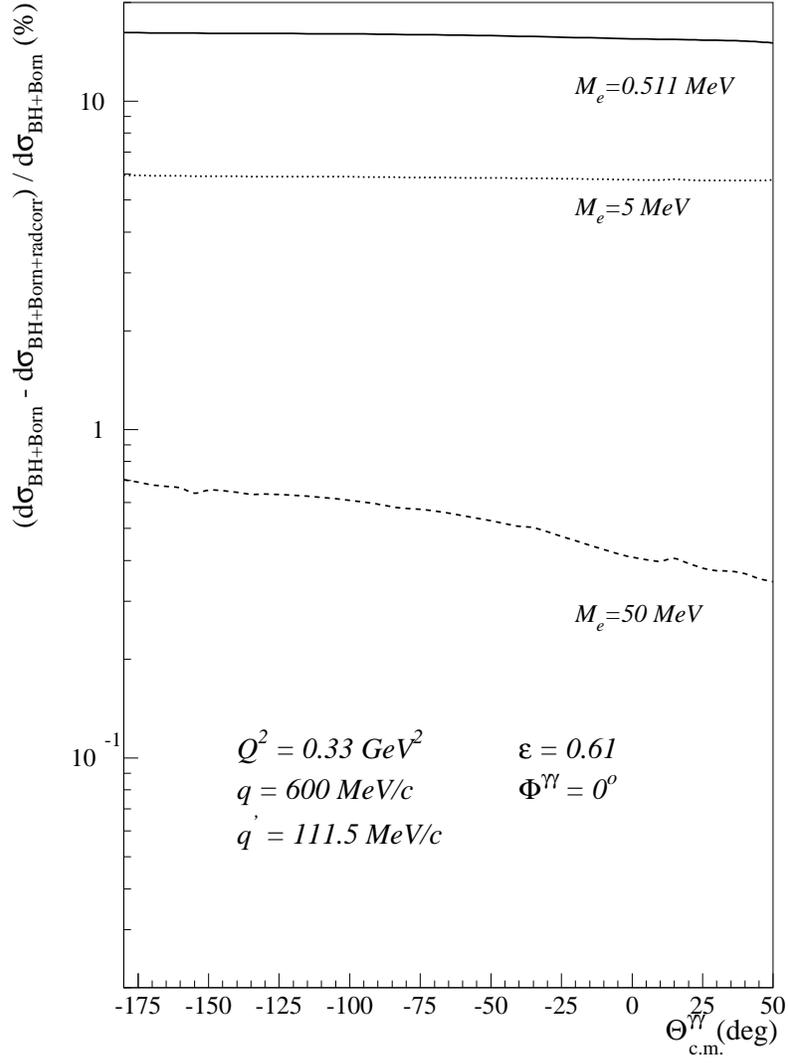} }
\end{center}
\centerline{\small \it  }
\vspace{-0.75cm}
\caption{Mass dependence of the virtual radiative corrections for MAMI kinematics.}
\label{fig:mass}
\end{figure}

Thanks to the analytical calculation of the first integration in the
Feynman parameter integrals under study, 
singularities on the real axis have been removed and the two
remaining integrations can then be performed numerically using the
Gauss-Legendre method. In the implementation of this algorithm the
major difficulty consisted in finding the suitable binnig of the
integration domain and to determine the number of points per bins. 
A detailed study of the integrated functions has been performed 
to estimate the width and
amplitude of the sharp variations close to the ends of the
domain. In this paragraph we discuss various checks of the 
precision of our results as well as their numerical stability. 
\newline
\indent
A strong cross check of the reliability of our calculations is
the exact agreement between two programs developed in paralell
\cite{david,domi}. Both of them use the same numerical method but they
have been coded independently using in most cases different
decomposition of the terms and different order in the integration
variables, which checks the symmetry in the permutation of $x_1,x_2$ and
$x_3$ variables. Comparison at each intermediate stage of the calculation 
also excludes any missprints in the writing of the 
quite extensive expressions.
\newline
\indent
Besides this agreement between two independent programs, the 
next requirement is the numerical convergence of the calculations. 
Figs.~\ref{fig:conv_mami} and \ref{fig:conv_cebaf} show results 
obtained for typical MAMI and JLab kinematics respectively. 
Beyond a certain density of
integration bins and points per bin, the numerical instabilities are
brought down to few $10^{-4}$ of the lowest order cross section. This
accuracy is far below all the other theoretical uncertainties related
to the performed approximations or experimental knowledge of the form
factors (of the order of 1\%). 
Nevertheless this kind of very good convergence is useful since
numerical instabilities can be amplified in the coherent sum of all
the diagrams or when computing higher energy kinematics. 
In the case of the deeply virtual Compton scattering, we have checked 
that one has to double the number of integration points, to get the
same numerical precision.  
\newline
\indent
Some features of the electromagnetic interaction itself can also be used to
check further the validity of our results. Let's consider the total
amplitude of the sum of all the virtual radiative correction
diagrams (Fig.~\ref{fig:radcorr}). Denoting the Lorentz index
associated with the real photon vertex by $\mu$, this amplitude can be written
as the scalar product $T^\mu\epsilon^*_\mu$ where $\epsilon^*$ stands
for the polarization vector of the real photon with four-vector 
$q^{'}$ and where $T^\mu$ represents the electromagnetic current. 
The gauge invariance of
electromagnetism implies $T^\mu q^{'}_\mu=0$ and provides us with a powerful
test of our calculations. Since our numerical accuracy is finite, we
cannot get exactly zero. Therefore, we rather define a quantity 
compared to which the scalar product $T^\mu q^{'}_\mu$ has to be small. 
A natural quantity is the product of the norms of the two Lorentz vectors. The
gauge invariance criteria thus becomes a test of the smallness of the
following dimensionless ratio~:
\begin{equation}
\frac{\left|T^\mu q^{'}_\mu \right|^2}{\left|T^\mu T^\dagger_\mu
  \right|(q^{'0})^2}\ll 1
\label{eq:gauge_criteria}
\end{equation}
This ratio is shown in Fig.~\ref{fig:gauge_inv} as a function of the
angle between $q$ and $q^{'}$. The gauge invariance is verified by the
fact that the smallest ratio (solid curve) stays in the range
$[10^{-4},10^{-6}]$ and is obtained when the complete set of diagrams
with analytical+numerical terms is included in $T^\mu$.
\newline
\indent
As a last consistency check, we investigated the mass dependence of the
virtual radiative corrections. The relative effect in the BH + Born
cross section is illustrated in Fig.~\ref{fig:mass} for different
values of the mass of the lepton. For this test we kept track of the mass
dependence in all the kinematical variables. We observe 
that when increasing the lepton mass (at fixed lepton kinematics), the
effect of the radiative corrections rapidly decreases, which reflects 
the suppression of photon emission by a heavy particle.


\subsection{Radiative corrections at the proton side and 
two-photon exchange corrections}
\label{sec:protonside}

In section \ref{sec:virtual} - \ref{sec:int}, we calculated the
radiative corrections to the $e p \to e p \gamma$ reaction,
corresponding with the diagrams of Figs.~\ref{fig:radcorr} and
\ref{fig:brems}. They are the virtual radiative
corrections at the lepton side, the vacuum polarization contributions
and the soft-photon emission from the lepton. These can be calculated
model-independently as has been shown above. 
Although these corrections are the dominant ones  
(when $Q^2 >> m^2$, leading to large logarithms), 
we want to estimate in this section how large are the virtual
radiative corrections at the proton side, the two-photon exchange
corrections (direct and crossed box diagrams) and the soft-photon
emission from the proton. Generally, the radiative corrections from
the proton side are typically suppressed compared with those from
the electron, due to the much larger mass of the proton. 
However, to calculate the first order
radiative corrections to the $e p \to e p \gamma$ reaction 
which originate from the proton side, one needs a model for the VCS
process. We do not aim in this paper, to calculate these corrections
within a given model. However, 
to provide some quantitative estimate, we will follow
the results of \cite{MTj99}, where the corrections at the proton
side were studied for elastic scattering.   
\newline
\indent
The $Z$-dependent corrections originate from the interference
between soft-photon emission from the electron and from the proton,
and from the two-photon exchange contributions (direct and crossed box
diagrams). Both processes contain IR divergences, which cancel in
their sum at the cross section level. 
The interference between the soft-photon emission from the electron
and from the proton can be calculated
along the same lines as in appendix~\ref{app:elast5} for the electron 
(neglecting form factor effects in the soft-photon limit). 
For the two-photon exchange contributions, 
the calculation is dominated by those regions in the integration 
where one of the two exchanged photons is soft. 
Therefore, one can evaluate the
rest of this amplitude by taking the momentum of either of the two exchanged
photons to be zero. In this approximation, the original amplitude
factorizes and one can follow the
derivation of \cite{MTj99}, where this same calculation has been performed
for elastic scattering. Therefore, the $Z$-dependent radiative
corrections can be estimated in the soft-photon limit by the same
correction factor of Eq.~(\ref{eq:deltaz}) as for elastic scattering. 
\newline
\indent
The $Z^2$-dependent corrections originate from the soft-bremsstrahlung
from the proton and from the proton vertex corrections. 
In \cite{MTj99}, these corrections have also been calculated for elastic
scattering. For the soft-photon emission, one can again factorize the
orginal amplitude, so that the same correction factor is obtained 
for the $e p \to e p \gamma$ reaction as for elastic scattering. 
The proton vertex correction has been split 
in \cite{MTj99} into two parts. The first part contains entirely the IR
divergence, which cancels with the IR divergence from 
soft-photon emission from the proton, 
and in which the original amplitude factorizes. 
The second term in the proton vertex correction depends on the nucleon
structure (form factor dependence for elastic scattering) and will
be different when going from elastic scattering to the $e p
\to e p \gamma$ reaction. For elastic scattering, this  
structure dependent term was however
found \cite{MTj99} to be quite small, except when going to very
large $Q^2$ (much larger than $M_N^2$). When staying in the few
GeV$^2$ region, this correction was calculated in \cite{MTj99} to be well below
1\%. Therefore, we approximate the $Z^2$
dependent correction to the $e p \to e p \gamma$ reaction by the 
structure-independent term of Eq.~(\ref{eq:deltazsqr0}), 
as calculated in \cite{MTj99}, and will neglect in the following the structure 
dependent term.


\section{Radiative tail for elastic scattering and VCS}
\label{sec:radtail}

Besides the knowledge of the virtual radiative corrections and the
soft-photon emission contributions to the $e p \to e p \gamma$
reaction, which were studied in section \ref{sec:first}, the accurate 
determination of the $e p \to e p \gamma$ cross section from measured
spectra also implies the knowledge of the radiative tail. 
The radiative tail consists of the photon emission processes where a
semi-hard photon (with energy not very small compared with e.g. the
lepton energies) is radiated from the electron (or proton). 
\newline
\indent
The radiative tail to elastic or inelastic lepton-nucleon scattering 
has been the subject of numerous studies in the 
literature \cite{max64a,max64b,mo69,max87}. 
The elastic radiative tail also makes a sizeable contribution to 
the cross sections for deep-inelastic lepton-nucleon scattering
(see e.g. \cite{ABS86}). 
\newline
\indent
One should notice that the distinction between the soft-photon
emission and the radiative tail is not a fundamental one, the latter
being just the extension of photon emission processes to higher energies. 
Although the formulas given in this paper  
for the real radiative corrections can 
in principle be extended and applied to higher energies 
(e.g. Eqs.~(\ref{eq:radcorrtot}-\ref{eq:radcorrtotexp}) 
for elastic scattering), in some cases
the characteristics of the experimental detection apparatus can be such
that the cut in $E_e'^{el} - E_e'$ (elastic case) or 
in the missing mass $M_x^2$ ($M_{m1}^2$ or $M_{m2}^2$ for the VCS case) cannot 
be cleanly defined, because the apparatus can have
a changing acceptance as a function of $E_e'^{el} - E_e'$ or  $M_x^2$,
introducing a bias in the radiative tail. 
Therefore, it is useful to consider the radiative tail separately and 
to generate it in a Monte Carlo simulation. 
In doing such a simulation, it can be very helpful to have a
``recipe'', because it is a way to fold 
radiative effects with acceptance functions and other effects
(e.g. multiple scattering, energy loss by collision, 
external radiative effects). In the literature such ``recipes'' were
quite often presented. Many of them are based on one or another
version of the peaking approximation, introduced originally by 
{\it Schiff} \cite{Schiff52}. In the peaking approximation, 
the photon is radiated along either the initial or final electron
directions, i.e. the direction of the electron is not changed while radiating, 
only its energy is changed.  
\newline
\indent
Below, we start by giving such a recipe,
based on the formulas presented in this paper. 
What one essentially needs for a Monte Carlo simulation is an
electron energy loss distribution due to real internal radiative
effects. For each event one can then sample in such a distribution, 
both for the incoming and the outgoing electron. 
We next give a comparison between such a method based on the peaking
approximation, with an exact numerical calculation of the radiative
tail. We show to what extent the full calculation validates the
approximate method for the case of elastic electron-nucleon scattering,  
and show that this method is realistic enough to apply it next to the   
calculation of the radiative tail in the case of the VCS.

\subsection{Energy loss distribution for real internal radiative effects} 

The details of the calculation of the real radiative corrections can be found
in appendix~\ref{app:elast}. It is discussed there how the 
real internal radiative corrections give rise to a 
correction factor $e^{\delta_R}$ 
to the cross section. The part of $\delta_R$ giving
rise to the radiative tail 
(when differentiating $\delta_R$ with respect to the electron energy
loss) is the first term of Eq.~(\ref{eq:radcorrreal}), 
which contains the maximal energy of the emitted photon $\Delta E_s$,
which is defined as in Eq.~(\ref{eq:deltae}).
The correction factor $e^{\delta_R}$ can be written as 
the product of a number of factors, of which the
first one is given by~:
\begin{equation}
\left( {{ (\Delta E_s)^2} \over { \tilde E_e \tilde E_e^{'}}}
\right)^a \;,
\label{eq:radtail1}
\end{equation}
where $a$ is given by (see Eq.~(\ref{eq:radcorrreal}))~:
\begin{equation}
a \,=\, \frac{\alpha_{em}}{\pi} \left[ \ln\left(\frac{Q^2}{m^2}
  \right)-1 \right] \, , 
\label{eq:formula_a}
\end{equation}
and where the tilded quantities in 
Eq.~(\ref{eq:radtail1}) are expressed in the c.m. system of (soft
  photon + recoiling proton) as explained in appendix~\ref{app:elast5}.  
Because in a simulation it is more straightforward 
to apply radiative effects in the {\it lab}, we express 
Eq.~(\ref{eq:radtail1})  in {\it lab} quantities, 
by using Eq.~(\ref{eq:deltae}), which yields~:
\begin{equation}
\left( {{ (\eta\Delta E_e')^2} \over {  E_e  E_e^{'}}} \right)^a \;,
\label{eq:radtail2}
\end{equation}
where $\Delta E_e'= E_e'^{el} - E_e'$. 
Introducing furthermore the quantity 
$\Delta E_e=\eta^2 \Delta E_e'$, we can write Eq.~(\ref{eq:radtail2})
as~: 
\begin{equation}
\left( {{ (\eta\Delta E_e')^2} \over {  E_e  E_e^{'}}} \right)^a 
=\left(\frac{ \Delta E_e \Delta E_e'} {  E_e  E_e'} \right)^a 
=\left(\frac{ \Delta E_e}{E_e} \right)^a 
\left(\frac{\Delta E_e'} {E_e'} \right)^a \;.
\label{eq:radtail3}
\end{equation}
The energy changes $\Delta E_e$ ($\Delta E_e'$) can be interpreted as
the energy losses of the incoming (outgoing) electron due to radiation 
before (after) the scattering process respectively. We can then interpret 
the factor  $\left(\frac{ \Delta E_e}{E_e}\right)^a$
as the fraction of incoming electrons which have 
lost an energy between 0 and $\Delta E_e$, after being subject 
to real internal radiation in an equivalent radiator with thickness
$a$. The factor $\left(\frac{ \Delta E_e'}{E_e'}\right)^a$ has a
similar interpretation, but then on the outgoing electron side 
\footnote{Note that when applying Eq.~(\ref{eq:radtail3}) to the
  radiative tail, i.e. when considering the emission of a photon whose
  energy is not very small compared with the electron energies, we
  calculate $E_e'^{el}$ in the formula for $\Delta E_e'$, using the
  elastic scattered energy corresponding with an initial electron
  which has radiated and whose energy is given by $E_e - \Delta E_e$. In
  the soft-photon limit this difference disappears.}. 
Given this interpretation, if one uses a 
$\Delta E$ distribution $I_{int}(E,\Delta E,a)$, which satisfies~: 
\begin{equation}
\int_{0}^{\Delta E}I_{int}(E,\Delta E,a)d(\Delta E)
=\left( \frac{\Delta E}{E}\right)^{a} \;,
\label{eq:radtail4}
\end{equation}
then it is clear that by sampling such a distribution in a Monte Carlo 
simulation, the correction factor is correctly obtained. 
The distribution $I_{int}$, which has this property is given by~:
\begin{equation}
I_{int}(E,\Delta E,a)=\frac{a}{\Delta E}\left(\frac{\Delta E}{E}
\right)^{a} \;,
\label{eq:radtail5}
\end{equation}
and is normalized to 1~:
\begin{equation}
\int_{0}^{E}I_{int}(E,\Delta E,a)d(\Delta E)=1 \;.
\label{eq:radtail6}
\end{equation}

\subsection{Evaluation of the radiative tail and comparison with an
  exact numerical calculation for elastic electron-proton scattering} 

\begin{figure}
\begin{center}
\leavevmode
\hbox{
\epsfysize=17.0truecm
\epsfbox{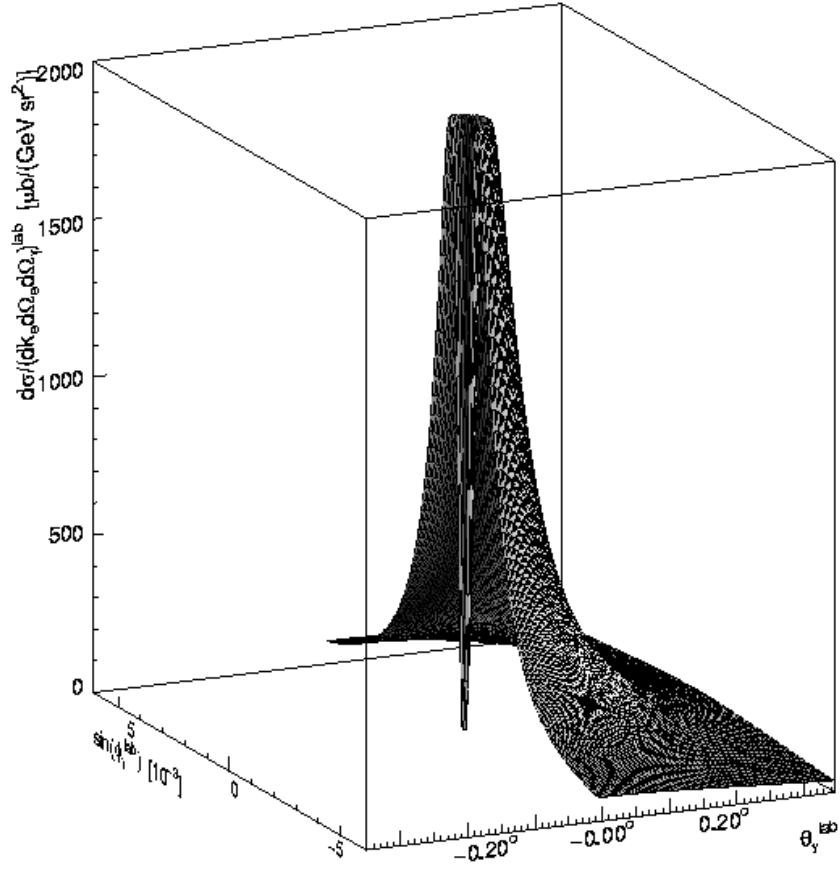} }
\end{center}
\centerline{\small \it  }
\caption{Detail of the cross section for photon emission from an
  electron (Bethe-Heitler cross section), when the photon 
is emitted around the incoming electron direction. The electron
kinematics correspond with : $E_e$ = 855.0 MeV, $E_e'$ = 621.4 MeV, 
$\theta_e$ = 52.18$^o$.}
\label{fig:zoom_bh}
\end{figure}

\begin{figure}
\begin{center}
\leavevmode
\hbox{
\epsfysize=17.0truecm
\epsfbox{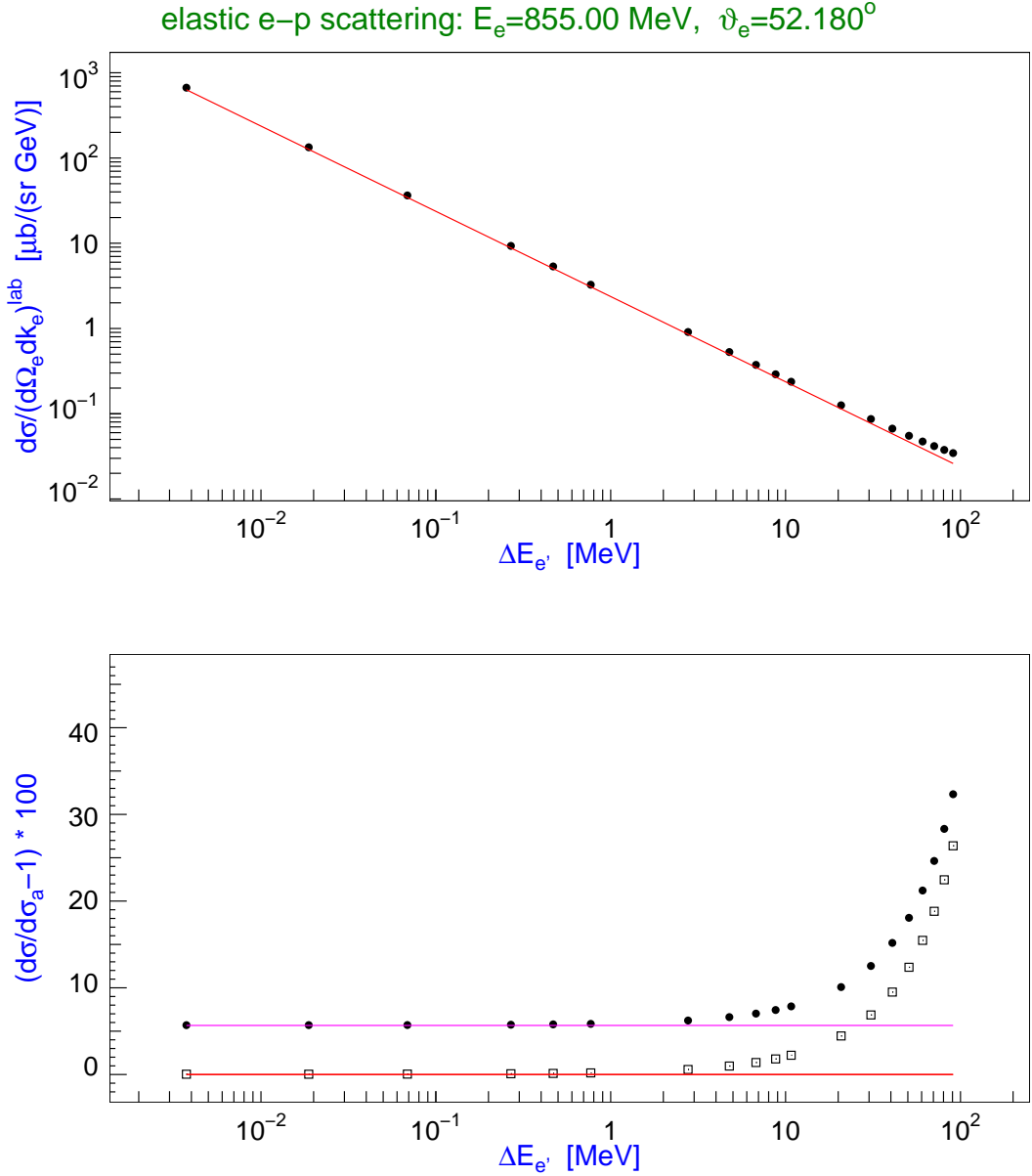} }
\end{center}
\centerline{\small \it  }
\caption{Radiative tail of elastic electron-proton scattering. Upper
  plot : fully numerical calculation (black points) 
compared with the 1/$\Delta E_e'$ dependence of the soft photon result
(straight line). Lower plot : deviation between 
the full calculation, when only radiation from the
electron is included (open diamonds) and when both radiation from
electron and proton are taken into account (black points), with the
soft photon result (straight lines). See text for details.}
\label{fig:radtail_elastic}
\end{figure}

Given the above distribution, a method for introducing a radiative
tail due to internal radiation in a Monte Carlo simulation 
for elastic electron scattering suggests itself~: \\
i) For the incoming electron, sample an energy loss $\Delta E_e$ using
the distribution (\ref{eq:radtail5}) with $E=E_e$ 
the incoming electron energy.\\
ii) Apply elastic electron scattering using the reduced electron 
energy $E_e-\Delta E_e$, and if the cross section behavior 
is implemented in the simulation, use the 
elastic scattering cross section at the reduced electron energy. 
After the elastic scattering process, the outgoing electron
has an energy $E_e'^{el}$.\\
iii) For the outgoing electron, sample an energy loss $\Delta E_e'$
using the distribution (\ref{eq:radtail5}) with $E=E_e'^{el}$. 
The final electron energy is now $E_e'^{el}-\Delta E_e'$.
\newline
\indent
To calculate the equivalent radiator thickness $a$ of 
Eq.~(\ref{eq:formula_a}), one needs the 
value of $Q^2$, which one can in principle only calculate after the 
complete process has taken place. However, one
can show that the above procedure reproduces the correction factor 
(\ref{eq:radtail2}) with a very good accuracy already by calculating
the value of $Q^2$ with elastic electron scattering kinematics. 
\newline
\indent
It is intuitively clear that the above procedure, 
in the case where a constant cross section is used, will
reproduce the correction factor of Eq.~(\ref{eq:radtail2}). 
In case the actual elastic scattering cross section behavior
is implemented, the cross section ``walk'' with the incoming electron energy
is taken into account. 
Remark that the above procedure implies an electron energy loss both 
at the incoming and the outgoing electron sides.  
\newline
\indent
The discussed method implies, however, the assumption of a strict alignment
of the bremsstrahlung photons in the direction of the radiating
leptons, which is known as the (angular) peaking approximation. 
The strength on the other hand is found by integrating the correct angular 
shape in the soft photon limit, as done in appendix~\ref{app:elast5}.
To test the validity of this approximate procedure, we performed a fully 
numerical calculation of the radiative tail 
for elastic electron-proton scattering. 
It consists of integrating 
over the photon phase space in the diagrams where a photon is emitted from
an electron (cfr. BH diagrams of Fig.~\ref{fig:vcstree} (a) and (b)), 
as well as the diagrams where a photon is emitted from
the nucleon (cfr. Born diagrams of Fig.~\ref{fig:vcstree} (c) and (d)).    
In doing so, we nowhere neglect
the photon momentum $l$, in contrast to the 
calculation of appendix~\ref{app:elast5} in the soft-photon limit, where
this momentum is neglected with respect to the lepton momenta.  
For fixed electron kinematics, 
the angular phase space of the soft photon is covered by a grid with
about 225000 points, chosen with increased density in the peak regions 
in order to keep the point-to-point change of the cross section
smaller than 10 \%. Attention has to be paid right in the middle of the 
peaks where the cross section drops very rapidly to (practically) zero 
within the characteristic angle $m/E_e$, as shown in
Fig.~\ref{fig:zoom_bh}. More details on this numerical integration can
be found in \cite{janf}. 
The result of this integration is the absolute cross section of the
radiative tail, differential in the outgoing electron's momentum and
angles. It is shown by the points in Fig.~\ref{fig:radtail_elastic}  
for $E_e=$ 855.0 MeV and $\theta_e$ = 52.18$^o$. 
The energy of the outgoing electron is then determined by 
$E_e' = E_e'^{el}- \Delta E_e'$. The points are compared with the analytical 
result in the soft-photon limit, obtained by differentiating the
expression of Eq.~(\ref{eq:radcorrreal}) for $\delta_R$ - for 
photon emission from the electron - with respect to 
$\Delta E_e'$. This gives a strict $\Delta E_e'^{-1}$ behaviour, 
yielding the cross section $\sigma_a \equiv \sigma_{Born}\, a / \Delta E_e'$
where the proportionality factor $a$ is given as in Eq.~(\ref{eq:formula_a}).  
The soft-photon formula gives thus 
a straight line when both the cross section and $\Delta E_e'$ are
presented on a logarithmic scale. The deviation can be seen in the
lower plot of Fig.~\ref{fig:radtail_elastic}. 
From the keV-region up to about 1~MeV for $\Delta E_e'$, 
the deviation is less then $10^{-3}$ which can be taken as an upper
limit for the error of the numerical integration procedure. 
This agreement demonstrates that the soft-photon approximation holds to
very good precision in this region. 
For higher values of $\Delta E_e'$, a raise of the 
photon emission cross section is observed as is expected due to 
the change of kinematics leading to a lower momentum transfer to the
proton, and to a resulting ``walk'' of the cross section.
We also show on the lower plot of Fig.~\ref{fig:radtail_elastic} the
result when both radiation from the electron and proton are
considered. For better presentation, both
results are normalized to the cross section $\sigma_a$ for soft-photon
emission from the electron, as defined above.
\newline
\indent
In Fig. 10, we compare for two kinematics the exact numerical
calculation of the radiative tail with the approximate method of the Monte
Carlo simulation as discussed above. The simulation has been investigated
by running it with and without the cross section behaviour (dipole form
factors assumed), and the ratio between the two versions is presented
by the lines, the outer lines representing the statistcal accuracy. 
One notices that the increase of the radiative tail is reproduced, but somewhat
overestimated compared with the exact calculation.

\subsection{Application to virtual Compton scattering}

The above procedure can also be applied to VCS, as long as the 
angular peaking approximation is
used, i.e. the electron does not change its direction while losing 
energy by internal real radiation.
Indeed, Eq.~(\ref{eq:radcorrrealbis}) 
is completely similar to the elastic case, when expressing it in the
c.m. system of either (soft photon + outgoing nucleon) or (soft-photon
+ outgoing photon) depending on how the $e p \to e p \gamma$ reaction
is measured, as explained in section \ref{sec:real}. 
After exponentiation, one can apply a factorization 
completely similar as in Eq.~(\ref{eq:radtail3}). 
Because under the assumption of the angular peaking approximation 
$\Delta E_s/\tilde E$ is constant under a
Lorentz transformation, we obtain the property that the shape 
of the distribution (\ref{eq:radtail5}) is system independent, 
only its endpoint value $E$ changes. As a result, 
one can apply the distribution of Eq.~(\ref{eq:radtail5}) 
in the {\it lab} for VCS, but then using {\it lab} values for 
$E_e$ and $E_e'$. For VCS, one certainly can have a changing
acceptance of the detection apparatus as a function of missing mass
(making a ``clean'' cut in missing mass on the data impossible),
so that generating a radiative tail in a Monte Carlo simulation with
the above described method is probably the best way to implement 
the radiative tail correction  to the data. Such a simulation was
implemented for the VCS experiments already performed at MAMI
\cite{mami} and at JLab \cite{jlab}, and 
will be fully described in a forthcoming paper \cite{LVH2000}.


\section{Results and discussion}
\label{sec:results}

\subsection{Elastic electron-proton scattering}

Before showing results for VCS, we briefly discuss first the effect of
the radiative corrections to elastic electron-proton scattering, in
order to have a point of reference. The radiative corrections to elastic
electron-proton scattering are presented in detail in
appendix~\ref{app:elast}. In Table~\ref{tab1}, we show for different
elastic kinematics (MAMI, JLab) the numerical values of the vertex correction 
($\delta_{vertex}$ of Eq.~(\ref{eq:el2})), 
the vacuum polarization correction ($\delta_{vac}$ of Eq.~(\ref{eq:el3})), 
and the real radiative correction at the electron side   
($\delta_R$ of Eq.~(\ref{eq:radcorrreal})). We also show the $Z$ and 
$Z^2$ dependent corrections, $\delta_1$ (Eq.~(\ref{eq:deltaz})) and 
$\delta_2^{(0)}$ (Eq.~(\ref{eq:deltazsqr0})) respectively, as derived
in the recent work of \cite{MTj99}. We omit here the small part in the
$Z^2$ dependent correction which depends on the particular model for
the nucleon structure (in the elastic case, the form factors), as can
be found in \cite{MTj99}. In Table~\ref{tab1}, we indicate the total
radiative correction $\delta_{tot}$ as the sum of all the different
contributions as in Eq.~(\ref{eq:radcorrtotz}). 
From Table~\ref{tab1}, we see that by far the largest contribution to
the radiative correction comes from the large logarithm and double
logarithm in $Q^2/m^2$ in the electron vertex correction. 
When evaluating the real radiative corrections
for $E_e'^{el} - E_e' = 0.01 \, E_e$, 
the total effect of the radiative correction is an upwards correction
of the data (for negative $\delta_{tot}$) of the order 20 - 25 \%. 
In the last column of Table~\ref{tab1} (denoted by EXP), we also
indicate the result when exponentiating all corrections except the
vacuum polarization contribution, which - as modification of the photon
propagator - is resummed as in Eq.~(\ref{eq:radcorrtotexp}). 
One sees that this can lead to differences of the order of 2 \%.
\newline
\indent
In an elastic scattering experiment, one measures an scattered
electron spectrum and has to evaluate the real radiative corrections
as a function of the cut $(E_e'^{el} - E_e')$ which one performs in
the spectrum. Dividing the measured cross section by the correction
factor ($1 + \delta_{tot}$) and plotting the result as function of 
$(E_e'^{el} - E_e')$, should then lead to a ``plateau'' behavior,
which demonstrates the consistency of the procedure (within a 
certain range of the value $(E_e'^{el} - E_e')$ where one knows the
radiative tail to sufficient accuracy). 
\newline
\indent
The determination of the elastic cross section for the kinematics
$E_e$ = 705.11~MeV, $\theta_e$ = 42.6$^o$ is shown in Fig.~\ref{fig:plateau}. 
The upper plot shows the
dE-spectrum of elastic data taken (during the beam time of the VCS
experiment) at MAMI. 
It is compared with the simulated spectrum (dashed line). 
On the lower plot, the ratio of the experimental spectrum integrated up
to the value $\Delta E_e'$,
to the simulation integrated also up to $\Delta E_e'$ 
is shown as function of the cutoff energy $\Delta E_e'$.
This gives the elastic cross section, which is seen to be stable below the
1\%-level over a long interval up to the cut by the acceptance of the
spectrometer. The slow descent for higher $\Delta E_e'$ 
indicates that the simulation overestimates slightly the radiative tail.

\subsection{VCS below pion production threshold}

We next turn to the $e p \to e p \gamma$ reaction below pion
threshold. It was discussed in section \ref{sec:lowest}, that the lowest order
(in $\alpha_{em}$) amplitude of the $e p \to e p \gamma$ process at
low outgoing photon energies ${\mathrm q}' \equiv | \vec q^{\, '} |$
is given by the BH + Born processes. The deviation from the BH + Born
amplitudes grows with ${\mathrm q}'$, and can be parametrized 
(at low ${\mathrm q}'$) in terms of six generalized polarizabilities
(GP's) of the nucleon, which are function of $Q^2$. 
\newline
\indent
A first VCS experiment has been performed at MAMI
\cite{mami}. It consisted of measuring the $e p \to e p \gamma$
reaction at five values of ${\mathrm q}'$ below pion threshold,
ranging from ${\mathrm q}'$ = 33 MeV/c to ${\mathrm q}'$ = 111.5 MeV/c.  
At the lowest value ${\mathrm q}'$ = 33 MeV/c, where the
polarizability effect is negligeably small, the measurement serves as
a check of the Low Energy Theorem (LET). The measured deviation as
function of ${\mathrm q}'$ can then be interpreted as the effect of
the GP's. It is clear that both to test the LET as well as to extract
the GP's from the measured deviation with respect to the BH + Born
result (which is expected to be of the order 10 - 20 \% at the highest
${\mathrm q}'$ value), 
it is a prerequisite to know very accurately how the result is
modified due to radiative corrections. 
\newline
\indent
In Fig.~\ref{fig:cross33_mami}, we first show the differential cross
section for MAMI kinematics at a low value ${\mathrm q}'$ = 33 MeV/c,
as function of the c.m. angle of the emitted real photon with respect
to the direction of the virtual photon. One sees from
Fig.~\ref{fig:cross33_mami} that the virtual radiative corrections
reduce the BH + Born result in these kinematics by about 16 \% (or
when applied to data, increase the uncorrected data by 16 \%). 
The real radiative corrections have to be estimated as function of the
cut which one performs in the missing mass spectrum. The VCS
experiments below pion threshold measure the $e p \to e p \gamma$
reaction by detecting the outgoing electron and proton, and
reconstruct the missing mass $M_{m2}$ as defined in
Eq.~(\ref{eq:missmass2}). In Fig.~\ref{fig:cross33_mami}, 
the real radiative corrections are shown  for a value of $\Delta E_s$ = 10 MeV,
where the soft-photon energy $\Delta E_s$ is determined from 
the cut in the missing mass according to Eq.~(\ref{eq:deltae2vcs}). 
For the small value ${\mathrm q}'$ = 33 MeV/c, the real radiative
correction depends only very little on the angle $\theta_{\gamma
  \gamma}$ (through the last terms on the {\it rhs} of  
Eqs.~(\ref{eq:vcstransf1b},\ref{eq:vcstransf2b})). 
For $\Delta E_s$ = 10 MeV, the real radiative correction $\delta_R$ 
is given by $\delta_R \approx$ - 0.025, which corresponds with 
increasing the uncorrected data by about 2.5 \%. 
For $\Delta E_s$ = 20 MeV, $\delta_R \approx$ + 0.02 
(reducing the uncorrected data by about 2 \%), 
and for $\Delta E_s$ = 30 MeV, $\delta_R \approx$ + 0.045 
(reducing the uncorrected data by about 4.5 \%).  
To determine the $e p \to e p \gamma$ cross section from the
measured missing mass spectra, one has to perform a 
consistency check by plotting the experimentally measured
(uncorrected) cross section divided by the radiative correction 
factor as function of the cut in the missing mass spectrum.  
In this way, one has to find a ``plateau'' behavior, as was
demonstrated before for elastic data. 
This consistency check was also performed 
on the VCS data measured at MAMI
\cite{mami}, and will be shown in a forthcoming publication \cite{LVH2000}. 
\newline
\indent
In Fig.~\ref{fig:cross112_mami}, we show the the differential cross
section for MAMI kinematics at the highest measured outgoing photon
energy~: ${\mathrm q}'$ = 111.5 MeV/c. The virtual radiative
corrections are mainly ${\mathrm q}'$ independent (for these rather
small values) and lead thus also here to a reduction of the BH + Born
result by about 16 \%. The real radiative corrections are again shown
for $\Delta E_s$ = 10 MeV, and exhibit a slight angular dependence. 
These corrections were applied to the data from the unpolarized MAMI
experiment of \cite{mami}. 
From the deviation of the radiatively corrected data 
and the BH + Born result, two
combinations of GP's have been extracted at $Q^2 \simeq$ 0.33 GeV$^2$ 
in \cite{mami}. 
\newline
\indent
An experiment below pion production threshold to measure the GP's at
higher $Q^2$ has also been measured at JLab \cite{jlab} and is under
analysis at the time of writing. In Fig.~\ref{fig:cross_jlab}, we show
how the BH + Born cross section is modified due to the virtual
radiative corrections. It is seen that for the JLab kinematics of 
Fig.~\ref{fig:cross_jlab}, the BH + Born result is reduced at the
backward angles by about 20 \% due to the virtual radiative corrections. 
\newline
\indent
The unpolarized VCS cross section below pion threshold provides three 
independent structure functions (when varying the value of
$\varepsilon$ in the experiment), which allows to extract three of the
six (lowest order) generalized nucleon polarizabilities. 
To extract the three remaining nucleon polarizabilities, one has to
resort to double polarization observables as discussed in~\cite{MVdh97}. 
In particular, double polarization observables
with polarized electron beam and with a polarized target 
(along either of the three axes), 
or alternatively by measuring the recoil nucleon polarization, 
provide three new observables to extract the 
three additional nucleon response functions
\cite{MVdh97,vcsreview}. 
In Fig.~\ref{fig:asymm_mami}, we show the double polarization
asymmetries for MAMI kinematics, by measuring the recoil polarization
components along the $z$-direction (virtual photon direction) or along
the $x$-direction (perpendicular to the virtual photon but parallel to
the scattering plane). One aims to extract the polarizability effect
in these observables from the deviation of the measured asymmetry and 
the BH + Born result (see e.g.~\cite{MVdh97} for an estimate of
this effect within a model calculation). Therefore, it is important to
know how much the BH + Born result is affected by the radiative
corrections before extracting the polarizability effect. It is 
seen in Fig.~\ref{fig:asymm_mami} that the effect of the radiative
corrections on the double polarization asymmetries nearly drops out in
the ratio (much less than 1 \% change of the asymmetries). 
At the low values of the outgoing photon energy ${\mathrm q}'$ 
(e.g. ${\mathrm q}' \simeq$ 33 MeV/c) where the
polarizability effect is very small, these asymmetries are also 
hardly affected by radiative corrections.  
Therefore, these asymmetries can also provide an 
independent check of the LET.  
An experiment to measure the VCS double polarization observables by
measuring the recoil nucleon polarization is planned at MAMI in the
near future \cite{loimami}.

\subsection{Deeply virtual Compton scattering}

Besides the low energy region, the VCS process is also studied in the
Bjorken regime, where $Q^2$ and $\nu = p.q/M_N$ are large, with  
$x_B = Q^2 / (2 M_N \nu)$ fixed. In this kinematical region, the
process is refered to as deeply virtual Compton scattering (DVCS). 
In the Bjorken regime, the DVCS amplitude factorizes into 
a perturbatively calculable hard scattering amplitude, and
into a non-perturbative part at the proton side, 
expressed in terms of so-called skewed parton distributions (SPD's)
which generalize the ordinary parton distributions. These SPD's
are new nucleon structure observables which one aims to extract by
measuring e.g. the exclusive $e p \to e p \gamma$ reaction in the
Bjorken regime. Similarly as was seen before in the threshold region,
the $e p \to e p \gamma$ reaction can have an important contribution
from the BH process, besides the DVCS process of
actual interest. However, the BH and DVCS contributions behave
differently as function of the lepton beam energy, as studied in
Refs. \cite{MVdh98,vcsreview,MVdh99}. In particular, at the lower beam
energies, such as e.g. available at JLab, the BH process dominates in
the forward direction over the DVCS process. 
In this region, the DVCS process becomes only 
measurable due to its interference with the BH process. 
In order to extract the DVCS process (and the nucleon structure
information) from its interference with the BH, 
it is therefore important to have good
knowledge of how the radiative corrections modify the BH amplitude.
\newline
\indent
In Fig.~\ref{fig:dvcs_jlab}, we show the $e p \to e p \gamma$ 
cross section in kinematics accessible at JLab, 
where such an experiment is planned  
\cite{bertindvcs}. 
The DVCS cross section is calculated by using the ansatz for the SPD's
of \cite{MVdh99}. It is seen from Fig.~\ref{fig:dvcs_jlab}, that the BH
indeed dominates over the DVCS cross section in these kinematics, and
that the DVCS cross section gets enhanced due to its interference with
the BH. One furthermore sees that the virtual radiative
corrections reduce the BH + DVCS cross section by about 23\% in these
kinematics. This is mainly due to the reduction of the BH process 
when including virtual radiative corrections. 
The real radiative corrections are shown in Fig.~\ref{fig:dvcs_jlab}
for a value $\Delta E_s$ = 0.1 GeV, which corresponds with a cut in
the recoiling hadronic missing mass spectrum 
(defined in Eq.~(\ref{eq:missmass1})) of 
$M_{m1}^2$ - $M_N^2 \simeq$ 0.21 GeV$^2$. 
\newline
\indent
In Ref.~\cite{vcsreview}, it was suggested that an exploratory study
of the DVCS process might be possible by studying the $e p \to e p
\gamma$ reaction with a polarized electron beam. The electron single
spin asymmetry (SSA) does not vanish out of plane and is only due to the
interference of the BH amplitude and the imaginary part of the DVCS
amplitude (i.e. the BH amplitude does not lead to a SSA, because
it is purely real). Therefore, one expects this SSA to be less
sensitive to radiative corrections on the BH amplitude. However, as the
BH amplitude enters the SSA linearly in the numerator, but
quadratically in the denominator (as in the unpolarized cross
section), one might wonder what is the residual effect of the
radiative corrections on this observable. 
In Fig.~\ref{fig:ssa_jlab}, we show the SSA for DVCS at JLab. 
One sees that the SSA gets only slightly reduced due to the 
radiative corrections. The reduction of the SSA amounts to
maximum 5\% of its value around 5$^o$, where the asymmetry reaches its
maximal value. Therefore, the SSA shows to be a rather ``clean''
observable for extracting the DVCS amplitude in a region where the BH
process dominates. Its measurement is also envisaged at JLab in the
near future \cite{bertindvcs}.


\section{Conclusions}
\label{sec:conclusion}

We studied in this work the first order QED radiative corrections to
the $e p \to e p \gamma$ reaction. The one-loop virtual radiative
corrections have been evaluated by a combined analytical-numerical
method. Several tests were shown to cross-check the numerical method used. 
Furthermore, it was shown how all IR divergences cancel when adding the
soft-photon emission processes. A fully numerical method was presented
for the photon emission processes where the photon energy is not very
small compared with the electron energies, which makes up the
radiative tail. Besides, we have also presented an approximate
calculation of the radiative tail, which was shown to be realistic
enough for use in a Monte Carlo simulation. 
\newline
\indent
We compared our results first to elastic electron-proton scattering. 
Subsequently, the results for the radiative corrections to the 
$e p \to e p \gamma$ reaction were shown both below pion
threshold and in the deeply virtual Compton scattering regime. 
\newline
\indent
Below pion threshold, our calculations were applied to the first
dedicated VCS experiment at MAMI, and show that the effect of the
radiative corrections results in an enhancement of the uncorrected
data by about 20 \% (or an equivalent reduction of the theory). 
VCS double polarization asymmetries where shown to be insensitive to
radiative corrections.
\newline
\indent
For the DVCS, we calculated radiative corrections for JLab kinematics
and found the virtual radiative corrections to lead to an enhancement
of the data by about 23 \%. The single spin asymmetry was shown to be
only slightly reduced by radiative corrections. 
\newline
\indent
Although we focussed here on the kinematical regimes of
ongoing or planned experiments, the present work can also 
serve as a tool in the analysis of future VCS experiments.

\section*{acknowledgments}

As this work grew over a period of a couple of years, 
we enjoyed discussions with many colleagues about the subject of this
work, in particular with N. d'Hose and P. Vernin,
about the application of radiative corrections to elastic and VCS data. 
We thank M. Distler and H. Merkel for their help in the numerical
implementation of the radiative tail calculation.  
Furthermore, we would like to thank P.A.M. Guichon for many discussions, 
which were at the origin of this work. We also want to acknowledge 
in particular very useful discussions and correspondence with
L. Maximon and are grateful for his continued interest in this work.
\newline
\indent   
This work was supported in part by the French Commissariat \`a 
l'Energie Atomique (CEA), by the EU/TMR contract ERB FMRX-CT96-0008, 
by the Deutsche Forschungsgemeinschaft (SFB 443), by the French CNRS/IN2P3, 
and by the Fund for Scientific Research - Flanders (Belgium).

\newpage

\appendix


\section{Radiative corrections to elastic lepton-nucleon scattering 
using the dimensional regularization method for both UV and IR 
divergences.}
\label{app:elast}

In this appendix, we provide the reader with some details of the
derivation of the radiative corrections to elastic lepton scattering
at one-loop level. In our derivation, we use the dimensional
regularization procedure to regularize both ultraviolet and infrared
divergences. 

After a short introduction of the renormalization method, we 
calculate subsequently 
the vertex diagram at the lepton side (Fig.~\ref{fig:elasticdia}(a)), 
the lepton self-energy diagram (Fig.~\ref{fig:elasticdia}(b)), 
the vacuum polarization diagram (Fig.~\ref{fig:elasticdia}(c)), 
and give an analytical result, without approximations, for the soft
photon emission at the lepton side (Fig.~\ref{fig:elasticdia} (d) and (e)). 
We compare our results with other derivations found in the
literature. At the end we collect the results to correct the elastic 
lepton-nucleon scattering cross sections and discuss the role of the 
radiative corrections at the proton side and the two-photon exchange
corrections by referring to the recent work of Ref.~\cite{MTj99}. 
In this appendix, we use the same notations as explained in
section~\ref{sec:lowest}.

\subsection{Renormalization method}
\label{app:elast1}
In calculating QED radiative corrections in this work, 
we are using the BPHZ renormalization 
method (as explained e.g. in Ref.\cite{chengli}), which consists of 
replacing in the unrenormalized Lagrangian all bare quantities by 
renormalized ones. For QED, the bare Lagrangian is given 
by (we are using the conventions of {\it Bjorken and Drell} \cite{BD64} 
in this work) 
\begin{equation}
{\mathcal{L}_B} = \bar \Psi_B {\left(
i \gamma^\mu \partial_\mu - m_B \right)} \Psi_B 
\;-\; {1 \over 4} F_{B \; \mu \nu} {F_B}^{\mu \nu} 
\;-\; e_B \, \bar \Psi_B \gamma^\mu \Psi_B A_{B \; \mu} \;,
\label{eq:bareL}
\end{equation}
where the bare field tensor $F_B^{\mu \nu}$ is given by 
\begin{equation}
F_B^{\mu \nu} \;=\; \partial^\mu \; A_B^{\nu} \;-\; \partial^\nu \; A_B^{\mu} \;.
\end{equation}
The renormalization of the theory amounts in redefining the bare quantities 
in terms of renormalized (i.e. physical observable) ones : 
\begin{eqnarray}
&&\Psi_B \; = \; Z_2^{1/2} \; \Psi \;, 
\hspace{2cm} {A_B}^\mu \; = \; Z_3^{1/2} \; A^\mu \nonumber \;, \\
&&m_B \; = \; Z_m \; m \;, 
\hspace{2.2cm} e_B \; = \; Z_g \; e \;.
\label{eq:renormcts}
\end{eqnarray}
In Eq.~(\ref{eq:renormcts}), the renormalized finite quantities are 
$\Psi, A^\mu, m$ and $e$. A theory in which all divergences can be 
absorbed into renormalization constants such as $Z_2, Z_3, Z_m$ and $Z_g$ 
in Eq.~(\ref{eq:renormcts}), is called {\it multiplicatively renormalizable}. 
This procedure leads to a decomposition of the 
QED Lagrangian of Eq.~(\ref{eq:bareL}) 
into 
\begin{equation}
{\mathcal{L}_B} \; = \; {\mathcal{L}_R} + {\mathcal{L}_{CT}}\;,
\end{equation}
where ${\mathcal{L}_R}$ represents the renormalized Lagrangian in 
terms of the physical (finite) quantities 
\begin{equation}
{\mathcal{L}_R} = \bar \Psi {\left(
i \gamma^\mu \partial_\mu - m \right)} \Psi 
\;-\; {1 \over 4} F_{\mu \nu} F^{\mu \nu} 
\;-\; e \, \bar \Psi \gamma^\mu \Psi A_{\mu} \;,
\label{eq:renormL}
\end{equation}
and where ${\mathcal{L}_{CT}}$ is called the counterterm Lagrangian 
\begin{equation}
{\mathcal{L}_{CT}} = (Z_2 - 1) \, \bar \Psi i \gamma^\mu \partial_\mu \Psi 
\;-\; (Z_2 Z_m - 1) \bar \Psi m \Psi 
\;-\; (Z_3 - 1) {1 \over 4} F_{\mu \nu} F^{\mu \nu} 
\;-\; (Z_1 - 1) e \, \bar \Psi \gamma^\mu \Psi A_{\mu} \;.
\label{eq:countertermL}
\end{equation}
In Eq.~(\ref{eq:countertermL}), the vertex renormalization constant $Z_1$ 
is defined as $Z_1 = Z_g Z_2 Z_3^{1/2}$. 
For a renormalizable theory such as QED, all divergences obtained 
by calculating loop diagrams with the renormalized Lagrangian 
${\mathcal{L}_R}$ are cancelled by the corresponding contributions 
in the counterterm Lagrangian ${\mathcal{L}_{CT}}$. 
It will be shown below how the QED renormalization constants 
are calculated to order 
$O(e^2)$ by calculating the vertex diagram, the lepton self-energy diagram 
and the photon polarization diagram at the one-loop level.
\newline
\indent
As QED is a gauge invariant theory, 
we will simplify all calculations in this work by using the Feynman gauge.

\subsection{Vertex diagram}
\label{app:elast2}
The on-shell photon-lepton-lepton vertex is represented by 
\begin{equation}
M^{\mu}_{v}\;=\;\bar u( k^{'} , h' )\;
\left[\;-\,i e \, \Lambda^\mu \left( {k^{'}, k} \right) \right]
\; u( k , h )\;,
\label{eq:a_v1}
\end{equation}
and the on-shell vertex of Eq.~(\ref{eq:a_v1}) can be parametrized as
\begin{equation}
\bar u( k^{'} , h' )\;
\Lambda^\mu \left( {k^{'}, k} \right) \; u( k , h )\;
=\;\bar u( k^{'} , h' )\;
\left[ \left( {1 \,+\,F(Q^2) }\right) \,\gamma^\mu\;
-\;G(Q^2)\,i \sigma^{\mu \nu} {{q_\nu} \over {2 m}} \right] \; 
u( k , h )\;,
\label{eq:a_v3}
\end{equation}
where $q = k - k^{'}$. 
\newline
To order $O(e^2)$ , the vertex $\Lambda^\mu$ (corresponding with 
Fig.~\ref{fig:elasticdia}(a)) is given by 
\begin{equation}
\Lambda^{\mu} \left( {k^{'}, k} \right)\;=\;\gamma^\mu \;
{-\,ie^2} \, \mu^{4 - D}\, \int {{{d^Dl} \over {\left( {2\pi }
\right)^D}}}\;{{\gamma ^\alpha \left( {\not k^{'}+\not l+m}
\right)\gamma ^\mu \left( {\not k+\not l+m} \right)\gamma _\alpha } \over
{\left[ {l^2} \right]\,\left[ {l^2+2l.k^{'}} \right]\,\left[ {l^2+2l.k}
\right]\,}} + O\left( {e^4} \right) \;,
\label{eq:a_v2}
\end{equation}
where a mass scale $\mu$ (renormalization scale) has to be introduced when 
passing to $D \neq 4$ dimensions in order to keep the coupling constant 
dimensionless. 
It is immediately seen by power counting 
that in four dimensions ($ D = 4$), the one-loop integral in Eq.~(\ref{eq:a_v2}) contains 
an ultraviolet ($l \rightarrow \infty$) logarithmic divergence 
and an infrared ($l \rightarrow 0$) logarithmic divergence. 
To subtract the divergent parts (by the corresponding counterterms) 
of expressions such as Eq.~(\ref{eq:a_v2}), one has to regularize them first. 
\newline
\indent
We follow in this work the dimensional regularization procedure 
to regularize both ultraviolet and infrared divergences. 
The dimensional regularization method amounts in calculating 
loop diagrams in $D$ dimensions. 
Physical observables are obtained by letting $D \rightarrow 4$ at the end. 
To obtain an integral which is ultraviolet convergent, one has to 
take $D < 4$, or $\epsilon_{UV} \equiv 2 - D /2 > 0$ in expressions such as 
Eq.~(\ref{eq:a_v2}). 
To obtain an integral which is infrared convergent, one has to 
take $D > 4$, or $\epsilon_{IR} \equiv 2 - D /2 < 0$. 
The two different limits show that care has to be taken with the limit 
$D \rightarrow 4$, which means that the parts in Eq.~(\ref{eq:a_v2}) 
that are infrared divergent and the parts that are 
ultraviolet divergent have to be 
separated and in the corresponding terms, two different limits have to 
be taken when one approaches $D = 4$. 
Although the dimensional regularization scheme has been applied originally 
to ultraviolet divergent expressions as it respects the symmetries 
of the theory (in particular the gauge symmetry for a gauge theory), 
it has also been applied in a few works to regularize infrared divergences 
\cite{gastmans73,marciano75}.  
\newline
\indent
When working out the integral in Eq.~(\ref{eq:a_v2}), one obtains 
after some algebra the following expressions 
for $F(Q^2)$ and $G(Q^2)$ to order $O(e^2)$ :   
\begin{eqnarray}
F(Q^2)\,=\,{e^2 \over {\left( {4\pi } \right)^2}}\,
&&\left\{ \left[ {{1 \over {\varepsilon _{UV}}}-\gamma_E 
+\ln \left( {{{4\pi \mu^2} \over {m^2}}} \right)} \right] 
+ \left[ {{1 \over {\varepsilon _{IR}}}-\gamma_E 
+\ln \left( {{{4\pi \mu^2} \over {m^2}}} \right)} \right] 
\,{{v^2+1} \over v}\,\ln \left( {{{v+1} \over {v-1}}} \right)  
\right. \nonumber\\
&&+{{v^2+1} \over {2v}}\;\ln \left( {{{v+1} \over {v-1}}}\right)\;
\ln \left( {{{v^2-1} \over {4v^2}}}\right) \,
+\, {{2 v^2 + 1} \over {v}} \,\ln \left( {{{v+1} \over {v-1}}} \right) \;
\nonumber\\
&&\left. +{{v^2+1} \over v}\,\left[ {Sp\left( {{{v+1} \over {2v}}} \right) - 
Sp\left( {{{v-1} \over {2v}}} \right)} \right] \right\} \;,
\label{eq:a_v4}
\end{eqnarray}
and
\begin{equation}
G(Q^2)\,=\,{e^2 \over {\left( {4\pi } \right)^2}}\;
{{v^2-1} \over {v}}\;\ln \left( {{{v+1} \over {v-1}}}\right)\;,
\label{eq:a_v5}
\end{equation}
where $v$ is given by 
\begin{equation}
v^2 \; \equiv \; 1 + {{4 m^2} \over Q^2} \;, 
\label{eq:v}
\end{equation}
with $Q^2$ = - $q^2$ $>$ 0.
In Eq.~(\ref{eq:a_v4}), $\gamma_E$ represents the Euler constant, 
and the Spence (or dilogarithmic) function is defined by 
\begin{equation}
Sp(x) \;\equiv\; - \int\limits_0^x {dt\,{\ln(1 - t) \over t}}\;.
\label{eq:spence}
\end{equation} 
\indent
From Eq.~(\ref{eq:a_v5}), the one-loop radiative correction to the 
electron magnetic moment follows as 
\begin{equation}
\mu \;=\; {e \over {2 m}} \, \left( 1 + G(Q^2\,=\,0) \right) \;=\;
{e \over {2 m}} \, \left( 1 + {\alpha_{em} \over {2 \pi}} \right) \;,
\end{equation}
which is the result first obtained by {\it Schwinger} \cite{schwinger48}. 
\newline
\indent
To remove the UV divergence from the vertex correction Eq.~(\ref{eq:a_v4}), 
one has to determine the vertex renormalization constant 
$Z_1$ of Eq.~(\ref{eq:countertermL}). 
$Z_1$ is determined by requiring that the total vertex 
\begin{equation}
\tilde{\Lambda}^\mu \;=\; \Lambda^\mu \;+\;(Z_1\,-\,1)\,\gamma^\mu \;,
\label{eq:a_v6}
\end{equation}
defines the physical electron charge at $Q^2 = 0$, i.e. 
\begin{eqnarray}
\label{eq:a_v7}
&&Z_1\,=\, 1 \,-\, F(Q^2\,=\,0) \nonumber\\
&&= 1 \,- {e^2 \over {\left( {4\pi } \right)^2}} 
\left\{ \left[ {{1 \over {\varepsilon _{UV}}}-\gamma_E 
+\ln \left( {{4\pi \mu^2} \over {m^2}} \right) } \right] 
 \,+\,2 \left[{1 \over {\varepsilon _{IR}}} -\gamma_E 
+\ln \left( {{4\pi \mu^2} \over {m^2}} \right) \right] 
\,+\, 4 \right\} 
\,+  O(e^4) .
\end{eqnarray}
It is seen that the vertex renormalization constant $Z_1$ 
contains besides the UV divergence also an IR divergence. 
The renormalized vertex of Eq.~(\ref{eq:a_v6}), is determined 
by the vertex correction function $F(Q^2) - F(Q^2 = 0)$ 
which is given to first order in $\alpha_{em}$ 
(where $\alpha_{em} = e^2 / 4 \pi$) by the expression 
\begin{eqnarray} 
F(Q^2) - F(Q^2 = 0) \;=\;&&
{\alpha_{em} \over {2 \pi}} \left\{ 
\left[ { {1 \over {\varepsilon _{IR}}} - \gamma_E 
+\ln \left( {{{4\pi \mu^2} \over {m^2}}} \right)\,} \right] .
\left[ {{{v^2+1} \over {2 v}}\,\ln \left( {{{v+1} \over {v-1}}} \right) \,-\, 1}
\right]  \right.  \nonumber\\
&&\hspace{.5cm}+{{v^2+1} \over {4v}}\;\ln \left( {{{v+1} \over {v-1}}}\right)\;
\ln \left( {{{v^2-1} \over {4v^2}}}\right) \,
+\, {{2 v^2 + 1} \over {2v}} \,\ln \left( {{{v+1} \over {v-1}}} \right) \,-2\,\;\nonumber\\
&&\hspace{.5cm}+\left. {{v^2+1} \over {2v}}\,
\left[ {Sp\left( {{{v+1} \over {2v}}} \right) - 
Sp\left( {{{v-1} \over {2v}}} \right)} \right] \right\} \;,
\label{eq:a_v8}
\end{eqnarray}
\indent
The expression for the vertex correction function 
$F(Q^2) - F(Q^2 = 0)$, 
which was calculated here 
using the dimensional regularization method for both the UV and IR 
divergences, agrees with the ones derived in many textbooks 
(see e.g. Eq.~(47.52) of Ref.\cite{akhiezer} where a 
full derivation is given). 
The correspondence with the calculations which use a finite photon mass 
($\lambda$) as IR regulator is found to be 
\begin{equation}
{1 \over {\epsilon_{IR}} } \,-\, \gamma_E \,+\, 
\ln \left( {{4 \pi \mu^2} \over {m^2}} \right)
\;\longleftrightarrow \ln {\lambda^2 \over m^2}\;.
\end{equation}
\indent
In the ultrarelativistic limit ($Q^2 >> m^2$), the vertex correction 
$F(Q^2) - F(Q^2 = 0)$ can be found from Eq.~(\ref{eq:a_v8}) to be given by 
\begin{eqnarray}
F(Q^2) - F(Q^2 = 0)\,\stackrel{Q^2 >> m^2}{\longrightarrow} \;
{{\alpha_{em}} \over {2 \pi}}\;
&&\left\{ 
\left[  {1 \over {\varepsilon _{IR}}} - \gamma_E 
+\ln \left( {{4\pi \mu^2} \over {m^2}} \right)\, \right] \;.\;
\left[ \ln \left( {{Q^2} \over {m^2}} \right)  - 1 \right] \right. \nonumber\\
&&\left. +\, \left(  {3 \over 2}  \ln \left( {{Q^2} \over {m^2}} \right) 
 - 2 \right)  
+ { \left( { {- {1 \over 2} } \ln^2 \left( {{Q^2} \over {m^2}}
      \right) \,+\, {{\pi^2} \over 6} } \right) } \right\} .
\label{eq:vertexhighs}
\end{eqnarray}
It is seen from Eq.~(\ref{eq:vertexhighs}), that the finite part of the 
vertex correction at high $Q^2$ is dominated by a 
quadratic logarithmic term.

\subsection{Lepton self-energy diagram}
\label{app:elast3}

The free lepton propagator (for a lepton with four-momentum $k$)
\begin{equation}
S^o\left( k \right) \;=\; {{\not k + m} \over {k^2 - m^2 + i \epsilon}} \;,
\label{eq:a_self1}
\end{equation}
is modified through the lepton self-energy $\Sigma \left( k \right)$, 
to the full lepton propagator 
\begin{equation}
S \left( k \right) \;=\; S^o\left( k \right) \;+\; S^o\left( k \right) \,
\Sigma\left( k \right) \; S\left( k \right) \;.
\label{eq:a_self2}
\end{equation}
To first order $O(e^2)$, the lepton self-energy 
(corresponding with Fig.\ref{fig:elasticdia}(b)) is given by 
\begin{equation}
-i\, \Sigma \left( k \right)\;=\;
-e^2 \;\mu^{4 - D}\; \int {{{d^D l} \over {\left( {2\pi } \right)^D}}}\;
{{\gamma ^\alpha \left( {\not k +\not l+m} \right) \gamma _\alpha } \over
{\left[ {l^2} \right]\,\left[ {(k+l)^2 - m^2} \right]\, }} \;.
\label{eq:a_self3}
\end{equation}
By power counting, it is seen that the integral of Eq.~(\ref{eq:a_self3}) 
contains a linear UV divergence but is IR finite in the limit 
$D \rightarrow 4$. 
The integral of Eq.~(\ref{eq:a_self3}) can be worked out and yields 
\begin{eqnarray}
\Sigma \left( k \right)
\;=\;-\,{e^2 \over {\left( {4\pi } \right)^2}}\;
&&\left\{ \left[ {{1 \over {\varepsilon _{UV}}}-\gamma_E 
+\ln \left( {{{4\pi \mu^2} \over {m^2}}} \right)\,} \right] 
\; \left( {\not k - 4 m} \right) \right.\nonumber\\
&&+\not k \; \left[ {1\,+\,{1 \over {\tilde k^2}}\,+\,
{{1+\tilde k^2} \over {\left( \tilde k^2 \right)^2}} 
(1-\tilde k^2) \, \ln \left(1-\tilde k^2 \right) } \right] \nonumber\\
&&\left. +\,2m \; \left[ {-3\,-\,{{2} \over {\tilde k^2}} 
(1-\tilde k^2) \, \ln \left(1-\tilde k^2 \right)} \right] \right\}\;, 
\nonumber\\
\label{eq:a_self4}
\end{eqnarray}
where $\tilde k^2 \;=\; k^2 / m^2$. 
\newline
\indent
To remove the UV divergence from the self-energy Eq.~(\ref{eq:a_self4}), 
one has to determine the renormalization constants $Z_2$ and $Z_m$ 
from Eq.~(\ref{eq:countertermL}). This counterterm contribution leads to the 
renormalized self-energy 
\begin{equation}
\tilde \Sigma \left( k \right) \;=\; \Sigma \left( k \right) \;-\;
(Z_2 - 1) \not k \;+\;(Z_2 Z_m - 1) \,m \;.
\label{eq:a_self5}
\end{equation}
Inserting Eq.~(\ref{eq:a_self5}) into Eq.~(\ref{eq:a_self2}) and 
developing $\Sigma \left( k \right)$ as a Taylor series expansion 
around $\not k = m$ yields for inverse of the total lepton propagator 
\begin{eqnarray}
S^{-1} \,&=&\, (\not k - m) 
\left[ 1 \,-\, {{d \Sigma} \over {d \not k}}
{\Bigg |}_{\not k = m} \,+\, (Z_2 - 1)\right] 
\;+\; \left[\, (1 - Z_m) Z_2 m \,-\, \Sigma(\not k = m) \, \right] \nonumber\\
&&+ \; O\left( (\not k - m)^2\right) \;.
\label{eq:a_self6}
\end{eqnarray}
Requiring that the total propagator $S$ has a pole at $\not k = m$ with 
residue 1, determines the renormalization constants $Z_2$ and $Z_m$ as 
\begin{eqnarray}
Z_2 \;=\; 1 + {{d \Sigma} \over {d\not k}}{\Bigg |}_{\not k = m}\;, \\
(1 - Z_m) Z_2 m \;=\;\Sigma(\not k = m) \;.
\label{eq:a_self7}
\end{eqnarray}
Using the first order expression of Eq.~(\ref{eq:a_self4}) for the 
lepton self-energy, yields 
\begin{eqnarray}
\label{eq:a_self8}
Z_2\;=\;1 \,-\,{e^2 \over {\left( {4\pi } \right)^2}}\;
&&\left\{ \left[ {{1 \over {\varepsilon _{UV}}}-\gamma_E 
+\ln \left( {{4\pi \mu^2} \over {m^2}} \right)\,} \right] \right. \nonumber\\
&&\left. +\;2\;\left[{1 \over {\varepsilon _{IR}}} -\gamma_E 
+\ln \left( {{4\pi \mu^2} \over {m^2}} \right)\, \right]
\;+\; 4 \right\} \; + \; O(e^4)\;, \\
Z_2 \, Z_m\;=\;1 \,-\,{e^2 \over {\left( {4\pi } \right)^2}}\;
&&\left\{  4\, \left[ {{1 \over {\varepsilon _{UV}}}-\gamma_E 
+\ln \left( {{4\pi \mu^2} \over {m^2}} \right)\,} \right] \right. \nonumber\\
&&\left. +\;2\;\left[{1 \over {\varepsilon _{IR}}} -\gamma_E 
+\ln \left( {{4\pi \mu^2} \over {m^2}} \right)\, \right]
\;+\; 8 \right\} \; + \; O(e^4)\;. 
\label{eq:a_self9}
\end{eqnarray}
Remark that although the unrenormalized lepton self-energy $\Sigma(k)$ of 
Eq.~(\ref{eq:a_self3}) is IR finite, 
the lepton field renormalization constant $Z_2$ contains an infrared 
divergence for the derivative of $\Sigma$ that appears 
in its definition (see Eq.~(\ref{eq:a_self7})). 
Furthermore, a comparison of the first order expressions for 
the lepton field renormalization constant $Z_2$ (Eq.~(\ref{eq:a_self8})) 
with the vertex renormalization constant $Z_1$ (Eq.~(\ref{eq:a_v7})) 
shows that they are the same (It is known as a Ward identity and 
can be shown to hold to all orders as a consequence 
of the gauge invariance of QED).
\newline
\indent
Finally, using the expressions of Eqs.(\ref{eq:a_self8}),(\ref{eq:a_self9}), 
the renormalized lepton self-energy to first order in $\alpha_{em}$ 
is given by 
\begin{eqnarray}
\tilde \Sigma \left( k \right) \;=\;
-\,{{\alpha_{em}} \over {4 \pi}}\;
&&\left\{ \not k \left[ -\;2\;\left({1 \over {\varepsilon _{IR}}} -\gamma_E 
+\ln \left( {{4\pi \mu^2} \over {m^2}} \right)\, \right) - 3 + 
{1 \over {\tilde k^2}} + {{(1 - \tilde k^4)} \over {\tilde k^4}} 
\ln(1 - \tilde k^2) \right] \right. \nonumber\\
&&\left.\;- m \left[ -\;2\;\left({1 \over {\varepsilon _{IR}}} -\gamma_E 
+\ln \left( {{4\pi \mu^2} \over {m^2}} \right)\, \right) - 2 + 
{4 \over {\tilde k^2}} (1 - \tilde k^2) \ln(1 - \tilde k^2) \right] \right\}\;.
\label{eq:a_self10}
\end{eqnarray}
It is seen from Eq.~(\ref{eq:a_self10}) that for an on-shell lepton 
($\not k = m$), the renormalized lepton 
self-energy $\tilde \Sigma$ is exactly zero. Consequently, this correction 
has only to be applied for internal lepton lines.  

\subsection{Vacuum polarization diagram}
\label{app:elast4}

Starting form the free propagator of a photon with four-momentum $q$ 
(as stated before, we give all expressions in the Feynman gauge)   
\begin{equation}
D^{\mu \nu}_o (q) \;=\; { {- \,g^{\mu \nu}} \over {q^2} } \;,
\label{eq:pol1}
\end{equation}
the full photon propagator can be written as 
\begin{equation}
D^{\mu \nu} ( q) \;=\;D^{\mu \nu}_o (q) \;+\;
D^{\mu \kappa} (q) \; \Pi_{\kappa \lambda} (q) \; \;D^{\lambda \nu}_o (q) \;,
\label{eq:pol2}
\end{equation}
where $\Pi_{\kappa \lambda} (q)$ represents the vacuum polarization correction.
To order $O(e^2)$, the vacuum polarization 
(corresponding with Fig.\ref{fig:elasticdia}(c)) 
due to $l^+ \, l^-$ loops (with lepton $l = e, \mu, \tau$) is given by 
\begin{equation}
-i \; \Pi^{\mu \nu} (q) \;=\;
{-e^2} \mu^{4 - D}\, \int {{{d^Dl} \over {\left( {2\pi }
\right)^D}}}\; { {Tr \left\{ \gamma ^\mu \left( {\not l+\not q+m}
\right)\gamma ^\nu \left( {\not l+m} \right) \right\} } \over
{\left[ \left( l + q \right)^2\,-\,m^2 \right]\,
\left[ l^2 \,-\,m^2 \right] }} + O\left( {e^4} \right) \;.
\label{eq:pol3}
\end{equation}
The gauge invariance of QED leads to the relation 
$q^\kappa q^\lambda \; \Pi_{\kappa \lambda} (q)  = 0 $ 
(Ward-Takahashi identity). Consequently, the vacuum polarization correction 
can be written as 
\begin{equation}
\Pi_{\kappa \lambda} (q) \;=\; \left( -\,g_{\kappa \lambda} \; q^2 
\;+\;q_\kappa \; q_\lambda \right) \; \Pi \left( q^2 \right) \;,
\label{eq:pol4}
\end{equation}
where the function $\Pi \left( q^2 \right)$ is IR convergent and 
contains only a logarithmic UV divergence as can be seen from 
Eq.~(\ref{eq:pol3}).
\newline
\indent
Using Eq.~(\ref{eq:pol4}), the self-consistent relation 
for the full photon propagator (Eq.~(\ref{eq:pol2})) yields 
\begin{equation}
D^{\mu \nu} ( q) \;=\;{{-\,g^{\mu \nu}} \over {q^2 \; \left( 1 - \Pi(q^2)\right)}} 
\;+\; {\mathrm term \; in} \; q^\mu \;q^\nu \;,
\label{eq:pol5}
\end{equation}
where we don't have to specify the term in $q^\mu \;q^\nu$, as  
the photon propagator will be contracted with conserved currents on both sides 
so that this term in $q^\mu \;q^\nu$ 
will not contribute to physical observables. 
Evaluating the one-loop integral of Eq.~(\ref{eq:pol3}), one obtains 
\begin{equation}
\Pi(Q^2)\,=\,-{ {e^2} \over {(4 \pi)^2}} 
{4 \over 3} \left[ {{1 \over {\varepsilon _{UV}}}-\gamma_E 
+\ln \left( {{{4\pi \mu^2} \over {m^2}}} \right) } 
- \left( v^2 - {8 \over 3} \right)
\,+\, v   {{\left( v^2 - 3\right)} \over 2} \,
\ln\left( {{v + 1} \over {v - 1}}\right) \right] ,
\label{eq:pol6}
\end{equation}
where $v$ is given by Eq.~(\ref{eq:v}).
\newline
\indent
The UV divergent term in Eq.~(\ref{eq:pol6}) is removed by adding the 
counterterm in $Z_3$ of Eq.~(\ref{eq:countertermL}). This leads to 
the renormalized photon propagator 
\begin{equation}
\tilde D^{\mu \nu} ( q) \;=\;{{-\,g^{\mu \nu}} \over 
{q^2 \; \left( 1 - \tilde \Pi(q^2)\right)}} \;+\; {\mathrm term \; in}
\; q^\mu \;q^\nu \;,
\label{eq:pol7}
\end{equation}
where the renormalized photon polarization $\tilde{\Pi}$ is given by 
\begin{equation}
\tilde{\Pi}(Q^2)\;=\;{\Pi}(Q^2)\;-\;\left( Z_3 - 1\right) \;.
\label{eq:renormpol}
\end{equation}
Requiring that the renormalized photon propagator (Eq.~(\ref{eq:pol5})) 
has a pole at $q^2 = 0$ with residue 1, determines the 
renormalization constant $Z_3$ :
\begin{equation}
Z_3 \;=\; 1 + \Pi \left(q^2 = 0 \right) \;.
\end{equation}
Consequently, the renormalized finite photon polarization is found 
from Eqs.(\ref{eq:pol6}) and (\ref{eq:renormpol}) to be given by 
\begin{equation}
\tilde{\Pi}(Q^2)\;=\;{{\alpha_{em}} \over \pi}
 {1 \over 3} \left[ \left( v^2 - {8 \over 3} \right)
\; + \; v \; {{\left( 3 - v^2 \right)} \over 2} \;
\ln\left( {{v + 1} \over {v - 1}}\right) \right]\;,
\label{eq:photonpol}
\end{equation}
which agrees with the result derived in Ref.\cite{akhiezer}.

\subsection{Soft photon emission contributions}
\label{app:elast5}

The calculation of the one-loop vertex correction of Eq.~(\ref{eq:a_v2}) 
was seen to be both UV and IR divergent. The ultraviolet divergence 
was removed by renormalizing the fields and parameters of the theory. 
The remaining infrared divergences are cancelled at the cross 
section level by the soft bremsstrahlung contributions 
\cite{Bloch37,Jauch55}. 
\newline
\indent
In this bremsstralung process (see Figs.\ref{fig:elasticdia} (d) and (e)),  
an electron is accompanied by the emission of a soft photon of maximal energy 
$\Delta E_s$ (which is related to the detector resolution 
and is therefore much smaller than the electron 
energy which radiates this soft photon). 
To first order in $\alpha_{em}$ (relative 
to the Born cross section) the bremsstrahlung cross section amounts to
calculate a phase space integral of the form~: 
\begin{eqnarray}
d \sigma \sim   
{{d^3 \vec k^{\; '}_e} \over {\left( {2\pi }\right)^3 \, 2 E'_e}} 
{{d^3 \vec p_N^{\; '}} \over {\left( {2\pi }\right)^3 \, 2 E'_N}} 
{{d^3 \vec l} \over {\left( {2\pi }\right)^3 \, 2 {\mathrm l}}} 
&&\; (2 \pi)^4 \delta^4(k + p - k' - p' - l) \nonumber\\
&&\times\, |M_{BORN}|^2 \, \left( - e^2 \right)  
\left[{ {k^{'}_\mu \over {k^{'}.l}} - {k_\mu \over {k.l}}}
  \right] . 
\left[{ {k'^\mu \over {k^{'}.l}} - {k^\mu \over {k.l}}}
  \right] ,
\label{eq:softel0a}
\end{eqnarray}
where ${\mathrm l} \equiv | \vec l|$ denotes the soft photon energy,
and where $M_{BORN}$ denotes the Born amplitude for elastic 
lepton-nucleon scattering.   
In Eq.~(\ref{eq:softel0a}), 
terms in the soft photon momentum were neglected compared with 
the electron momenta $k$ and $k^{'}$, except in the denominators 
of the lepton propagators where they matter.  
\newline
\indent
If one performs an experiment where the outgoing electron 
is detected, and where the
recoiling proton remains undetected (i.e. if one measures a single arm
electron spectrum), one eliminates in Eq.~(\ref{eq:softel0a}) the
integral over $\vec p_N^{\;'}$ with the momentum conserving
$\delta$-function, which gives~:
\begin{eqnarray}
d \sigma \sim   
{{d^3 \vec k^{\; '}_e} \over {\left( {2\pi }\right)^3 \, 2 E'_e}} 
{{d^3 \vec l} \over {\left( {2\pi }\right)^3 \, 2 {\mathrm l}}} 
\, {1 \over {2 E'_N}} \, &&(2 \pi) 
\delta\left(E_e + E_N - E'_e - \sqrt{(\vec q + \vec p_N - \vec l)^2 + M_N^2}
- {\mathrm l} \right) \nonumber\\
&&\times\, |M_{BORN}|^2 \, \left( - e^2 \right)  
\left[{ {k^{'}_\mu \over {k^{'}.l}} - {k_\mu \over {k.l}}}
  \right] . 
\left[{ {k'^\mu \over {k^{'}.l}} - {k^\mu \over {k.l}}}
  \right] .
\label{eq:softel0b}
\end{eqnarray}
Due the energy conserving $\delta$-function in
Eq.~(\ref{eq:softel0b}), the integration volume for the soft photon
has a complicated ellipsoidal shape in the {\it lab} system. 
In order for the soft-photon phase space integration volume 
to be spherical, one has to perform
the calculation in the c.m. system ${\mathcal S}$ of the (recoiling nucleon +
soft-photon), as discussed in \cite{Tsai61}. 
The system ${\mathcal S}$ is thus defined by~: 
$\vec p_N^{\; '} + \vec l = \vec q + \vec p_N = 0$. 
In the system ${\mathcal S}$, 
the energy conserving delta function is independent of the soft-photon
angles, and the maximal soft photon energy is isotropic. 
The integral over the soft-photon momentum (up to some maximum value
$\Delta E_s$) can then be performed independently from the integration
over the soft photon emission angles. 
If $\Delta E_s$ is sufficiently small, one can furthermore 
neglect the soft photon energy with respect to the other energies 
in the $\delta$-function, and perform the integral over
the electron momentum $|\vec k_e^{\; '}|$ in Eq.~(\ref{eq:softel0b}). 
The integration over the outgoing electron momentum eliminates the
$\delta$-function, which implies the elastic scattering
constraint. This yields then for the differential cross section with
respect to the outgoing electron angles, the following correction due
to soft bremsstrahlung~:
\begin{equation}
\left( {{d \sigma} \over {d \Omega_e^{'}}} \right)_{REAL\,SOFT\,
  \gamma} =  
\left( {{d \sigma} \over {d \Omega_e^{'}}} \right)_{BORN} 
\left( -\,e^2 \right)  
\int {{{d^3 \vec l} \over {\left( {2\pi }\right)^3 \, 2 {\mathrm l}}}} 
\left[{ {k^{'}_\mu \over {k^{'}.l}} - {k_\mu \over {k.l}}}
  \right] . 
\left[{ {k'^\mu \over {k^{'}.l}} - {k^\mu \over {k.l}}}
  \right] ,
\label{eq:softel1}
\end{equation}
where the soft-photon phase space integral is performed in
the system ${\mathcal S}$, in which the integration volume is spherical. 
We will denote in the following the external kinematics in the
system ${\mathcal S}$ by tilded quantities 
($\tilde E_e, \tilde E_e', \tilde E_N, \tilde E'_N$) 
to distinguish them from the {\it lab} quantities, which we denote
by untilded quantities ($E_e, E_e', E_N \equiv M_N, E'_N$). 
To make the transformation between the system ${\mathcal S}$ and the 
{\it lab} system, we first introduce the 
missing four-momentum $p_m \equiv p_N' + l$. 
The system ${\mathcal S}$ is defined by $\vec p_m = \vec 0$, and the soft
photon limit implies $p_m^0 \approx M_N$.    
We can then easily express in the system ${\mathcal S}$, 
the energies for the external particles in the elastic scattering
process, in terms of {\it lab} quantities~:
\begin{eqnarray}
\tilde E_e &\approx& {{k \cdot p_m} \over {M_N}} = 
{1 \over M_N} k \cdot (p + q) = {1 \over {M_N}} (M_N E_e - Q^2/2) = E_e' \, , 
\label{eq:elastransf1} \\
\tilde E_e' &\approx& {{k' \cdot p_m} \over {M_N}} = 
{1 \over M_N} k' \cdot (p + q) = {1 \over {M_N}} (M_N E_e' + Q^2/2) = E_e \, , 
\label{eq:elastransf2} \\
\tilde E_N &\approx& {{p \cdot p_m} \over {M_N}} = 
{1 \over M_N} p \cdot (p + q) = M_N + E_e - E_e' = E_N' \, , 
\label{eq:elastransf3} 
\end{eqnarray}
where the elastic scattering condition ($Q^2 = 2 M_N (E_e - E_e')$)
has been used in the last step in 
Eqs.~(\ref{eq:elastransf1},\ref{eq:elastransf2}).  
The angle $\tilde \theta_e$ in the frame ${\mathcal S}$ is obtained
from $k. k' = \tilde E_e \tilde E_e' (1 - \cos \tilde \theta_e) 
= E_e E_e' (1 - \cos \theta_e)$, which shows (using
Eqs.~(\ref{eq:elastransf1},\ref{eq:elastransf2})) 
that in the soft-photon limit, this angle is the same as in the 
{\it lab} system, i.e. $\cos \tilde \theta_e = \cos \theta_e$. 
\newline
\indent
The integral of Eq.(\ref{eq:softel1}) extends up to 
a maximal soft-photon energy $\Delta E_s$ in the system ${\mathcal S}$, 
which is expressed in terms of the {\it lab} quantities $E_e$ and $E_e'$, by
using~:
\begin{equation}
(p' + l)^2 - M_N^2 
= (p + k - k')^2 - M_N^2
= 2 p \cdot (k - k') + (k - k')^2 \; , 
\label{eq:derivedeltae1}
\end{equation}
which leads (for soft-photon energies, i.e. keeping only terms of
first order in $\Delta E_s$) to 
\begin{eqnarray}
2 M_N \Delta E_s &\approx& 2 M_N (E_e - E_e') 
- 4 E_e E_e' \sin^2 \theta_e/2 \;, \nonumber\\
&=& 2 M_N (E_e - E_e') 
- 2 M_N ( E_e - E_e'^{el}) E_e' / E_e'^{el} \;.
\label{eq:derivedeltae}
\end{eqnarray}
All quantities on the {\it rhs} of Eq.~(\ref{eq:derivedeltae})
are in the {\it lab}, and the 
elastic scattering condition has been used in the last line 
($E_e'^{el}$ denotes the elastic scattered electron 
{\it lab} energy, to distinguish it from $E_e'$). 
From Eq.~(\ref{eq:derivedeltae}), one determines then 
$\Delta E_s$ in terms of {\it lab} quantities 
from the scattered electron spectrum through 
\begin{equation}
\Delta E_s = \eta \left( E_e'^{el} -  E_e' \right) \;,
\label{eq:deltae}
\end{equation}
where the recoil factor $\eta$ is given by $\eta = E_e / E_e'^{el}$.
\newline
\indent
Deviations from the soft-photon emission formula Eq.~(\ref{eq:softel1})
will show up when $\Delta E_s$ is not very small compared with the 
lepton momenta in the process. 
The emission of such a semi-hard photon is what is usually referred to
as the radiative tail. Although the distinction is somewhat arbitrary,
one can always split the integral for photon emission 
into two parts, one by integrating up to a small value 
$\Delta E_s$, where the soft-photon approximation in writing down
Eq.~(\ref{eq:softel1}) holds, and a second integral, starting from
this small (but non-zero) value of $\Delta E_s$ up to the energy where
one performs the cut in the spectrum.
This second integral is finite and can be performed numerically. Such
a numerical calculation of the radiative tail
without approximations is presented in section~\ref{sec:radtail}. 
In the present section, we give an analytical result for the 
soft-photon (i.e. small $\Delta E_s$) integral 
of Eq.~(\ref{eq:softel1}), without any further
approximations (Remark that in \cite{mo69} only an approximate
evaluation of Eq.~(\ref{eq:softel1}) has been given). 
\newline
\indent
As is immediately seen by power counting, 
the integral in Eq.~(\ref{eq:softel1}) has a logarithmic 
IR divergence, corresponding with the emission of photons with zero energy. 
To demonstrate the cancellation with the IR divergence 
of the vertex diagram as 
stated above, one has to regularize the integral of Eq.~(\ref{eq:softel1}). 
In this work this is performed by also using dimensional
regularization. The soft photon integral 
is then evaluated in $D - 1$ dimensions 
($D \rightarrow 4$ corresponds to the physical limit). 
One now sees that it is extremely advantageous to have a
spherical integration volume, in order to evaluate the integral for
dimensions $D \neq 4$. 
Before continuing the integral of Eq.~(\ref{eq:softel1}) into 
$D - 1$ dimensions, 
the integration limits for $l$ have to be made dimensionless, which leads 
in the dimensional regularization scheme 
to the introduction of the same scale $\mu$ in Eq.~(\ref{eq:softel1b}) 
as was introduced when changing the dimension of the 
virtual photon loop integral of Eq.~(\ref{eq:a_v2}). 
This leads then in $D - 1$ dimensions, to the bremsstrahlung integral~: 
\begin{eqnarray}
I\;=\;
\;-\;e^2 \; \int^{{\mathrm l} < \Delta E_s / \mu} 
{{{d^{D - 1} l} \over {\left( {2\pi }\right)^{D - 1} \, 2 \, {\mathrm l}}}} 
\left[{ {k^{'}_\alpha \over {k^{'}.l}}\,-\,{k_\alpha \over {k.l}}}\right] \,.
\, \left[{ {k'^\alpha \over {k^{'}.l}}\,-\,{k^\alpha \over {k.l}}}\right]\;.
\label{eq:softel1b}
\end{eqnarray}
\indent 
The integral in Eq.~(\ref{eq:softel1b}) is worked out by introducing 
polar coordinates in $D - 1$ dimensions. To define the polar angle in  
the interference term of Eq.~(\ref{eq:softel1b}), a Feynman parametrization 
is performed. This leads for $I$ to the expression~:
\begin {eqnarray}
I\;=\;
&e^2& \, \int_0^{{\mathrm l} < \Delta E_s / \mu} 
{{d\,{\mathrm l}} \over {\left( {2\pi }\right)^{D - 1} } }\, 
{{{\mathrm l}^{D - 2}} \over {2 \, {\mathrm l}^3 } } \nonumber\\   
&\times& \, \int_{D - 2} d \Omega_l
\, \left\{ {{k . k^{'}} \over {\tilde E_e \, \tilde E_e^{'}}} \,
\int_{-1}^{+1} \, dy \, {1 \over {\left( 1 - \tilde{\vec\beta_y}\,.\hat{l} 
\right)^2} } 
-{{ ( 1 - \tilde{\beta_e}^2 )} \over 
{\left( 1 - \tilde{\vec\beta_e} \,.\,\hat{l} \right)^2} }
- {{ ( 1 - \tilde{\beta_e^{'}}^2 )} \over 
{\left( 1 - \tilde{\vec\beta_e^{\, '}} \,.\,\hat{l} \right)^2} } \right\} ,
\label{eq:softel2}
\end{eqnarray}
where $\hat l$ is the unit-vector along the soft photon direction, 
$\tilde \beta_e \equiv | \tilde{\vec\beta_e} |$, 
$\tilde \beta_e^{'} \equiv | \tilde{\vec\beta_e^{\, '}} |$ 
are the incoming and outgoing electron velocities 
(in the system ${\mathcal S}$) respectively 
and where $\tilde \beta_y  \equiv | \tilde{\vec\beta_y} |$ with
\begin{eqnarray}
&&\tilde{\vec\beta_e} \equiv {{\tilde{\vec{k}_e}} \over {\tilde E_e}}\;,  
\hspace{1cm}
\tilde{\vec\beta_e^{\, '}} \equiv {{\tilde{\vec{k}_e^{\, '}}} \over 
{\tilde E_e^{'}} } \;, \nonumber\\
&&\tilde{\vec\beta_y} \equiv \tilde{\vec{\beta}_e} \, {1 \over 2} (1 + y)\;+\;
\tilde{\vec\beta_e^{\, '}} \, {1 \over 2} (1 - y) \;.
\label{eq:softel2b}
\end{eqnarray}
The integrals over ${\mathrm l}$ 
and the azimuthal angular integral (over $D - 2$ 
dimensions) can be performed immediately which yields~: 
\begin{eqnarray}
I\;=\;&e^2& \left[ {(2 \pi)^{2 \epsilon_{IR}} \over {(2 \pi)^3} } 
\left( {{\Delta E_s} \over \mu} \right)^{- 2 \epsilon_{IR}} {1 \over {- 4 \epsilon_{IR}}} \right] 
\,.\, \left[ {{2 \pi} \over {\pi^{\epsilon_{IR}}} }
{1 \over {\Gamma (1 - \epsilon_{IR})} }\right] \nonumber\\
&&\times \left\{ {{k . k^{'}} \over {\tilde E_e \, \tilde E_e^{'}}} \;
\int_{-1}^{+1} dy \; \int_{-1}^{+1} dx \; 
{{\left( 1 - x^2\right)^{- \epsilon_{IR}}} \over 
{\left( 1 - \tilde{\beta_y} \, x \right)^2} } \right. \;\nonumber\\
&&\hspace{.5cm} \left. -\; \left( 1 - \tilde{\beta_e}^2 \right) \; 
\int_{-1}^{+1} dx \; 
{{\left( 1 - x^2\right)^{- \epsilon_{IR}}} \over 
{\left( 1 - \tilde{\beta_e} \, x \right)^2} }
\;-\; \left( 1 - {\tilde{\beta_e}^{'\, 2}} \right) \; \int_{-1}^{+1} dx \; 
{{\left( 1 - x^2\right)^{- \epsilon_{IR}}} \over 
{\left( 1 - \tilde{\beta_e}^{'} \, x \right)^2} }
\right\} \;, 
\label{eq:softel3}
\end{eqnarray}
The IR divergent term and the finite term are obtained by developing 
the polar angular integral in Eq.~(\ref{eq:softel3}) as 
\begin{equation}
\int_{-1}^{+1} dx \; {{\left( 1 - x^2\right)^{- \epsilon_{IR}}} \over 
{\left( 1 - \beta \, x \right)^2} } \;=\;
\int_{-1}^{+1} dx \; { 1 \over {\left( 1 - \beta \, x \right)^2} }
\;-\;\epsilon_{IR} \int_{-1}^{+1} dx \; { { \ln \left( 1 - x^2\right) } \over 
{\left( 1 - \beta \, x \right)^2} } \; + O\left( \epsilon_{IR}^2 \right) \;.
\label{eq:softel4}
\end{equation}
Performing the integrations in Eq.~(\ref{eq:softel4}) (the second integral 
in Eq.~(\ref{eq:softel4}) is simplified by making the substitution 
$ x \rightarrow u = \beta / (1 - \beta x)$ ) yields 
\begin{equation}
\int_{-1}^{+1} dx \; {{\left( 1 - x^2\right)^{- \epsilon_{IR}}} \over 
{\left( 1 - \beta \, x \right)^2} } \;=\;
 {2 \over {1 - \beta^2}} \, 
- \epsilon_{IR} {2 \over {1 - \beta^2}} 
\left[ \ln 4 + {1 \over \beta} \ln {{1 -\beta} \over {1 + \beta}} \right]  
\; + O\left( \epsilon_{IR}^2 \right) \;.
\label{eq:softel5}
\end{equation}
Consequently, the IR divergent term and the finite term of the integral $I$ 
are obtained by using Eq.~(\ref{eq:softel5}) in Eq.~(\ref{eq:softel3}) 
and by developing all other factors also to order $\epsilon_{IR}$~: 
\begin{eqnarray}
I \;=\; - {{e^2} \over {4 \pi^2}} \, 
&&\left\{ \left[ - {1 \over {\epsilon_{IR}}} + \gamma_E - \ln {{4 \pi \mu^2} \over {m^2}}  
+ \ln {{4 (\Delta E_s)^2} \over {m^2}}  \right]
\;\left[ 1 - {1 \over 2} \left( 1 \,-\, \tilde\beta_e \, \tilde\beta_e^{'} \, 
\cos \tilde\theta_e \right) \;I_y^{(1)} \right] \right. \nonumber\\
&&\left.+\;\left[ {1 \over {2 \tilde{\beta_e}}} 
\ln {{1 - \tilde\beta_e} \over {1 + \tilde\beta_e} } \;+\;
{1 \over {2 \tilde\beta_e^{'}}} 
\ln {{1 - \tilde\beta_e^{'}} \over {1 + \tilde\beta_e^{'}} } \;-\; 
{1 \over 2} \left( 1 \,-\, \tilde\beta_e \, \tilde\beta_e^{'} \, 
\cos \tilde\theta_e \right) \; I_y^{(2)} \right] \right\} \;,
\label{eq:softel6}
\end{eqnarray}
where the remaining Feynman parameter integrals $I_y^{(1)}$ and $I_y^{(2)}$ 
are given by
\begin{eqnarray}
&&I_y^{(1)} \;\equiv\; \int_{-1}^{+1} dy \; 
{1 \over {1 - \tilde\beta_y^2}} \;, \nonumber\\
&&I_y^{(2)} \;\equiv\; \int_{-1}^{+1} dy \; 
{1 \over {\tilde\beta_y \left(1 - \tilde\beta_y^2 \right)}} 
\; \ln {{1 - \tilde\beta_y} \over {1 + \tilde\beta_y}} \;,
\label{eq:softel7}
\end{eqnarray}
and where $\tilde\beta_y$ is given by Eq.~(\ref{eq:softel2b}).
The integral $I_y^{(1)}$ in Eq.~(\ref{eq:softel7}) 
can be performed easily and yields 
\begin{equation}
I_y^{(1)} \;=\;{{2 \tilde E_e \tilde E_e^{'}} \over {m^2}} 
\,{{v^2 - 1} \over {2v}}\, \ln\left( {{v + 1} \over {v - 1}}\right) \;, 
\label{eq:softel8}
\end{equation}
with $v$ as defined in Eq.~(\ref{eq:v}).
To obtain an analytical formula for the integral $I_y^{(2)}$ is much harder 
but was performed in Ref.\cite{decalan91}, which we 
checked \footnote{Note that the relevant formula quoted in
  Ref.~\cite{decalan91} contains some typing errors.} 
and which yields the result~: 
\begin{eqnarray}
I_y^{(2)} \;=\;{1 \over {|\tilde{\vec \beta_e} - \tilde{\vec \beta_e^{'}}}|  
\, \tanh \alpha} \,
&&\left\{ \left[ - 2 \ln(2) + {1 \over 2} \ln({\sinh}^2 \alpha - {\sinh}^2 \phi_1) \right] \,
\ln {{\sinh \alpha + \sinh \phi_1} \over {\sinh \alpha - \sinh \phi_1}} \right. \nonumber\\
&&-\;\ln(\sinh \alpha + \sinh \phi_1) \, \ln {{\sinh \alpha - \sinh \phi_1} \over {4 \sinh^2 \alpha}} \nonumber\\
&&+2\, \ln \left[ e^{- \alpha} {{e^{\alpha} + e^{\phi_1}} \over {e^{- \alpha} + e^{\phi_1}} } \right] 
\; \ln { {\cosh \alpha + \cosh \phi_1} \over {\cosh \alpha - \cosh \phi_1}} \nonumber\\
&&-\;2 \Phi \left[ {{\sinh \alpha + \sinh \phi_1} \over {2 \,\sinh \alpha }}\right] 
\;+\;\Phi \left[ \left( {{e^{\alpha} - e^{\phi_1}} \over {e^{\alpha} + e^{\phi_1}} }\right)^2 \right] \nonumber\\
&&\left. -\;\Phi \left[ \left( {{e^{\phi_1} - e^{- \alpha} } \over {e^{\phi_1} + e^{- \alpha}} }\right)^2\right] 
\;-\; \left[ \phi_1 \longrightarrow \phi_2 \right] \right\} \;,
\label{eq:softel9}
\end{eqnarray}
where $\alpha$, $\phi_1$ and $\phi_2$ are given by~: 
\begin{eqnarray}
&&\cosh \alpha \;=\; 
{{|\tilde{\vec \beta_e} - \tilde{\vec \beta_e^{\, '}}|} 
\over {\tilde\beta_e \tilde\beta_e^{'} \sin \tilde\theta_e}} 
\hspace{.5cm} (\alpha > 0) \;, \nonumber\\
&&\cosh \phi_1 \;=\; \tilde\beta_e \cosh \alpha \;, 
\hspace{2.5cm} \sinh \phi_1 \;=\; 
{{- \tilde\beta_e \tilde\beta_e^{'} \cos \tilde\theta_e + \tilde\beta_e^2} 
\over {\tilde\beta_e \tilde\beta_e^{'} \sin \tilde\theta_e}} \;, \nonumber\\
&&\cosh \phi_2 \;=\; \tilde\beta_e^{'} \cosh \alpha \;, 
\hspace{2.5cm} \sinh \phi_2 \;=\; {{\, \tilde\beta_e \tilde\beta_e^{'}
 \cos \tilde\theta_e - \tilde\beta_e^{'2}} 
\over {\tilde\beta_e \tilde\beta_e^{'} \sin \tilde\theta_e}} \;.
\label{eq:softel10}
\end{eqnarray}
The function $\Phi$ in Eq.~(\ref{eq:softel9}) is given by 
\begin{equation}
\Phi(x) \;\equiv\; - \int\limits_0^x {dt\,{\ln|1 - t| \over t}}\;.
\label{eq:spenceabs}
\end{equation} 
which agrees with the Spence function (Eq.~(\ref{eq:spence})) when $x < 1$. 
Compared with previous calculations in the literature, it was shown in 
Ref.\cite{decalan91} that this integral 
$I_y^{(2)}$ was approximated in Ref.\cite{Yennie61} and that the calculation 
of this integral in Ref.\cite{mork65} contains a factor two error. 
We also checked the analytical formula of Eq.~(\ref{eq:softel9}) by performing 
the integral of Eq.~(\ref{eq:softel7}) numerically. 
\newline
\indent
In the ultrarelativistic limit ($\tilde\beta_e, \tilde\beta_e^{'} 
\rightarrow 1$), 
the integral $I_y^{(2)}$ of Eq.~(\ref{eq:softel9}) reduces to 
\begin{eqnarray}
I_y^{(2)} \,\stackrel{\tilde\beta_e \approx 1 , \tilde\beta_e^{'} \approx 1}
{\longrightarrow} {1 \over {2 \, \sin^2 {{\tilde\theta_e} \over 2}} }
&&\left\{ -{1 \over 2} \ln^2 \left( 1 - \tilde\beta_e^2\right)
- {1 \over 2} \ln^2 \left( 1 - \tilde\beta_e^{'2} \right) 
+ \ln 4 \, \ln \left( 1 - \tilde\beta_e^{2} \right)
+ \ln 4 \, \ln \left( 1 - \tilde\beta_e^{'2} \right) \right. \nonumber\\
&&+\,4 \,\left( \ln^2 \left( \sin {\tilde\theta_e \over 2} \right) - \ln^2 2
\right) 
\;-\;2 \, \ln \left( \cos^2 {\tilde\theta_e \over 2} \right) \, 
\ln \left( \sin^2 {\tilde\theta_e \over 2} \right) \nonumber\\
&&\left. -\, {{\pi^2} \over 3} \,-\,2 \; 
Sp\left( \sin^2 {\tilde\theta_e \over 2}\right) \right\}\;.
\label{eq:softel11}
\end{eqnarray}
\newline
\indent
Putting all pieces together, the result for the bremsstrahlung cross section 
accompanying elastic electron scattering is obtained as
\begin{eqnarray}
&&\left( {{d \sigma} \over {d \Omega_e^{'}}} \right)_{REAL\,SOFT\, \gamma} \nonumber\\
&&= \left( {{d \sigma} \over {d \Omega_e^{'}}} \right)_{BORN}  
\left\{ {\alpha_{em} \over \pi}  
 \left[ - {1 \over {\varepsilon _{IR}}}+\gamma_E 
-\ln \left( {{4\pi \mu^2} \over {m^2}} \right)\, \right]   
\left[ {{{v^2+1} \over {2 v}}\,\ln \left( {{{v+1} \over {v-1}}}
    \right) - 1} \right] + \delta_R \right\} ,
\label{eq:softel}
\end{eqnarray}
where the finite part $\delta_R$ of the real radiative corrections is given by
\begin{eqnarray}
\label{eq:radcorrreal1}
\delta_R &=&{\alpha_{em} \over \pi} \; 
\left\{  \ln \left( {{4 (\Delta E_s)^2} \over {m^2}} \right) \;  
\left[ {{{v^2+1} \over {2 v}}\,\ln \left( {{{v+1} \over {v-1}}} \right) 
\,-\, 1} \right] \right. \nonumber\\ 
&&\left. \hspace{.8cm} - \, {1 \over {2 \tilde\beta_e}} \, 
\ln \left( {{1 - \tilde\beta_e} \over {1 + \tilde\beta_e}}\right)
- {1 \over {2 \tilde\beta_e^{'}}} \, 
\ln \left( {{1 - \tilde\beta_e^{'}} \over {1 + \tilde\beta_e^{'}}}\right) 
+  {1 \over 2} 
\left( 1 - \tilde\beta_e \, \tilde\beta_e^{'} \, \cos \tilde\theta_e \right) 
\,  I_y^{(2)} \right\} , \\
&\stackrel{Q^2 >> m^2}{\longrightarrow}& \, {\alpha_{em} \over \pi} 
\left\{  \ln \left( {{4 (\Delta E_s)^2} \over {m^2}} \right) 
\left[ \ln \left( {Q^2 \over m^2} \right) \,-\, 1 \right] 
\,- {1 \over 2 } \, \ln \left( {{1 - \tilde\beta_e^2} \over 4}\right)
\,-\, {1 \over 2} \, \ln \left( {{1 - \tilde\beta_e^{'2}} \over 4}\right) 
\right. \nonumber\\
&&\hspace{.8cm} -{1 \over 4} \ln^2 \left( 1 - \tilde\beta_e^2\right)
- {1 \over 4} \ln^2 \left( 1 - \tilde\beta_e^{'2} \right) 
+ \ln 2 \, \ln \left( 1 - \tilde\beta_e^{2} \right)
+ \ln 2 \, \ln \left( 1 - \tilde\beta_e^{'2} \right) \nonumber\\
&&\hspace{.8cm}\left. +\; 2 \left( \ln^2 \left( \sin {\tilde\theta_e \over 2}
  \right) - \ln^2 2 \right) 
\,-\, {{\pi^2} \over 3} \,+\, Sp\left( \cos^2 {\tilde\theta_e \over 2}\right) 
\right\} , \label{eq:radcorrreal2} \\
&=&  \, {\alpha_{em} \over \pi} \left\{
\ln \left( {{ (\Delta E_s)^2} \over { \tilde E_e \tilde E_e^{'}}} \right)  
\left[ \ln \left( {Q^2 \over m^2} \right) \,-\, 1 \right] \right. \nonumber\\
&&\hspace{.8cm}\left. - {1 \over 2} \ln^2 \left( {{\tilde E_e} 
\over {\tilde E_e^{'}}}  
\right) \,+\, {1 \over 2} \ln^2 \left( {Q^2 \over m^2} \right)
\,-\, {{\pi^2} \over 3} \,+\, Sp\left( \cos^2 {\tilde \theta_e \over 2}\right) 
\right\} \;,
\label{eq:radcorrreal}
\end{eqnarray}
where the expression of Eq.~(\ref{eq:radcorrreal2}) 
in the $Q^2 >> m^2$ limit has been rewritten in
Eq.~(\ref{eq:radcorrreal}) to allow comparison with other expressions
found in the literature. 
\newline
\indent
Finally to evaluate $\delta_R$, 
we have to express the quantities in the system ${\mathcal S}$ 
in terms of {\it lab} quantities. 
The relations given in
Eqs.~(\ref{eq:elastransf1},\ref{eq:elastransf2})) 
yield for elastic scattering : 
$\tilde E_e = E_e'$, $\tilde E_e' = E_e$, and 
$\cos \tilde \theta_e = \cos \theta_e$. From the formula for
$\delta_R$ (e.g. Eq.~(\ref{eq:radcorrreal}), one then sees that one
formally obtains exactly the same expression in terms of the 
{\it lab} quantities $E_e, E_e', \theta_e$. 
The quantity $\Delta E_s$ is calculated from the cut in the electron
spectrum, using the expression of Eq.~(\ref{eq:deltae}).
\newline
\indent
A comparison of expressions Eqs.~(\ref{eq:softel}),(\ref{eq:radcorrreal1}) 
with the literature, 
shows that the same result is obtained as in Ref.\cite{decalan91}.
A comparison with the expression used by {\it Mo and Tsai} \cite{mo69}
will be given in the next section when we add the vertex correction and 
soft photon emission contribution, because only their sum is IR finite 
(and thus independent of the IR regularization procedure used).

\subsection{Elastic lepton-nucleon scattering}

In this section, we bring together the first order radiative
corrections at the lepton side (lepton vertex and soft bremsstrahlung from
the lepton) and the photon polarization correction to correct the 
elastic lepton-nucleon scattering cross section. 
As was shown in the previous sections, these corrections can be
calculated model-independently. In the next section, we discuss the
additional radiative corrections to the lepton-proton cross section,
which originate from the proton side (proton vertex correction,
soft bremsstrahlung from proton and two-photon exchange corrections).  
To calculate these corrections at the proton side, 
a model for the off-shell (or half off-shell) $\gamma NN$
vertex is needed however, and which is therefore to some extent
model-dependent. For this latter part, we will refer to the 
recent work of Ref.~\cite{MTj99}. 
\newline
\indent
The elastic lepton scattering cross section, corrected 
to first order in $\alpha_{em}$ for the lepton vertex 
contribution and for the photon polarization contribution, 
is given by 
\begin{eqnarray}
&&\left( {{d \sigma} \over {d \Omega_e^{'}}} \right)_{VIRTUAL \, \gamma} 
\,\approx\, {\left( {{d \sigma} \over {d \Omega_e^{'}}} \right)}_{BORN} \; 
{1 \over {\left( 1 - \tilde{\Pi}(Q^2) \right)^2}} \;
\left( {1 \;+\; 2 \, \left\{ {F(Q^2) - F(Q^2 = 0)} \right\} } \right)
\nonumber\\
=&& \left({ { {d \sigma} \over {d \Omega_e^{'}} } }\right) _{BORN}
  \;{1 \over {\left( 1 - \tilde{\Pi}(Q^2) \right)^2}} \; \nonumber\\
&&\times \left( 1 + {\alpha_{em} \over \pi} 
\left[ { {1 \over {\varepsilon _{IR}}} - \gamma_E 
+\ln \left( {{{4\pi \mu^2} \over {m^2}}} \right)\,} \right] .
\left[ {{{v^2+1} \over {2 v}}\,\ln \left( {{{v+1} \over {v-1}}}
  \right) \,-\, 1} \right] + \delta_{vertex} \right) \;,
\label{eq:el1}
\end{eqnarray}
where the finite part $\delta_{vertex}$ of the lepton vertex correction is 
found from Eq.~(\ref{eq:a_v8}) to be given by 
\begin{eqnarray}
\delta_{vertex} &&\;=\; {\alpha_{em} \over \pi} \; 
\left\{ {{v^2+1} \over {4v}}\;\ln \left( {{{v+1} \over {v-1}}}\right)\;
\ln \left( {{{v^2-1} \over {4v^2}}}\right) \,
+\, {{2 v^2 + 1} \over {2v}} \,\ln \left( {{{v+1} \over {v-1}}} \right) \,-2\, \right. \nonumber\\
&&\hspace{2cm}\left. +{{v^2+1} \over {2v}}\,
\left[ {Sp\left( {{{v+1} \over {2v}}} \right) - 
Sp\left( {{{v-1} \over {2v}}} \right)} \right] \right\} \;, \nonumber\\
&&\stackrel{Q^2 >> m^2}{\longrightarrow} \; {\alpha_{em} \over \pi} 
\left\{ {3 \over 2}  \ln \left( {{Q^2} \over {m^2}} \right) 
\,-\, 2   
\,-\,{1 \over 2}  \ln^2 \left( {{Q^2} \over {m^2}} \right)
\,+\, {{\pi^2} \over 6}   \right\} \;.
\label{eq:el2}
\end{eqnarray}
\indent
In writing down Eq.~(\ref{eq:el1}) to first order in $\alpha_{em}$, 
the contribution of the anomalous magnetic moment 
term $G\left( Q^2 \right)$ in the vertex correction
Eq.~(\ref{eq:a_v3}) has been dropped. 
This contribution vanishes in the ultrarelativistic limit 
( $Q^2 >> m^2$ ) as can be seen from Eq.~(\ref{eq:a_v5}).
The first term in the last line of Eq.~(\ref{eq:el2}) 
corresponds with the vertex correction 
term quoted by {\it Mo and Tsai} (Eq.~( II.5) of Ref.\cite{mo69}).
\newline
\indent
The finite part of the photon polarization correction,
$\delta_{vac} \equiv 2 \, \tilde{\Pi}(Q^2)$, 
follows from Eq.~(\ref{eq:photonpol}) as  
\begin{eqnarray}
\label{eq:el3a}
\delta_{vac} \, &=& \, {\alpha_{em} \over \pi} \; 
{2 \over 3} \left\{ \left( v^2 - {8 \over 3} \right)
\; + \; v \; {{\left( 3 - v^2 \right)} \over 2} \; 
\ln\left( {{v + 1} \over {v - 1}} \right) \right\} \;, \\
&\stackrel{Q^2 >> m^2}{\longrightarrow}& \; {\alpha_{em} \over \pi} \; 
{2 \over 3} \left\{ -{5 \over 3} \;+\;  \ln\left( {Q^2 \over m^2} \right) \right\} \;,
\label{eq:el3}
\end{eqnarray}
which agrees with the expression quoted by 
{\it Mo and Tsai} (Eq.~( II.4) of Ref.\cite{mo69}).
To evaluate the vacuum polarization due to $\mu^+ \mu^-$ 
and $\tau^+ \tau^-$ pairs at intermediate $Q^2$, one has to use 
Eq.~(\ref{eq:el3a}) instead of the limit of Eq.~(\ref{eq:el3}). 
\footnote{Note that an incorrect expression is used in \cite{Walker94}
  for the vacuum polarization contribution due to $\mu^+ \mu^-$ pairs 
(Eq.~(A5) in their paper).} 
\newline
\indent
When adding the real (Eq.~(\ref{eq:softel})) and 
virtual (Eq.~(\ref{eq:el1})) radiative corrections 
at the lepton side, 
one verifies that the IR divergent parts exactly cancel. The remaining 
finite contribution is given to first order in $\alpha_{em}$ by 
\begin{equation}
\left( {{d \sigma} \over {d \Omega_e^{'}}} \right)_{VIRTUAL \, \gamma} \,+\,
\left( {{d \sigma} \over {d \Omega_e^{'}}} \right)_{REAL \, SOFT \,
  \gamma} \,=\,\left( {{d \sigma} \over {d \Omega_e^{'}}} \right)_{BORN} 
\, \left( 1 + \delta_{vac} + \delta_{vertex} + \delta_{R}  
\right) \;, 
\label{eq:radcorrtot}
\end{equation}
where $\delta_{vac}$, $\delta_{vertex}$ and $\delta_{R}$ are 
given by Eqs.~(\ref{eq:el3a}), (\ref{eq:el2}), and 
(\ref{eq:radcorrreal1})-(\ref{eq:radcorrreal}) respectively.
Bringing the three contributions together, leads to the 
expression (in the $Q^2 >> m^2$ limit)
\begin{eqnarray}
\delta_{vac} + \delta_{vertex} + \delta_R 
=  {\alpha_{em} \over \pi} 
&&\left\{ \; \ln \left( {{(\Delta E_s)^2} \over {E_e E_e^{'}}} \right) 
\left[ \ln \left( {Q^2 \over m^2} \right) \,-\, 1 \right] \right. \nonumber\\
+&&\left. {{13} \over 6}  \ln \left( {{Q^2} \over {m^2}} \right) 
- {{28} \over 9}
- {1 \over 2} \ln^2 \left( {{E_e} \over {E_e^{'}}}  \right) 
\,-\, {{\pi^2} \over 6} \,+\, Sp\left( \cos^2 {\theta_e \over 2}\right) 
\right\} ,
\label{eq:sumrealvertex}
\end{eqnarray}
where $\Delta E_s$, which is 
the maximum soft-photon energy in the c.m. system of
(recoiling proton + soft-photon), is determined as in
Eq.~(\ref{eq:deltae}), when applying this formula 
to the scattered electron spectrum. 
\newline
\indent
We can compare Eq.~(\ref{eq:sumrealvertex}) with the recent
calculation of {\it Maximon and Tjon} \cite{MTj99}, where this
calculation was also performed (using a finite photon mass to
regularize the IR divergences) without doing any
approximations. Comparing Eq.~(\ref{eq:sumrealvertex}) with their 
$Z$-independent term ($Z$ being the hadron charge) - i.e. when not
considering radiative corrections at the proton side or two-photon
exchange contributions at this point - we find exactly the same
result. As was noted in Ref.~\cite{MTj99}, the last two terms of 
Eq.~(\ref{eq:sumrealvertex}) were omitted by {\it Mo and Tsai} \cite{mo69}. 
\newline
\indent
We can approximately take into account the higher order radiative
corrections by exponentiating the first order 
vertex and real radiative corrections. This is strictly true only
for the IR divergent part of the vertex correction 
and soft photon emission contribution, and 
was demonstrated in Refs.~\cite{Bloch37,Yennie61} (see
e.g. Refs.~\cite{Weinberg95,Muta} for pedagogical derivations).  
The application of this exponentiation procedure also 
to the finite part consists of an approximation which can be checked
by comparing the result with the first order formula of 
Eq.~(\ref{eq:radcorrtot}).
For the photon polarization contribution, we iterate the first order
vacuum polarization contribution of Eq.~(\ref{eq:el3a}) to all orders
(resumming all vacuum bubbles of the type of 
Fig.~\ref{fig:elasticdia} (c)) by keeping the photon self-energy 
in the denominator as in Eq.~(\ref{eq:el1}). 
Remark that a resummation of the first order vacuum polarization 
contribution does {\it not} lead to an exponentiated form. 
Assuming exponentiation for the finite parts of the vertex 
and soft photon emission contributions - as occurs for their IR
divergent pieces - leads then to the radiative correction formula  
\begin{equation}
\left( {{d \sigma} \over {d \Omega_e^{'}}} \right)_{VIRTUAL \, \gamma} \;+\;
\left( {{d \sigma} \over {d \Omega_e^{'}}} \right)_{REAL \, SOFT \,
  \gamma} \;=\;\left( {{d \sigma} \over {d \Omega_e^{'}}} \right)_{BORN} 
\; {{e^{ \, \delta_{vertex} \; +\; \delta_{R} } } \over 
{\left( 1 - \delta_{vac} / 2 \right)^2} } \;. 
\label{eq:radcorrtotexp}
\end{equation}

\subsection{Radiative corrections at the proton side and two-photon
  exchange contributions}

In the previous sections, we considered radiative corrections to
elastic electron scattering originating solely from the electron side
(vertex correction and bremsstrahlung) and from the vacuum
polarization. These corrections, which are the dominant ones, can be
calculated model independently and follow from QED. To calculate the
first order radiative corrections originating from the proton side
(proton vertex correction, bremsstrahlung from proton and direct and
crossed two-photon exchange contributions), one needs a model for the
internal structure of the nucleon because one requires knowledge of
off-shell (or half off-shell) $\gamma NN$ vertices. This model
dependence will become important if one aims at a precision of
electron scattering experiments at the 1 \% level. 
To quantify the magnitude of those effects, we refer to the 
recent work of {\it Maximon and Tjon} \cite{MTj99}, where an
initial study was performed of the size of internal structure effects.  
\newline
\indent
In Ref.~\cite{MTj99}, the proton current was taken to have the usual
on-shell form and form factors were included in the calculation. 
The calculation of Ref.~\cite{MTj99} goes beyond previous works
\cite{Tsai61,mo69}, as the proton vertex correction and the bremsstrahlung
from the proton where calculated without approximations within the given
model for the proton current. In the calculation of the direct and
crossed box diagrams (two-photon exchange contributions), a 
less drastic approximation was made in \cite{MTj99} as in 
\cite{Tsai61} (where those box diagrams where only calculated in the
soft-photon approximation). 
\newline
\indent
The calculation of Ref.~\cite{MTj99} yields then the correction
formula for elastic electron scattering :
\begin{equation}
\left( {{d \sigma} \over {d \Omega_e^{'}}} \right)_{TOTAL} 
\,=\,\left( {{d \sigma} \over {d \Omega_e^{'}}} \right)_{BORN} 
\, \left( 1 + \delta_{vac} + \delta_{vertex} + \delta_{R} 
+ Z \, \delta_1 + Z^2 \, (\delta_2^{(0)} + \delta_2^{(1)} 
\right) \;, 
\label{eq:radcorrtotz}
\end{equation}
where $\delta_{vac}$, $\delta_{vertex}$ and $\delta_R$ are given as
above (Eq.(\ref{eq:sumrealvertex})). The terms in
Eq.~(\ref{eq:radcorrtotz}) proportional to $Z$ (hadron charge) and
$Z^2$ contain the corrections from the proton side. 
The correction $\delta_1$, proportional to $Z$, was calculated in
Ref.~\cite{MTj99} as  
\begin{equation}
\delta_1 
=  {2 \alpha_{em} \over \pi} 
\left\{ \; \ln \left( {{4 (\Delta E_s)^2} \over {Q^2 \, x}} \right) 
\ln \eta \,+\, Sp\left( 1 - {{\eta} \over {x}}\right) 
 \,-\, Sp\left( 1 - {1 \over {\eta \, x}}\right)  \right\} ,
\label{eq:deltaz}
\end{equation}
where $\Delta E_s$ and $\eta$ are given as in Eq.~(\ref{eq:deltae}) 
and where the variable $x$ is defined by 
\begin{equation}
x = {{(Q + \rho)^2} \over {4 M_N^2}} \;, \hspace{2cm}
\rho^2 = Q^2 + 4 M_N^2 \;, 
\label{eq:rho}
\end{equation}
The correction proportional to $Z^2$ was split into two parts in
Ref.~\cite{MTj99}. The contribution $\delta_2^{(0)}$, independent of
the nucleon form factors was calculated in Ref.~\cite{MTj99} as :  
\begin{eqnarray}
\delta_2^{(0)} 
&=&  {\alpha_{em} \over \pi} 
\left\{ \, \ln \left( {{4 (\Delta E_s)^2} \over {M_N^2}} \right) 
\left( {{E'_N} \over {|\vec p_N^{\, '}|}} \ln x - 1 \right) \,+\, 1
\right. \nonumber\\
+&&\left.  {{E'_N} \over {|\vec p_N^{\, '}|}} 
\left( - {1 \over 2} \ln^2 x - \ln x \, \ln \left( {{\rho^2} \over {M_N^2}} 
\right) + \ln x - Sp \left( 1 - {1 \over {x^2}}\right)
+ 2 Sp \left( - {1 \over x} \right) + {{\pi^2} \over 6}
\right) \right\},
\label{eq:deltazsqr0}
\end{eqnarray}
where $\rho$ is defined as in Eq.~(\ref{eq:rho}), and 
where $E'_N$ ($|\vec p_N^{\, '}|$) are the {\it lab} energy (momentum) of 
the recoiling nucleon. 
For the lengthier expression of $\delta_2^{(1)}$, which depends on
the nucleon form factors, we refer to Ref.~\cite{MTj99}.   


\newpage

\section{Treatment of singularities}
\label{app:b}

In the numerical calculation of the amplitudes 
for the virtual photon radiative corrections to the 
$e p \to e p \gamma$ reaction, we need to calculate two or three dimensional 
Feynman parameter integrals, as discussed in section~\ref{sec:int}. 
In the  integration over the first variable,
the numerator consists of polynomials and the denominators 
may have some structures
of the form
$ (\alpha ^\prime x + \beta ^\prime \pm i \varepsilon ^\prime )^n$, or
$        (\alpha ^\prime x^2 + \beta ^\prime x + \gamma ^\prime
       \pm i \varepsilon ^\prime)^n  $ with $n$=1,2.
Therefore, in the calculations, the following integrals appear :
\begin{equation}
\label{new1}
\hspace*{-0.5 truecm}
 \lim_{\varepsilon ^\prime \to 0^+}  \,
  \int _a ^ b
  \frac{x^m \ dx}
       {(\alpha ^\prime x + \beta ^\prime \pm i \varepsilon ^\prime )^n}  
   \qquad {\mathrm or } \qquad   
 \lim_{\varepsilon ^\prime \to 0^+}  \,
  \int _a ^ b
  \frac{x^m \ dx}
       {(\alpha ^\prime x^2 + \beta ^\prime x + \gamma ^\prime
       \pm i \varepsilon ^\prime)^n } \;.
\end{equation}
When the denominator has no singularities in the integration
range, it is, in principle,  easy to calculate these integrals
which have the form 
\begin{equation}
\label{new2}
  \int _a ^ b
  \frac{x^m \ dx}
       {(\alpha ^\prime x + \beta ^\prime )^n}  
   \qquad {\mathrm or } \qquad   
  \int _a ^ b
  \frac{x^m \ dx}
       {(\alpha ^\prime x^2 + \beta ^\prime x + \gamma ^\prime)^n } \;.
\end{equation}
Some recurrence relations for these integrals are known 
\cite {Dwight,Gradshteyn}, but for small values of $ \alpha ^\prime $ as
compared to $ \beta ^\prime $ or to $ \gamma ^\prime $, these relations
are unstable numerically. In these cases, we have used either a Taylor 
expansion or the usual  Gauss-Legendre integration method to get very accurate
results.

In the following part of this appendix, we give the relations
used when the denominators in Eq.~(\ref{new1})
have singularities in the integration 
range except in $a$ or $b$. The details are given elsewhere \cite{vdw}. 
The principle of the method is based on the following relation :

\begin{eqnarray}
\label{new2p}
 &&
 \lim_{\varepsilon  \to 0^+}  \,
  \int _a ^ b
  \frac{x^m \ dx}{(x-x_0 \pm i \varepsilon )^n } 
  \nonumber \\ [0.2 truecm]
 &&
 =
  \lim_{\varepsilon  \to 0^+} 
 \lim_{\eta  \to 0^+}
  \Biggl [ \ \
  \int _a ^{x_0 -\eta }   \frac{x^m \ dx}{(x-x_0 \pm i \varepsilon )^n }
  +
 \int _{x_0-\eta } ^{x_0 +\eta } \frac{x^m \ dx}{(x-x_0 \pm i \varepsilon )^n }
  \nonumber \\ [0.2 truecm]
 &&
 \hspace{ 6.5 truecm}
  +
   \int _{x_0 +\eta} ^b   \frac{x^m \ dx}{(x-x_0 \pm i \varepsilon )^n }
    \ \  \Biggr ] \;.
\end{eqnarray}

\noindent
Each integral  can be separated in a real part and an imaginary part and 
we can use for them the analytical expressions given in \cite {Dwight}.

Let us start with the case where the denominator is a polynomial of 
degree 1 in the integration variable. In that case, there is only one 
singularity for $ x_0=- {\beta ^\prime}/{\alpha ^\prime} $ and the 
sign of the imaginary part will depend on the sign of $ \alpha ^\prime$. 
For $ n=1 $ and $\alpha ^\prime > 0 $, we have
\begin{equation}
\label{new3}
 \lim_{\varepsilon ^\prime \to 0^+}  \,
  \int _a ^ b
  \frac{x^m \ dx}
       {\alpha ^\prime x + \beta ^\prime \pm i \varepsilon ^\prime }  
 =
 \frac{1}{\alpha ^\prime} \
 \lim_{\varepsilon  \to 0^+}  \,
  \int _a ^ b
  \frac{x^m \ dx}{x-x_0 \pm i \varepsilon } \;,
  \qquad 
\varepsilon = \frac{\varepsilon ^\prime}{\alpha ^\prime} \, .
 \end{equation}   
When $\alpha ^\prime < 0 $, we have only to replace $ \pm i \varepsilon $
by $ \mp i \varepsilon $ in the right hand side of the Eq.~(\ref{new3}).
We now define the following quantities
\begin{eqnarray}
\label{new4}  
&&J_1 =  \frac{1}{2}
  \log  \frac{ (b-x_0)^2   }
             { (a-x_0)^2   } \;, \nonumber\\
&&J_n= \frac{1}{n-1} \Bigl [ \ (b-x_0)^{n-1} - (a-x_0)^{n-1} \ \Bigr ] \;,
\quad n \geq 2
\end{eqnarray}
to obtain the relations
\begin{eqnarray}
\label{new5}  
&&  \lim_{\varepsilon  \to 0^+}  \,
  \int _a ^ b
  \frac{  dx}{x-x_0 \pm i \varepsilon } 
 = J_1 \ \mp i \ \pi \;, \\ 
&& \lim_{\varepsilon  \to 0^+}  \,
   \int _a ^ b
  \frac{x\  dx}{x-x_0 \pm i \varepsilon } 
= J_2 + x_0 \ J_1 \ \mp \ i \  \pi x_0 \;, \\ 
&& \lim_{\varepsilon  \to 0^+}  \,
   \int _a ^ b
  \frac{x^2 \  dx}{x-x_0 \pm i \varepsilon } 
= J_3 + 2 x_0 \ J_2 + x_0 ^2 \ J_1 \ \mp \ i \pi x_0^2 \;, \\ 
&& \lim_{\varepsilon  \to 0^+}  \,
   \int _a ^ b
  \frac{x^3 \  dx}{x-x_0 \pm i \varepsilon } 
= J_4 + 3 x_0 \ J_3 + 3 x_0 ^2 \ J_2 + x_0 ^3 \ J_1  \ \mp \ i \pi x_0^3 \;, \\ 
&& \lim_{\varepsilon  \to 0^+}  \,
   \int _a ^ b
  \frac{x^4 \  dx}{x-x_0 \pm i \varepsilon } 
= J_5 + 4 x_0 \ J_4 + 6 x_0 ^2 \ J_3 + 4 x_0 ^3 \ J_2 + x_0 ^4 \ J_1
\ \mp \ i \pi x_0^4 \;. \\
&&\cdots \nonumber   
\end{eqnarray}

For $ n=2 $ and $\alpha ^\prime > 0 $, we have 
\begin{equation}\label{new6}
 \lim_{\varepsilon ^\prime \to 0^+}  \,
  \int _a ^ b
  \frac{x^m \ dx}
       {(\alpha ^\prime x + \beta ^\prime \pm i \varepsilon ^\prime )^2 }  
=
\frac{1}{ {\alpha ^\prime }^2 } \
 \lim_{\varepsilon  \to 0^+}  \,
  \int _a ^ b
  \frac{x^m \ dx}{(x-x_0 \pm i \varepsilon )^2 } \;.
\end{equation}

\noindent
When $\alpha ^\prime < 0 $, we have only to replace $ \pm i \varepsilon $
by $ \mp i \varepsilon $ in the right hand side of the Eq.~(\ref{new6}).
We next define the following quantities :
\begin{eqnarray}\label{new7}
I_0 & = &  \frac{1}{a-x_0} - \frac{1}{b-x_0} \;, \\ 
I_1 & = &  \frac{1}{2}  \log  \frac{ (b-x_0)^2   }
                                      { (a-x_0)^2   } \;,       \\        
I_n  & = &
 \frac{1}{n-1} \ \bigl [ \  (b-x_0)^{n-1}  - (a-x_0)^{n-1} \ \bigr ]\;,
\qquad \quad  n\geq 2 \;.
\end{eqnarray}
In terms of these quantities, the integrals of Eq.~(\ref{new1}) with 
$n = 2$ are given by 
\begin{eqnarray}\label{new8}  
&&  \lim_{\varepsilon  \to 0^+}  \,
  \int _a ^ b
  \frac{  dx}{(x-x_0 \pm i \varepsilon )^2}
  = I_0 \;, \\ 
&&  \lim_{\varepsilon  \to 0^+}  \,
  \int _a ^ b
  \frac{ x \ dx}{(x-x_0 \pm i \varepsilon )^2}
= I_1 + x_0 I_0 \mp i \pi \;, \\ 
&&  \lim_{\varepsilon  \to 0^+}  \,
  \int _a ^ b
  \frac{ x^2 \ dx}{(x-x_0 \pm i \varepsilon )^2}
= I_2 + 2 x_0 I_1 + x_0 ^2 I_0 \mp i \ 2 \pi x_0 \;, \\ 
&&  \lim_{\varepsilon  \to 0^+}  \,
  \int _a ^ b
  \frac{ x^3 \ dx}{(x-x_0 \pm i \varepsilon )^2}
= I_3 + 3x_0 I_2 + 3 x_0 ^2 I_1 + x_0 ^3 I_0 \mp i \ 3 \pi x_0 ^2 \;, \\ 
&& \cdots    \nonumber
\end{eqnarray}

We can notice that the real part of these integrals for
$n=1$ as well as for $ n=2$ can be derived from the binomial expansion
$ (x_0 + X)^m$. In the case of $ n=1$, the imaginary part is
proportional to $ \pi f(x_0) $ where $ f(x)$ is the numerator
of the integrand. For $ n=2$, it is straightforward
to show \cite{vdw} that the imaginary part is proportional to 
$ \pi f'(x_0) $.

When the form of the denominator is
$  {(\alpha ^\prime x^2 + \beta ^\prime x + \gamma ^\prime
       \pm i \varepsilon ^\prime)^n } $, 
i.e. a polynomial of degree 2 in the integration variable, 
it is always possible to come back to the preceding cases. When 
$ \alpha ^\prime > 0$, we have
\begin{equation}
\label{new9}
\ \hspace{-0.9 truecm}
 \lim_{\varepsilon ^\prime \to 0^+}  \,
  \int _a ^ b
  \frac{x^m \ dx}
       {(\alpha ^\prime x^2 + \beta ^\prime x + \gamma ^\prime
       \pm i \varepsilon ^\prime )^n }  
=
\frac{1}{ {\alpha ^\prime }^n } \
 \lim_{\varepsilon  \to 0^+}  \,
  \int _a ^ b
  \frac{x^m \ dx}{( x^2 + \beta x + \gamma  \pm i \varepsilon )^n } \;,
\end{equation}  
with the following definitions :
\begin{equation}
\label{new10}
\beta=\frac{\beta ^\prime}{\alpha ^\prime} \;,
\qquad \gamma= \frac{\gamma ^\prime}{\alpha ^\prime} \;,
\qquad \varepsilon = \frac{\varepsilon ^\prime }{\alpha ^\prime} \;.
\end{equation}
The integrand in Eq.~(\ref{new9}) has some singularities when 
$ \delta=\beta ^2 - 4 \gamma $ is positive. 
When $\alpha ^\prime < 0 $, we have only to replace $ \pm i \varepsilon $
by $ \mp i \varepsilon $ on the right hand side of Eq.~(\ref{new9}). 

It can be shown \cite{vdw} that 
\begin{equation}\label{new11}
\ \hspace{-0.9 truecm}
 \lim_{\varepsilon  \to 0^+}  \,
  \int _a ^ b
  \frac{ x^m \ dx }{ (x^2 + \beta x + \gamma \pm i \varepsilon )^n }
=
 \lim_{\tilde {\varepsilon } \to 0^+}  \,
  \int _a ^ b
\frac{ x^m \ dx }
     {(x-x_{+}^{R}+i\tilde {\varepsilon } )^n \ 
      ( x-x_{-}^{R}-i\tilde {\varepsilon } )^n } \;,
\end{equation}
with the definitions 
\begin{equation}\label{new12}
x_{+}^{R} = \frac{-\beta + {\sqrt{\delta } } }{2} \;,
\qquad
x_{-}^{R} = \frac{-\beta - {\sqrt{\delta } } }{2} \;,
\qquad
\tilde {\varepsilon } = \frac{2 \varepsilon}{\delta} \;.
\end{equation}
These integrals can be easily calculated using the decomposition of the
fraction into elementary fractions. For $n=1$, we obtain 
\begin{eqnarray}\label{new13}
 && 
\lim_{\varepsilon  \to 0^+}  \,
  \int _a ^ b
  \frac{ x^m \ dx }{ x^2 + \beta x + \gamma \pm i \varepsilon  }
  \nonumber \\ [0.2 truecm]
 &&  = \hspace{0.4 truecm}   
 \frac{1}{\sqrt{\delta}} \
 \lim_{\tilde {\varepsilon } \to 0^+}  \,
  \int _a ^ b
\frac{ x^m \ dx }
     {x-x_{+}^{R} \pm i\tilde {\varepsilon } }
\, \, 
- \frac{1}{\sqrt{\delta}} \
 \lim_{\tilde {\varepsilon } \to 0^+}  \,
  \int _a ^ b
\frac{ x^m \ dx }
     {x-x_{-}^{R} \mp i\tilde {\varepsilon } } \;,
\end{eqnarray}
and for $n=2$ 
\begin{eqnarray}
\label{new14}
&& \lim_{\varepsilon  \to 0^+}  \,
  \int _a ^ b
  \frac{ x^m \ dx }{ (x^2 + \beta x + \gamma \pm i \varepsilon )^2  }
  \nonumber \\ 
&&  = \hspace{0.4 truecm} 
 \frac{1}{\delta} \
 \lim_{\tilde {\varepsilon } \to 0^+}  \,
  \int _a ^ b
\frac{ x^m \ dx }
     {(x-x_{+}^{R} \pm i\tilde {\varepsilon })^2 }
\, \, 
-  \frac{2}{\delta ^{3/2} } \
 \lim_{\tilde {\varepsilon } \to 0^+}  \,
  \int _a ^ b
\frac{ x^m \ dx }
     {x-x_{+}^{R} \pm i\tilde {\varepsilon } }
  \nonumber \\ 
&&   \hspace{0.4 truecm} 
+ \ \frac{1}{\delta} \
 \lim_{\tilde {\varepsilon } \to 0^+}  \,
  \int _a ^ b
\frac{ x^m \ dx }
     {(x-x_{-}^{R} \mp i\tilde {\varepsilon })^2 }
\, \, 
+  \frac{2}{\delta ^{3/2} } \
 \lim_{\tilde {\varepsilon } \to 0^+}  \,
  \int _a ^ b
\frac{ x^m \ dx }
     {x-x_{-}^{R} \mp i\tilde {\varepsilon } }    \;. 
\end{eqnarray}


\newpage

\begin{table}
\caption{Radiative corrections to elastic electron-proton
  scattering for MAMI and JLab kinematics. 
First column : $E_e$ in GeV, second column : $\theta_e$ 
in deg, third column : $Q^2$ in GeV$^2$. See text for details on the
different contributions. The real radiative 
corrections are calculated with $(E_e'^{el} - E_e')$ = 0.01 $E_e$. 
The total radiative correction (to first order) is indicated by 
$\delta_{tot}$, and the exponentiated (EXP) result 
(except for the vacuum polarization contribution, see text) is
shown in the last column.}
\label{tab1}
\begin{center}
\begin{tabular}{ccc|ddddd|dd}
& & & & & & & & & \\
$E_e$ & $\theta_e$ & $Q^2$ & $\delta_{vertex}$ & $\delta_{vacpol}$
& $\delta_R$ & $\delta_1$ & $\delta_2^{(0)}$ & $\delta_{tot}$ & EXP. \\
& & & & & & & & & \\
\tableline
& & & & & & & & & \\
0.705 &40.66 &0.203 &-0.1673 &0.0208 &-0.0453 &-0.0067 &-0.0018 &-0.2003 &-0.2025 \\
0.855 &52.18 &0.418 &-0.1881 &0.0228 &-0.0245 &-0.0123 &-0.0034 &-0.2054 &-0.2087 \\
& & & & & & & & & \\
4.000 &15.43 &1.000 &-0.2149 &0.0254 &-0.0260 &-0.0046 &-0.0055 &-0.2255 &-0.2277 \\ 
4.000 &23.82 &2.000 &-0.2374 &0.0275 &0.0018 &-0.0107 &-0.0096 &-0.2285 &-0.2322 \\
4.000 &32.45 &3.000 &-0.2511 &0.0287 &0.0300 &-0.0180 &-0.0128 &-0.2232 &-0.2292 \\
4.000 &42.91 &4.000 &-0.2611 &0.0296 &0.0623 &-0.0265 &-0.0150 &-0.2106 &-0.2200 \\
& & & & & & & & & \\
6.000 &14.93 &2.000 &-0.2374 &0.0275 &-0.0097 &-0.0062 &-0.0089 &-0.2348 &-0.2371 \\
6.000 &19.40 &3.000 &-0.2511 &0.0287 &0.0092 &-0.0103 &-0.0121 &-0.2355 &-0.2390 \\
6.000 &23.96 &4.000 &-0.2611 &0.0296 &0.0284 &-0.0149 &-0.0146 &-0.2326 &-0.2376 \\
6.000 &28.95 &5.000 &-0.2689 &0.0303 &0.0490 &-0.0200 &-0.0166 &-0.2263 &-0.2334 \\
6.000 &34.76 &6.000 &-0.2754 &0.0308 &0.0718 &-0.0257 &-0.0181 &-0.2165 &-0.2261 \\
& & & & & & & & & \\
\end{tabular}
\end{center}
\end{table}

\newpage

\newpage

\newpage

\newpage

\newpage

\newpage

\newpage

\newpage

\newpage

\newpage

\begin{figure}
\begin{center}
\leavevmode
\hbox{
\epsfxsize=14.0truecm
\epsfbox{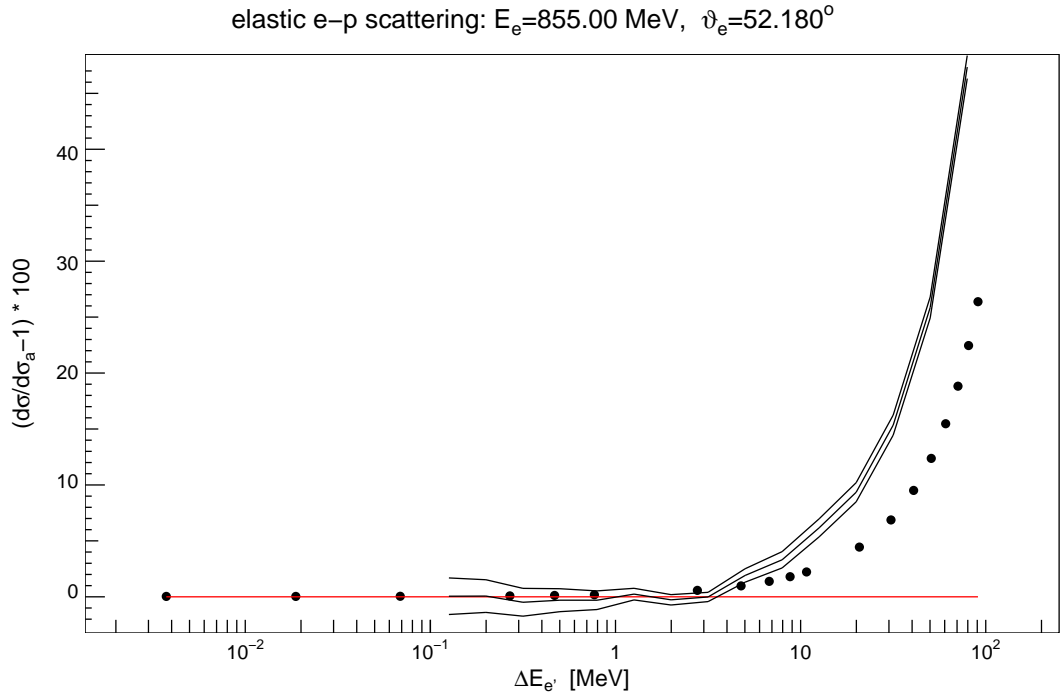} }
\end{center}
\centerline{\small \it  }
\vspace{-0.5cm}
\caption{Radiative tail of elastic electron-proton scattering at  
$E_e$ = 855.00 MeV and $\theta_e$ = 52.18$^o$. 
A comparison is shown between fully numerical calculation (indicated
by the points) and the simulation (curves, see text).}
\label{fig:radtail_compare}
\end{figure}

\begin{figure}
\begin{center}
\leavevmode
\hbox{
\epsfxsize=14.0truecm
\epsfbox{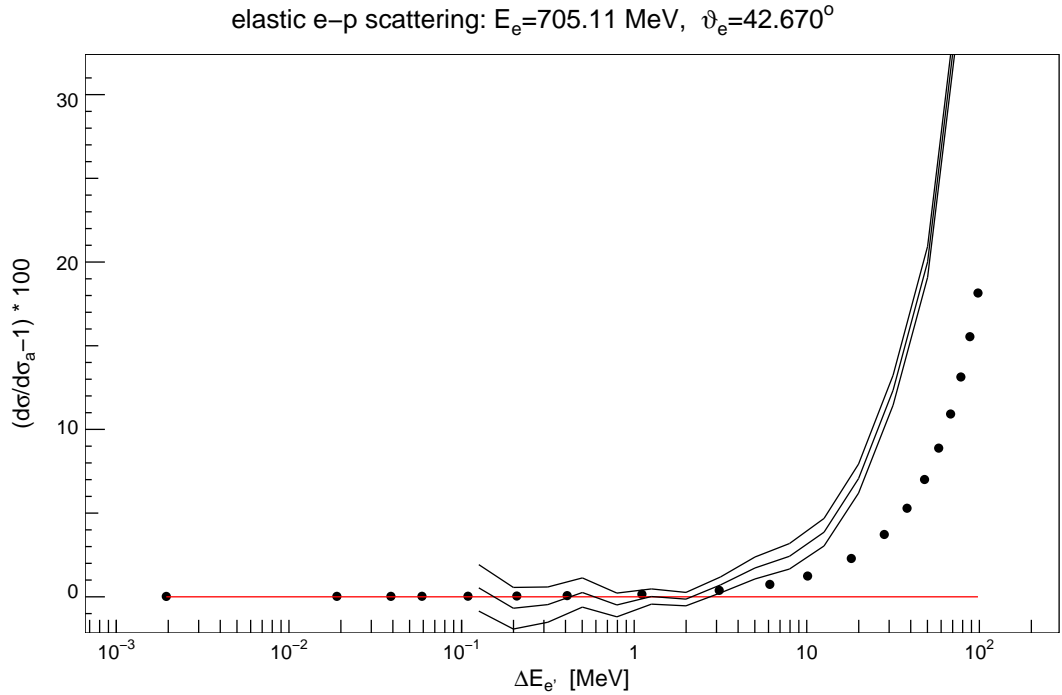} }
\end{center}
\centerline{\small \it  }
\vspace{-0.5cm}
\caption{Same as for Fig.~\ref{fig:radtail_compare}, but for 
elastic electron-proton scattering at 
$E_e$ = 705.11 MeV and $\theta_e$ = 42.67$^o$.} 
\label{fig:radtail_compare2}
\end{figure}

\newpage

\begin{figure}
\begin{center}
\leavevmode
\hbox{
\epsfysize=15.0truecm
\epsfbox{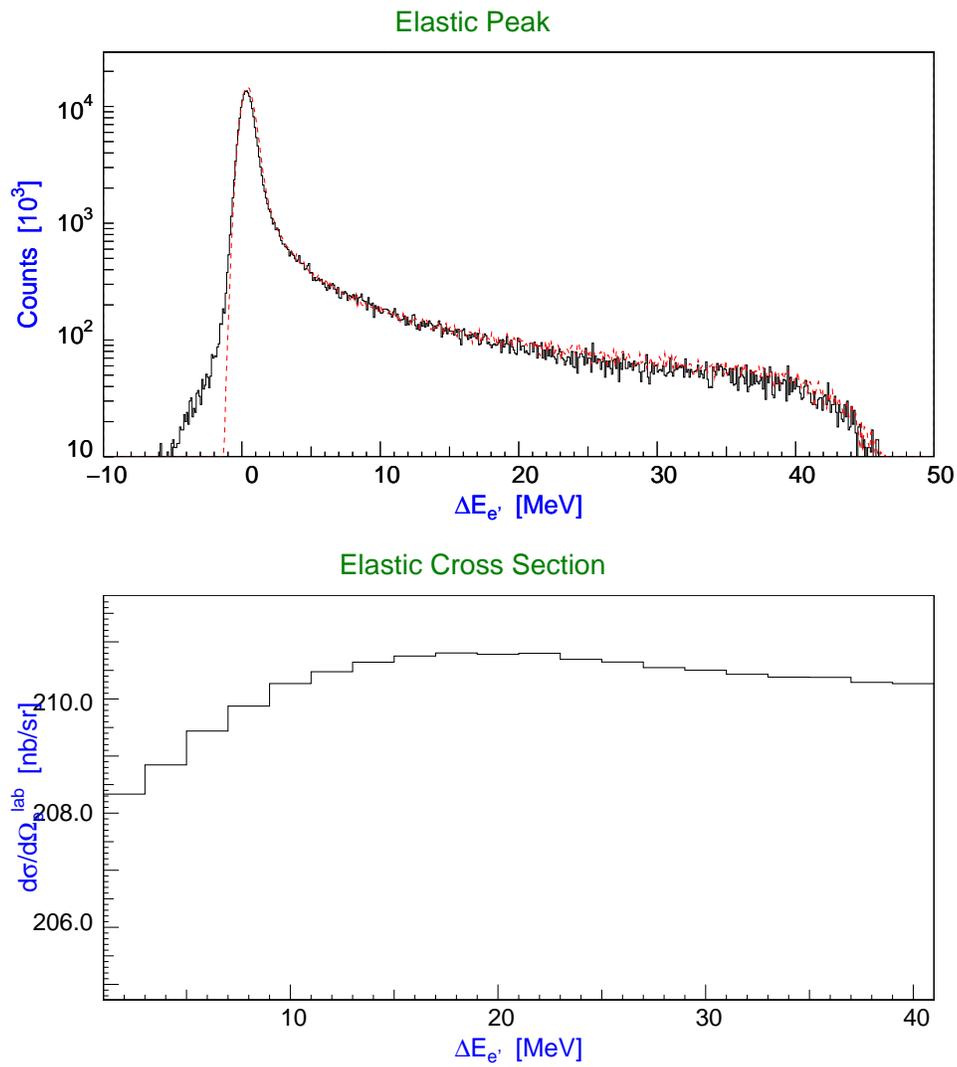} }
\end{center}
\centerline{\small \it  }
\vspace{-0.3cm}
\caption{Determination of the elastic cross section for the kinematics
$E_e$ = 705.11 MeV and $\theta_e$ = 42.6$^o$.}
\label{fig:plateau}
\end{figure}

\newpage

\begin{figure}[h]
\begin{center}
\leavevmode
\hbox{%
\epsfysize=15.0truecm
\epsfbox{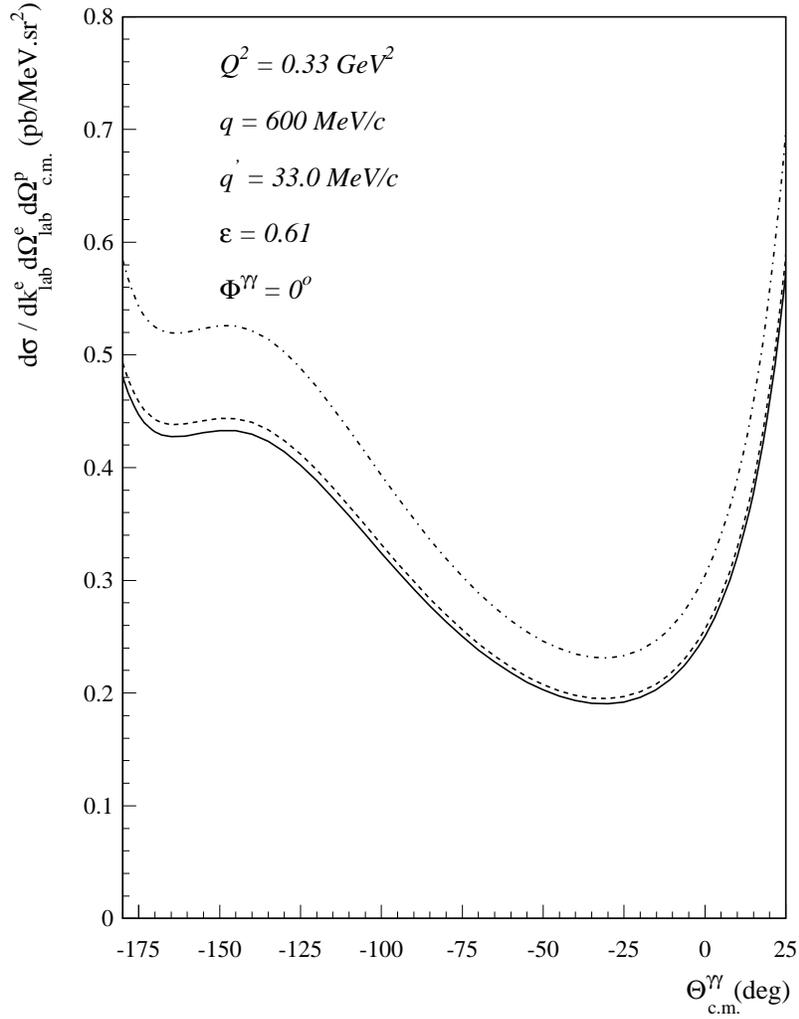} }
\end{center}
\centerline{\small \it  }
\vspace{-0.75cm}
\caption{Differential $e p \to e p \gamma$ cross section for MAMI 
kinematics at $q^{'}$ = 33 MeV/c. 
Dashed-dotted curve : BH + Born contribution, 
dashed curve : BH + Born + virtual radiative correction, 
full curve : BH + Born + total radiative correction. 
The real radiative correction is shown here for a maximal
soft-photon energy of $\Delta E_s$ = 10 MeV.}
\label{fig:cross33_mami}
\end{figure}

\newpage

\begin{figure}[h]
\begin{center}
\leavevmode
\hbox{
\epsfysize=15.0truecm
\epsfbox{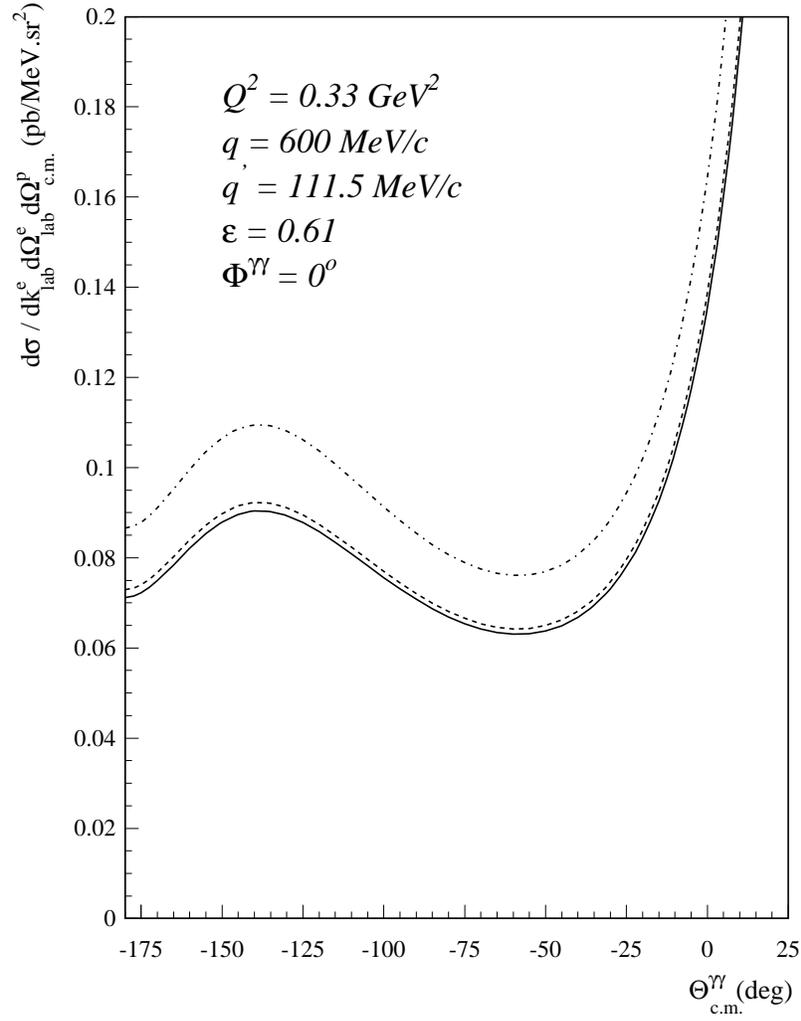} }
\end{center}
\centerline{\small \it  }
\vspace{-0.75cm}
\caption{Differential $e p \to e p \gamma$ cross section for MAMI 
kinematics at $q^{'}$ = 111.5 MeV/c. Curve conventions as in 
Fig.\protect\ref{fig:cross33_mami}. The real radiative correction 
is shown here for a maximal soft-photon energy of $\Delta E_s$ = 10 MeV.}
\label{fig:cross112_mami}
\end{figure}

\newpage

\begin{figure}[h]
\begin{center}
\leavevmode
\hbox{
\epsfysize=15.0truecm
\epsfbox{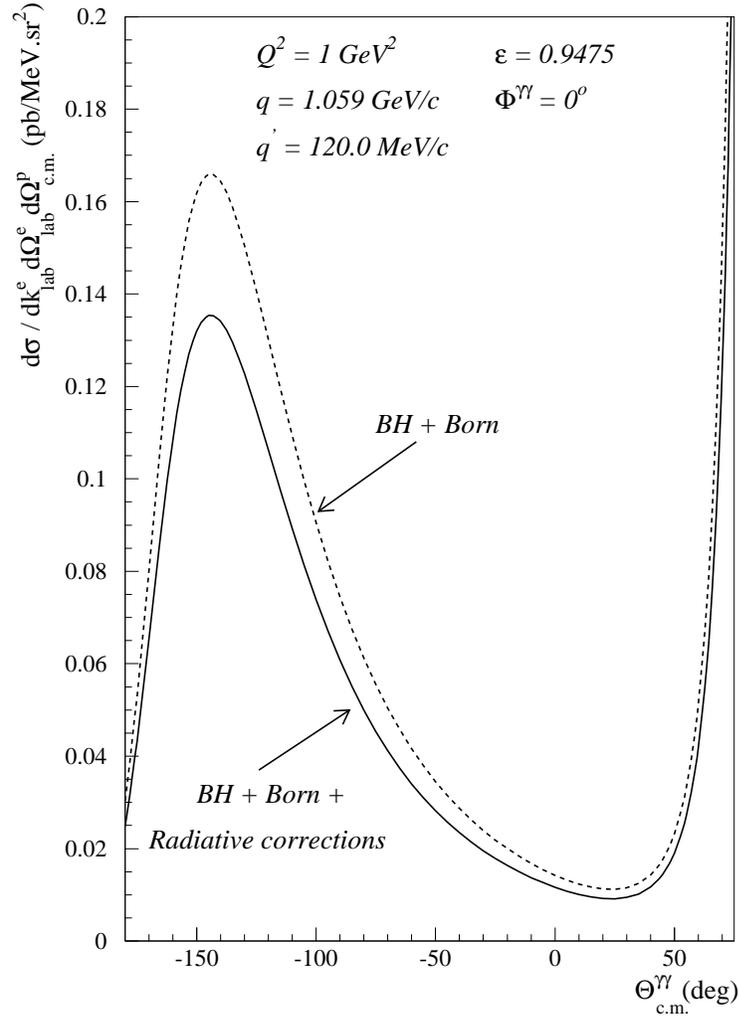} }
\end{center}
\centerline{\small \it  }
\vspace{-0.75cm}
\caption{Differential $e p \to e p \gamma$ cross section for JLab
  kinematics at $q'$ = 120 MeV/c. The BH + Born result is compared
  with the result including virtual radiative corrections.}
\label{fig:cross_jlab}
\end{figure}

\newpage

\begin{figure}[Hbtp]
\begin{center}
\leavevmode
\hbox{
\epsfysize=16.0truecm
\epsfbox{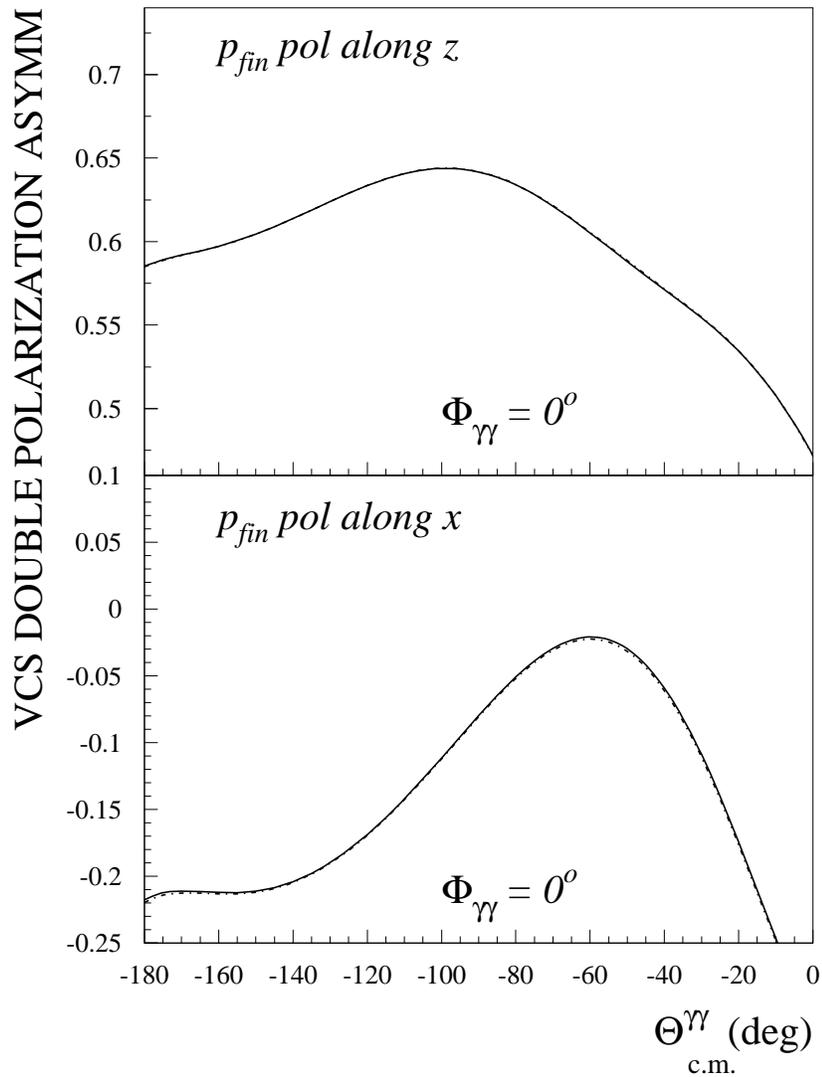} }
\end{center}
\centerline{\small \it  }
\vspace{-0.75cm}
\caption{Double polarization asymmetry for VCS with proton 
polarized along the z-axis (upper panel) 
or polarized along the x-axis (lower panel) for MAMI kinematics. 
Dashed-dotted curve : BH + Born, 
full curve : BH + Born + radiative corrections 
(both curves nearly coincide!).} 
\label{fig:asymm_mami}
\end{figure}

\newpage

\begin{figure}[Hbtp]
\begin{center}
\leavevmode
\hbox{%
\epsfysize=15.0truecm
\epsfbox{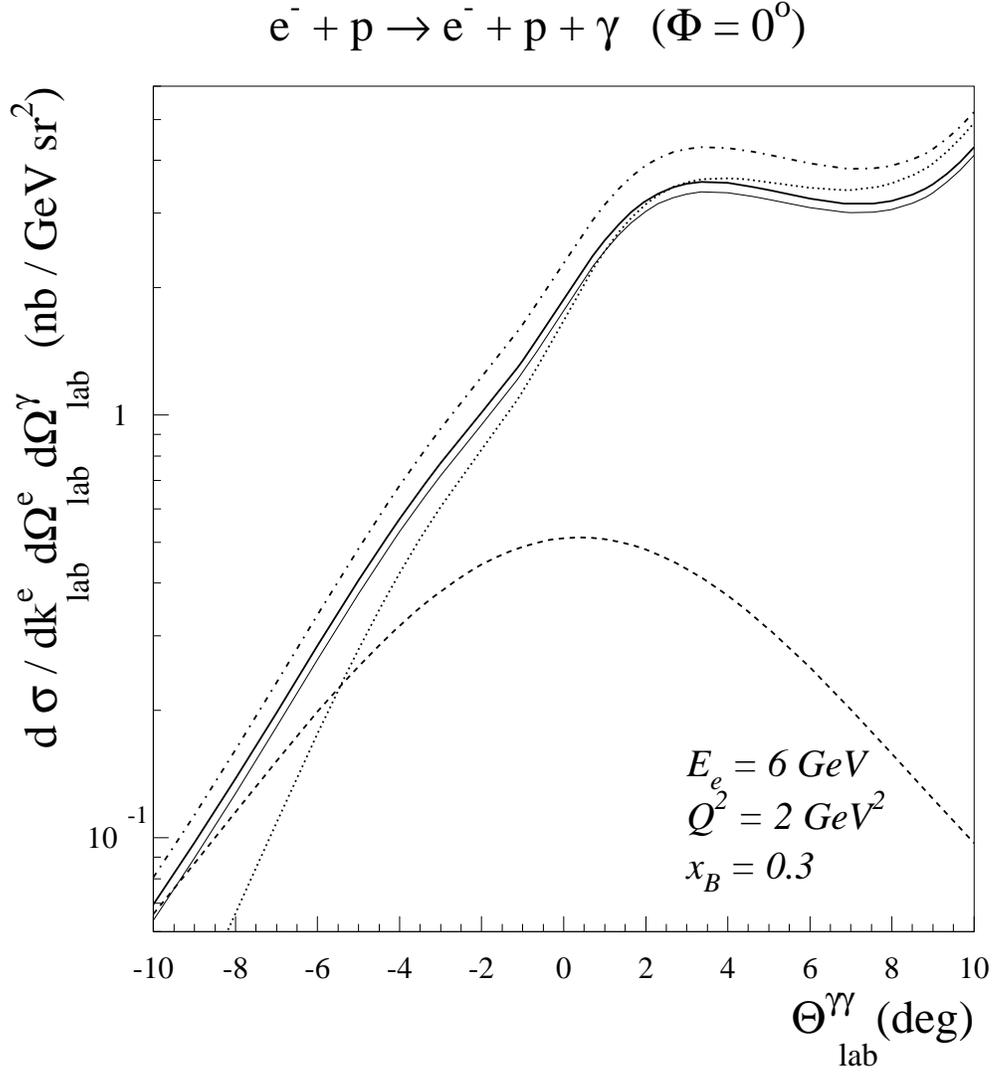} }
\end{center}
\centerline{\small \it  }
\vspace{-0.75cm}
\caption{Differential $e p \to e p \gamma$ cross section in {\it lab} 
: DVCS kinematics at JLab. 
Dotted curve : BH, dashed curve : DVCS, 
dashed-dotted curve : BH + DVCS, 
thin full curve : BH + DVCS + virtual radiative corrections. 
The thick full curve represents the BH + DVCS + virtual and real radiative
corrections, where the real radiative corrections are 
calculated with $\Delta E_s$ = 0.1 GeV, which corresponds with a cut
in the missing mass spectrum (Eq.~(\ref{eq:missmass1})) of  
$M_{m1}^2$ - $M_N^2 \simeq$ 0.21 GeV$^2$.}
\label{fig:dvcs_jlab}
\end{figure}

\newpage

\begin{figure}[Hbtp]
\begin{center}
\leavevmode
\hbox{%
\epsfysize=15.0truecm
\epsfbox{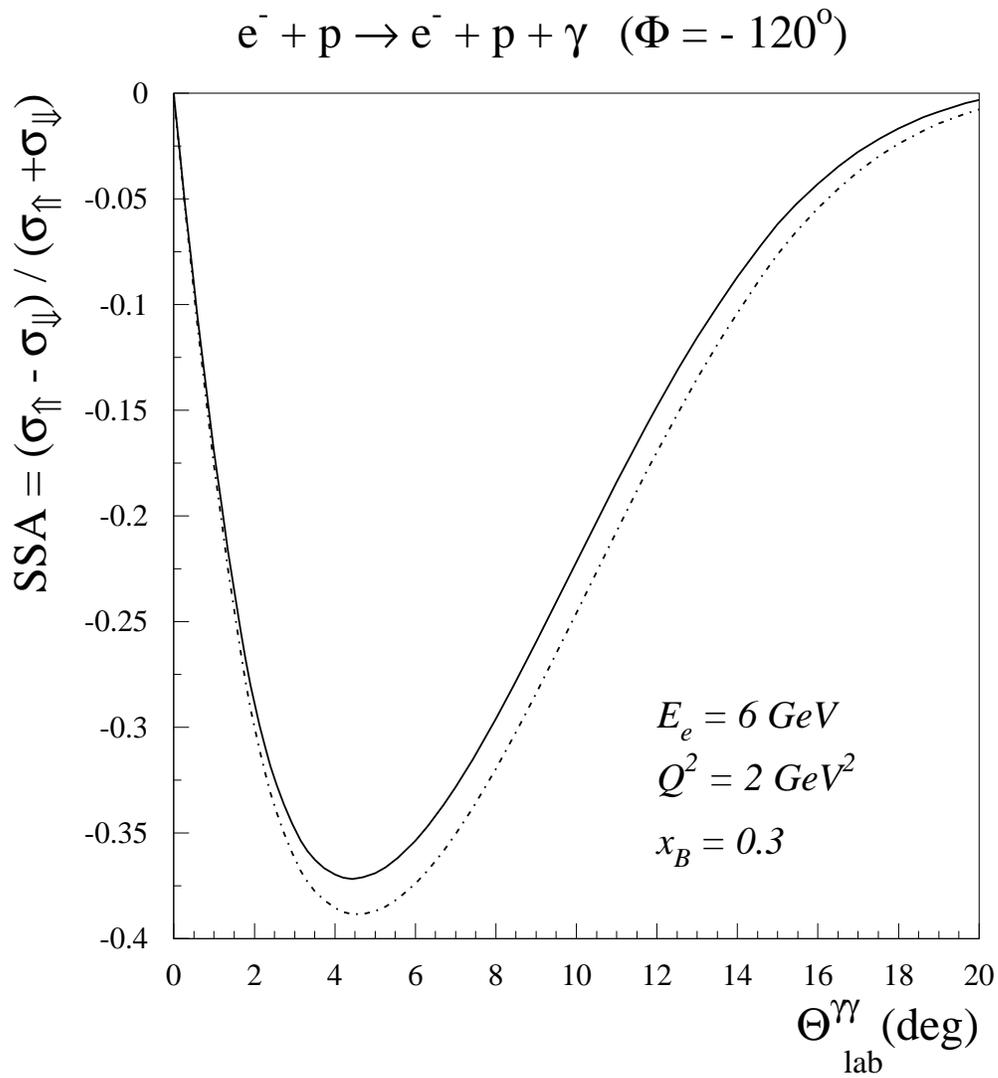} }
\end{center}
\centerline{\small \it  }
\vspace{-0.75cm}
\caption{Electron single spin asymmetry : DVCS kinematics at JLab. 
Dashed-dotted curve : BH + DVCS, 
full curve : BH + DVCS + radiative corrections.}
\label{fig:ssa_jlab}
\end{figure}

\newpage

\begin{figure}[Hbtp]
\begin{center}
\leavevmode
\hbox{%
\epsfysize=16.0truecm
\epsfbox{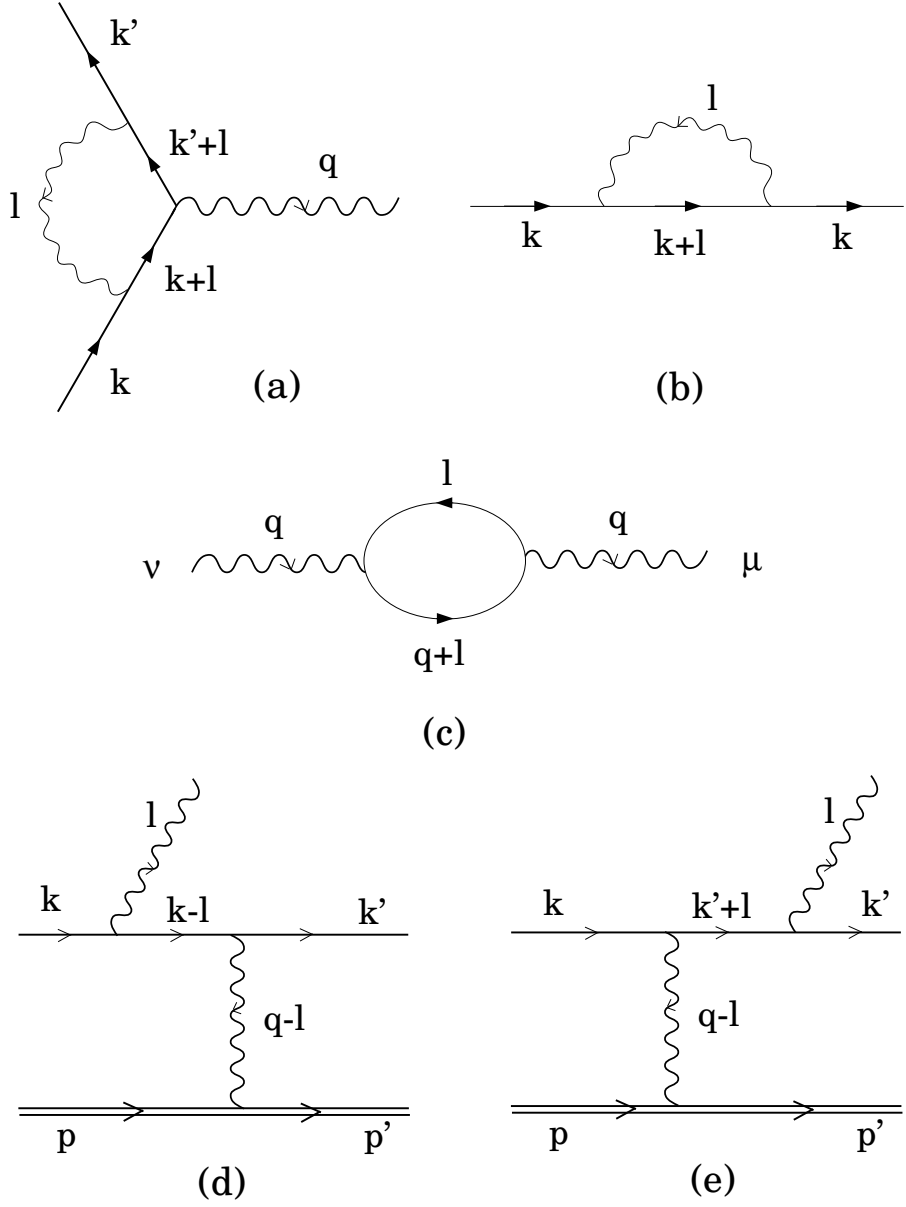} }
\end{center}
\centerline{\small \it  }
\vspace{-0.5cm}
\caption{First order virtual and real radiative correction processes : 
(a) vertex diagram, (b) lepton self energy diagram, (c) photon 
polarization diagram, (d) and (e) soft-photon emission contributions
to elastic lepton-nucleon scattering.}
\label{fig:elasticdia}
\end{figure}

\end{document}